\documentclass[aip,graphicx]{revtex4-1}
\pdfoutput=1

\usepackage[english]{babel}
\usepackage{graphicx}
\usepackage{epsf}
\usepackage{amstext}
\usepackage{amsmath}
\usepackage{amssymb}
\usepackage{subfigure}
\usepackage{flafter}
\usepackage{placeins}
\usepackage{appendix}
\usepackage{float}
\restylefloat{figure}

\begin{document}

\title{Geometrical aspects of the interaction between expanding clouds and
environment}



\author{F. Spineanu}
\email[]{florin.spineanu@euratom.ro}
\author{M. Vlad}
\author{D. Palade}
\affiliation{National Institute of Laser, Plasma and Radiation Physics Bucharest, Romania}


\date{\today}

\begin{abstract}
This work is intended to be a contribution to the study of the morphology of the rising
convective columns, for a better representation of the processes of entrainment and
detrainment. We examine technical methods for the description of the
interface of expanding clouds and reveal the role of \emph{fingering}
instability which increases the effective length of the periphery of the
cloud. Assuming Laplacian growth we give a detailed derivation of the
time-dependent conformal transformation that solves the equation of the 
\emph{fingering} instability. For the phase of slower expansion, the
evolution of complex poles with a dynamics largely controlled by the Hilbert
operator (acting on the function that represents the interface position)
leads to \emph{cusp} singularities but smooths out the smaller scale
perturbations.

We review the arguments that the rising column cannot preserve its integrity
(seen as compacity in any horizontal section), because of the penetrative
downdrafts or the incomplete repulsion of the static environmental air
through momentum transfer. Then we propose an analytical framework which is
adequate for competition of two distinct phases of the same system.

The methods exmined here are formulated in a general framework and can be easily adapted to particular cases of atmospheric convection.
\end{abstract}


\maketitle 

\section{Introduction}

In the atmospheric convection leading to cumulus clouds there is a
continuous exchange of heat and water (vapor, liquid) between the rising
column and the environmental air. This problem is complex and has been examined by observation, analytical theory and numerical simulation \cite{blyth}, \cite{Carpenter1998}, \cite{brethertonpark}, \cite{siebesmacuijpers}, \cite{houzeclouds}. The diversity of situations cannot be captured by a unique model.

The rate of exchange depends, besides many
other factors, on the geometry of the contact interface between the two
gaseous media. Part of the interaction with the environment takes place at
the periphery of the cloud, an interface that evolves and is subject to
geometric instabilities. In the expanding phase the small scale structure is
determined by instabilities of the \emph{fingering} type. Like in the case of the Hele-Shaw instability, one can map the physical problem on a time-dependent conformal transformation to a fixed complex line. At small
scale, there can be random fluctuations due to the background turbulence.
The effective length of contact between the cloud and the environment is
then much greater than the perimeter of \emph{hull} of the cloud in
horizontal plane. This enhances the transport between the two neighboring
gaseous media.

In the phase where the expansion slows down (the input from the rising flux
is reduced) the small scale profiles (resulted from \textquotedblleft
fingering\textquotedblright\ and random fluctuations) are smoothed out by a
process of attraction and alignement of the complex poles that define the solution. The
result is that the interface exhibits\emph{\ singularities} of a special
type, \emph{cusps}. They occur through coalescence of small scale
quasi-singular perturbations of the wrinkled interface. If the time
evolution leads to self-intersection of the interface, a parcel of
environmental air is simply swallowed into the expanding cloud.

\bigskip

Finally, inside the rising column there is interaction between the cloud air
and the environmental air that either remains by breaking the initially
compact rising column, or penetrates from above as downdrafts on long
vertical distances.

\bigskip

This last case requires few comments. There are various forms of contact
between the two gaseous media that are in relative motion. In a simplified descriptive
picture we can see the rising column as a stream of gas with specific
properties penetrating a volume of stationary gas (static environmental air)
with different properties. If the speed of the stream is high, there is
transfer of momentum from the stream to the the environmental air through
collisions at the front of contact, {\it i.e.} 
at the top of the rising column. Then the environmental air is deflected (pushed up and sidewise) and
the body of the column preserve its properties.

Alternatively, if the linear momentum of the rising column is small, then
the stream of the convective air cannot repulse through momentum transfer
the environmental air and there is easier interpenetration of the two gases. The rising stream is teared apart and
elements of the air of the column are interspersed between elements of the
environmental air and this occurs up to small spatial scales, of the
turbulence. Obviously, the exchange of heat and water (vapor and liquid) is
much more efficient in this case. This is the case of either shallow
convective events that are dissipated before becoming a buoyant column or
(and this is most interesting) of clouds in the last phase of their ascent
where the vertical advancement is slowed down and elements of the cloud are
dispersed in the environment.

For a comparison, the two cases have been found experimentally in the
expansion of a laser-generated plasma in a low density plasma \cite%
{gasinterpenetration}. With variation within a range of parameters the
interaction changes from interpenetration to formation of an interface.

These two situations are limiting cases. In general it is expected that the
rising convection columns that form the cumuli are characterized by strong
buoyancy that allows them to reach high altitude. Therefore they are closer
to the first case described above, where the stream is able to push off (out
of) its way the environmental air and the contact with this one mainly takes
place at the moving top and at the circumference. Even if the vertical
momentum of the rising column is high, the possibility to retain a compact
structure is improbable. More realistic is the expectation that the
initially compact column will break up into streams separated by irregular
but vertically-connected volumes (channels) of static environmental air. In
short, the breakup of the rising column leaves open spaces inside the
initially compact area and in these spaces there still is static
environmental air. The transfer of mechanical momentum from the rising cloud
air to the static environment is mediated by exchange of turbulent eddies,
therefore implicitely involves dilution of cloudy air and evaporative
cooling for the environmental air. This occurs mainly at the top of the
rising column. Parcels of the cooled environmental air will descend and will
push the still un-mixed environmental air present in the contiguous channels
remaining inside the broken column. The mass conservation requires the
environment to respond to the rise of the convective air by currents of
descending air, which takes place in a layer around the rising column but
also in the spaces left open after the breakup of the rising column.
Therefore, elements of rising air that are at a certain level height\ will
be in close contact with environmental air that, actually, originates at
levels of height that are much above the current level and are \emph{not}
yet mixed with cloud air. Then the \emph{mixing} will take place at various
heights. This picture is very close to what is observed in a classical
Reyleigh-Benard system in the transitory phase where the purely conductive
state is going to be replaced by the convective state. This bifurcation is
preceded by emission of thermal streams (plumes) at the hot plate, which do
not have the chance to produce a full scale convection. However they are
able to determine thorugh mass and momentum conservation, generation of
opposite streams, originating at the cold plate and descending. There is no
mixing between the rising and descending plumes except at late stages, when
the convection span the whole volume. Parcels of environmental undiluted air, of large
dimensions (up to $500$ $\left( m\right) $) have been observed at all levels in the cloud.

\bigskip

This is the situation that we have in mind when we consider the possible
downdraft of environmental air not yet mixed with cloud air. A parcel of
environmental air can be pushed to descent by the request of mass and
momentum conservation, to compensate the rise of streams of rising cloud
air, in a structure of the broken column characterized by coexistance of
buoyancy-driven ascent and vertical channels of environmental air. We
therefore note that the latter may be not yet cooled by the evaporation of
the water after mixing with cloud air close to the top.

Therefore we propose to include in the physical picture the downdraft of
pure environmental air, not yet mixed with the cloud air, resulting from the
constraints of mass and momentum conservation. The downdrafts are located
inside the broken cloud column. In addition, after mixing, the new downdraft
will be cooled by evaporation and the mixed parcel will descent even more,
up to the cloud base.

During the rise there is a smooth change from the situation of strong
stream, specific to the initial stages and the soft slowing down,
characteristic of the phase where the column reaches the highest level. In
the first phase the contact between the convective column and the
environment takes place at the top, at the interface with non-mixed
downdraft and at the periphery and in the last phase there is less momentum
transfer and easier mutual interpenetration of parcels of air. The exchange
of heat and water is less efficient in the first phase and is highly
efficient in the last phase, which actually accelerates the process of loss
of buoyancy.

This approximative and descriptive picture suggests to treat separately the
two situations. In the first one, the exchanges between the convective column
and the environment require to examine the periphery. Close to the final
slowing down of the ascending column, the exchanges involve a large volume
where parcels of similar dimension of cloud and environmental air are
intermingled. Since the top of the column has been in contact with the
environment all along the rise, one must respresent the breakup of the
rising column into smaller columns and the presence, between them, of
columns belonging to the initially static environment or downdrafts
resulting from evaporative cooling of parcels of mixed air.

These two situations will be our subject.
For the problem of cloud periphery, we will discuss the possible evolutions of the
interface between the expanding gas and the surrounding air: the shape of
the interface, generated the fingering
instability and random fluctuations. As expanding front, we examine the formation of {\it cusp} singularities.

For the problem of breaking of the rising column, we will discuss the
statistics of phase competition. We find that, in this schematic
representation, based essentially on geometric aspects of competition of two
distinct phases of the same gas, it is relevant to discuss in terms of \emph{%
labyrinth} structured in the horizontal plane. 

Therefore we have simultaneously a problem of separation of phases and a
problem of interface dynamics. For tractability we divide the problem into
two different components: the interface dynamics for each component, as well
as for the large boundary circumscribing the full horizontal section of the
convective column, is investigated with methods of pole dynamics and/or
wrinkled advancing fronts; the phase separation is treated with method of
coupled lattice maps, in order to follow easily the breaking of the compact
raising column; and, separately.

\bigskip

\section{Geometry of the interface between the rising column and the
environment}

\subsection{The exterior interface}

\subsubsection{Introduction}

It is considered that the the circumference of the cloud at a fixed altitude 
$z$ has an important role in the exchange of heat and water vapors with the
environment. The turbulent diffusion sustained by random eddy exchanges
introduces environmental air into the mass of the cloud and there the
exchange of heat / vapor modifies the tendency of the column to rise. During
the phase of rise the edge of the convective column has a vertical motion relative to the
static environment, which creates a layer of vorticity at the interface.
There are two mechanisms that can affect this interface, depending on the
relative velocity and viscosity. The Kelvin-Hlemholtz instability can get a
positive growth and peripheric elements of rising column are rolled up to
create the known \textquotedblleft cats-eye\textquotedblright\ pattern in a
vertical-plane section, {\it i.e.} a ring vortex with vertical principal axis. Alternatively,
a combined lateral expansion and rise of the column can produce the ring
vortex at the head of the column, \emph{i.e.} due to the inertial resistance
of the static environment, a typical moshroom head (well known from Rayleigh
- Taylor instability). In both cases parcels of environmental air are
absorbed and entrained being surrounded by cloudy air, thus facilitating the
mixing \cite{govindarajan}.

If at the circumference the transport processes associated with the
entrainment depend on the area of contact between the cloud and the
environment then a careful representation of this area is necessary.
Equivalently, at fixed altitude $z$, a good representation of the geometry
of the interface cloud/environment is necessary. We will discuss small scale and respectively large scale structure. The small scale structure of the interface is generated by two processes: (1)
deterministic instabilities, like \emph{fingering}; (2) random perturbation,
related to the turbulence. On a large scale, the structure of the interface
can lead the cloud to incorporate (swallow) parcels of environmental air.
The mechanism originates in - and is a limiting form of, - the \emph{cusp}
singularity that is formed as an asymptotic organization of the small scale
quasi-discontinuities.

\subsubsection{The small scale structure: the \textquotedblleft
fingering\textquotedblright\ instability of the interface}

The interface between the air of a rising column and the environment air
shows a specific profile. In every horizontal plane there is a fluid (cloud air) expanding
into a static fluid (the environment). An universal model for such process
assumes that the velocity of the interface is derived from the gardient of a
scalar function that verifies the Laplace equation (is harmonic function).
Then the interface is subject to \textquotedblleft \emph{fingering%
\textquotedblright\ instability}. The role of the Laplacian field is played
by the pressure of the expanding gas. We need an analytical instrument to describe the small scale
breaking of the continuity of the derivative of the line of the interface.
This wiggled profile is the place where the exchange of heat and water vapor
takes place.

The two \ fluids (cloud and external environment) are assimilated with two
different phases separated by a moving interface $\Gamma \left( t\right) $,
a curve in the physical plane of coordinates $\left[ X\left( t\right)
,Y\left( t\right) \right] $. In a simplified representation, the interface
is a line that extends between $-\infty $ and $+\infty $, \emph{i.e.} it
separates two regions of the plane. The region I is the inside of the cloud,
limited by $\Gamma \left( t\right) $ and the region II is outside, the
environment. It is assumed that the velocity of the expanding fluid (the
cloud) is the gradient of a scalar function $P\left( X,Y\right) $.
\begin{equation}
\mathbf{v}_{n}=-\left( \mathbf{\nabla }P\right) _{n}\ \ \text{at the
interface }\Gamma  \label{eq101}
\end{equation}%
The subscript $n$ means projection of the vector $\mathbf{\nabla }P$ on the
normal at the interface $\Gamma $. The equation for the scalar function is $2D$ Laplace: $\Delta P=0$ in the
cloud region, \emph{i.e.}%
 in the lower part limited by $\Gamma \left(
t\right) $. This function is $P=0$ in the free (environment) region II. The physical source of expansion is the input of cloud air from
below the current position of the interface $\Gamma(t)$. This is represented as an asymptotic
condition for velocity: somewhere very far inside the cloud ($Y\rightarrow
-\infty $), the velocity of the air is a constant directed toward the
interface 
\begin{equation}
\mathbf{\nabla }P=\widehat{\mathbf{e}}_{Y}\ \text{for}\ \ Y\rightarrow
-\infty  \label{pasym}
\end{equation}%
In addition%
\begin{equation}
P=0\ \ \text{at the interface }\Gamma  \label{eq100}
\end{equation}%

\bigskip

\subsubsection{The time-dependent complex conformal transformation}

The evolution consists of changes in time of the curve $\Gamma \left(
t\right) \equiv \left[ X\left( t\right) ,Y\left( t\right) \right] $
representing the interface, \emph{i.e.} expansion of the boundary cloud/environment in any horizontal section of the cloud. The idea is to find a mapping between the physical plane $%
\left( X,Y\right) $ and the complex plane $z\equiv x+iy$. At every moment of time $t$,  the lower semi-plane in the mathematical complex plane $(x,y)$ is mapped to the space below the
interface $\Gamma \left( t\right) $, where $P$ verifies the Laplace equation. The evolution of the interface is then a set of conformal
transformations parametrized by time \cite{cmplxheleshaw}. The scalar function $P$, is defined as the real part of a new complex variable, whose imaginary part is a function $%
\Psi \left( Z\right) $ 
\begin{equation}
W\left( Z\right) =P\left( Z\right) +i\Psi \left( Z\right)  \label{eq102}
\end{equation}%
The two real functions $\Psi $ and $P$ are harmonically conjugated where $W$
is holomorphic.

\bigskip

The conformal map is the function $f$, 
$f:\mathbf{C\rightarrow C},\ Z\equiv X+iY=f\left( z,t\right)$ where 
$z=x+iy$. Since $f\left( z,t\right) $ maps the
lower half complex plane $\left( y<0\right) $ to the region under $\Gamma
\left( t\right) $, $Y<0$, its derivative $\frac{\partial f}{\partial z}$ should have no
singularities or zeros in the lower semi-plane. All of them must be in the
upper semi-plane. Translating Eq.(\ref{pasym}) it is expected that at very large distances on
the plane, relative to the interface, the variables will have very close
values%
\begin{equation}
f\left( z,t\right) \sim z\ \ \text{for}\ \ z\rightarrow x-i\infty
\label{eq103}
\end{equation}%
which corresponds to constant velocity of the incoming air, the source being
the air rising from below the horizontal plane of the current height.

\bigskip

Now the system is rewritten: the new function (instead of the pressure) is $%
W\left( z\right) $ and the new variables (instead of $\left( x,y\right) $ )
are $z$ and $\overline{z}$.%
\begin{eqnarray}
\frac{\partial W}{\partial \overline{z}} &=&0\ \ (W \text{\ is holomorphic})  \label{eq104} \\
\frac{\partial W}{\partial z} &=&-i\ \ \text{for}\ \ z\rightarrow -i\infty 
\text{\ at large distance from }\Gamma  \notag \\
\mathbf{Re}W &=&0\ \ \text{at}\ \ z=x-i0\text{\ (at the mapped interface)} 
\notag
\end{eqnarray}%
From Eq.(\ref{eq101})%
\begin{equation}
\mathbf{Re}\left[ \overline{n}\left( \frac{\partial f}{\partial t}+\frac{%
\overline{\left( \frac{\partial W}{\partial z}\right) }}{\overline{\left( 
\frac{\partial f}{\partial z}\right) }}\right) \right] =0  \label{eq105}
\end{equation}%
here $\overline{n}$ is a complex number associated to the normal versor at
the interface. The solution is%
\begin{eqnarray}
W\left( z\right) &=&-iz  \label{eq106} \\
&=&-if^{-1}\left( Z,t\right)  \notag
\end{eqnarray}%
from where the scalar function $P$ is%
\begin{equation}
P\left( X,Y\right) =\mathbf{Im}\left[ f^{-1}\left( X+iY,t\right) \right]
\label{eq107}
\end{equation}
The normal at the interface is%
\begin{equation}
\overline{n}=-i\frac{\frac{\partial f}{\partial z}}{\left\vert \frac{%
\partial f}{\partial z}\right\vert }\ \ \text{at }\ \ z=x-i0  \label{eq108}
\end{equation}
Then the equation describing the\emph{\ Laplacian growth} is%
\begin{eqnarray}
\mathbf{Im}\left( \frac{\partial f\left( z,t\right) }{\partial z}\overline{%
\frac{\partial f\left( z,t\right) }{\partial t}}\right)  &=&1\ \text{at}\ \
z=x-i0  \label{eq109} \\
&&\text{(for }z\text{ on the real }x\text{ axis, just below)}  \notag
\end{eqnarray}%
which is the \textquotedblleft Polubarinova - Galin\textquotedblright\  equation \cite{poncemineev0}, \cite%
{poncemineev1}, \cite{poncemineev2}, \cite{leechenHilbert}, \cite%
{cmplxheleshaw}, \cite{topologicaltransitions}.

\begin{figure}[h]
\includegraphics[height=10cm]{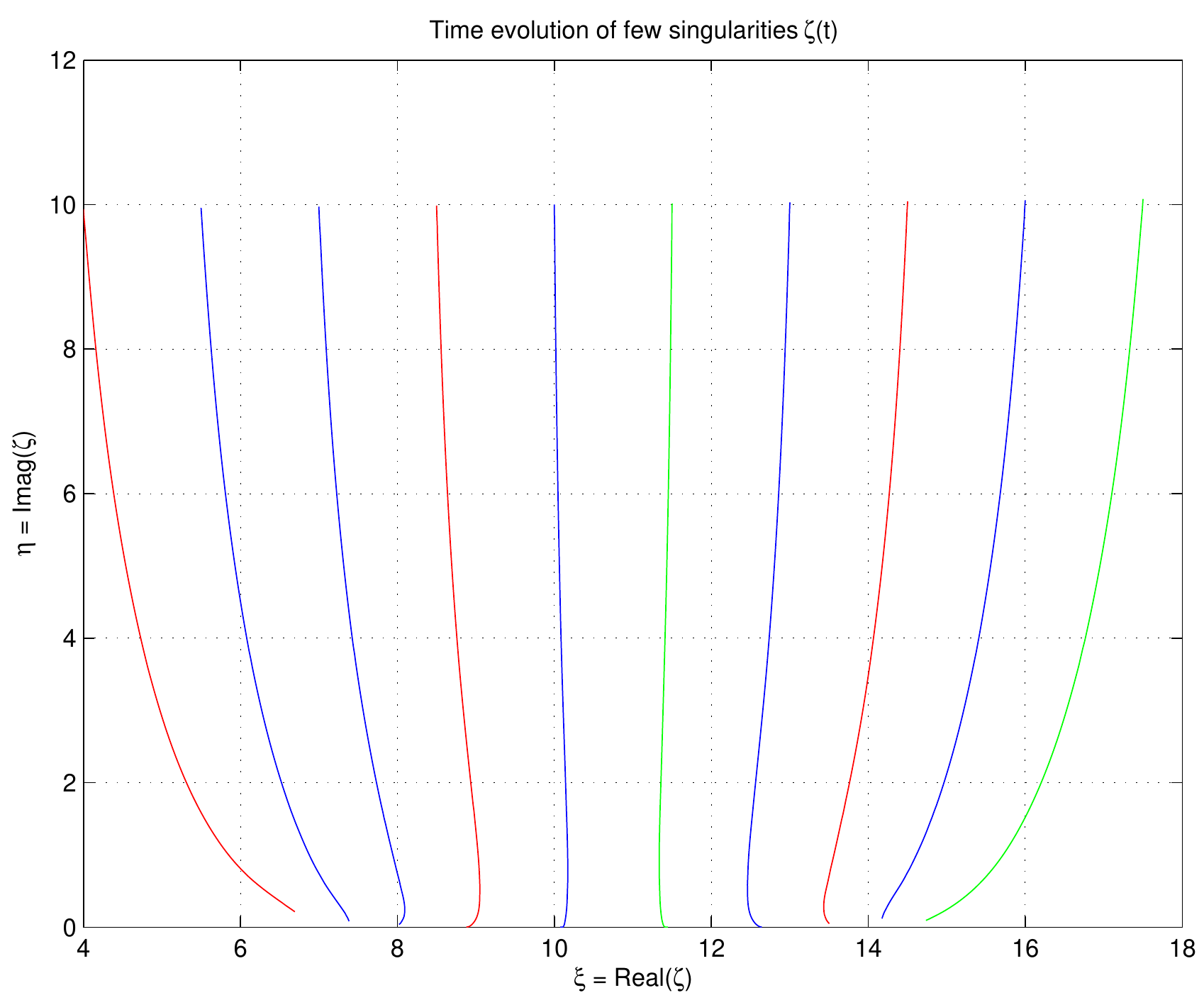}
\caption{The curves represent the time evolution of the positions of the $N+1$ singularities $\zeta_{k}(t)=[\xi_{k}(t),\eta_{k}(t)]$, ($k=1,N+1$ for $N=9$). The initial positions $\eta_{k}(t=0)$ are at in a small interval around $10$ and the real parts $\xi_{k}(t=0)$ are distributed on equal intervals.}
\label{fig1}
\end{figure}
%

\begin{figure}[h]
\includegraphics[height=10cm]{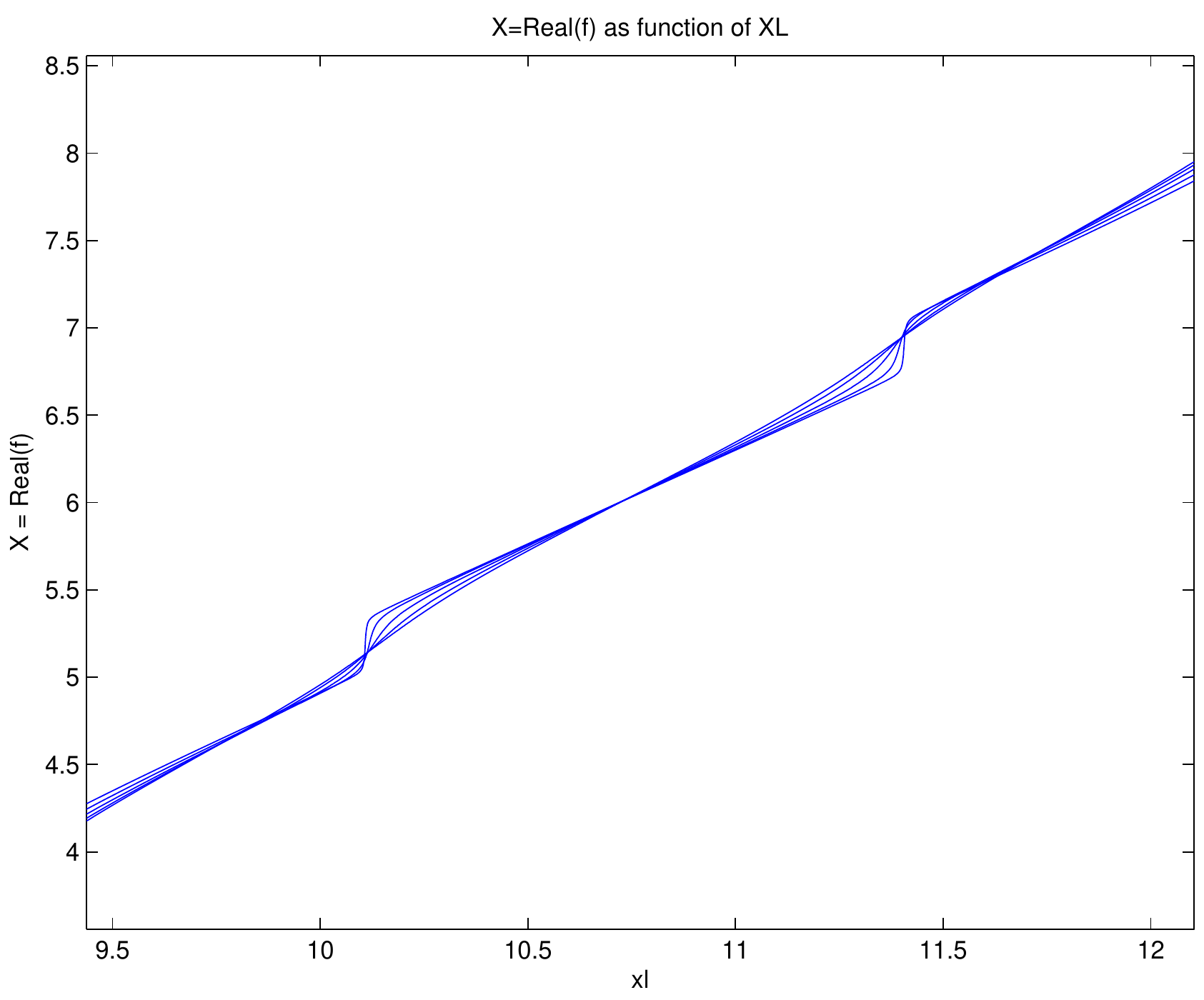}
\caption{The lines $X\left( t\right) \equiv \mathbf{Re}\left[ f\left( z,t\right) \right]
$ at $5$ moments of time, $t=90,...,95$. Since both real and imaginary parts
of $\zeta _{l}$ are not too close to $0$, the lines are smooth.}
\label{fig104}
\end{figure}

\begin{figure}[h]
\includegraphics[height=10cm]{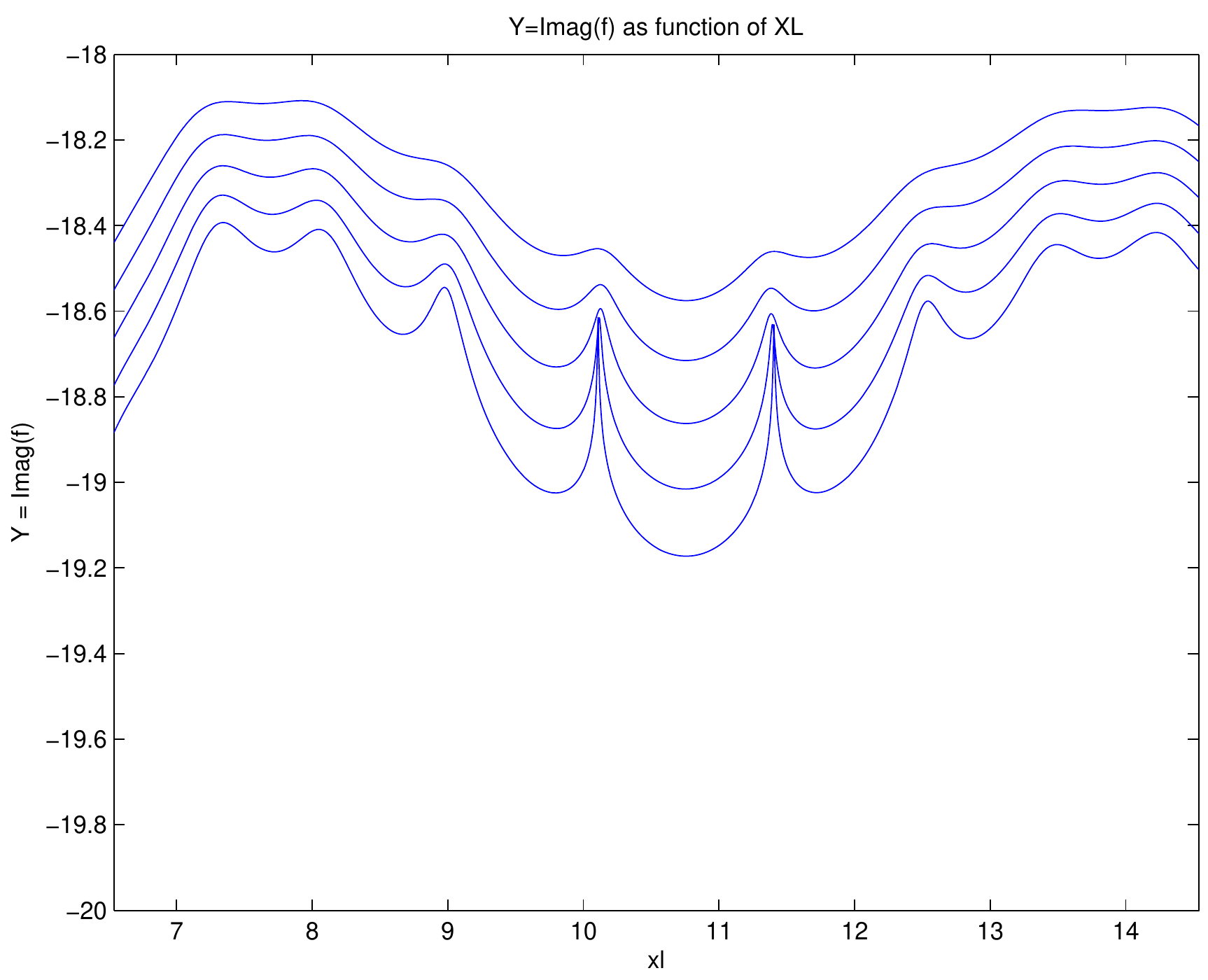}
\caption{Same as Fig\ref{fig104} for $Y\left( t\right) \equiv \mathbf{Im}\left[
f\left( z,t\right) \right] $.}
\label{fig105}
\end{figure}

\begin{figure}[h]
\includegraphics[height=10cm]{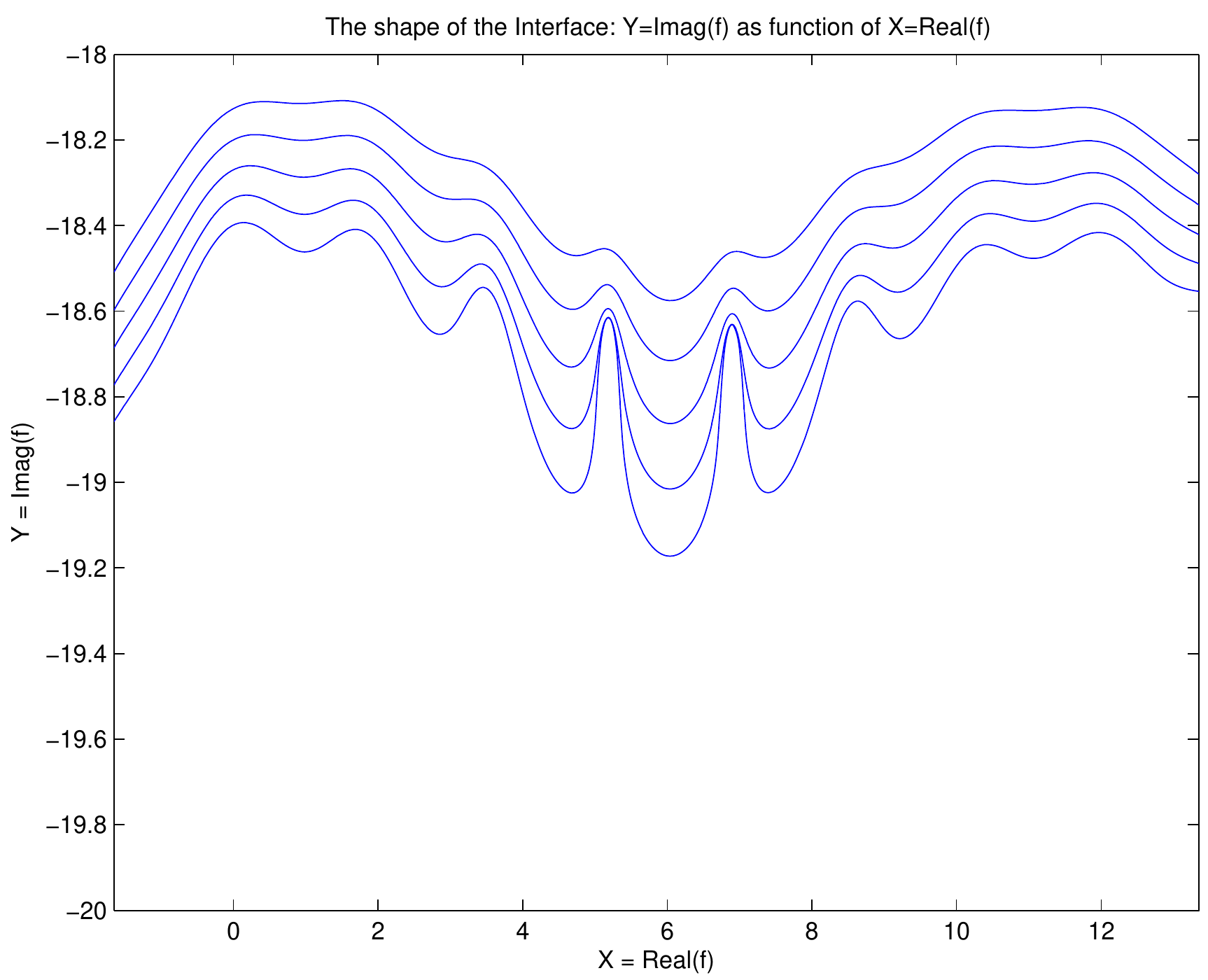}
\caption{The interface at $5$ moments of time, as in Figs\ref{fig104} and \ref{fig105}%
.}
\label{fig106}
\end{figure}

A particular form of the solution $f$ is  derived by Ponce Dowson and Mineev Weinstein in Refs. \cite%
{poncemineev0}, \cite{poncemineev1}, \cite{poncemineev2}.
Since $f$ is holomorphic in the lower complex half-plane a general expression is defined by choosing for $\frac{\partial f}{\partial z}$ a number of zeros and poles in the upper half-plane. This provides the explicit form of the mapping, at a fixed moment of time. Since we have assumed infinite extension of the interface, a class of solutions is%
\begin{equation}
f_{\inf }\left( z,t\right) =z-it-i\sum\limits_{l=1}^{N+1}\alpha _{l}\log 
\left[ z-\zeta _{l}\left( t\right) \right]  \label{eq110}
\end{equation}%
where 
\begin{equation}
\alpha _{l}\equiv \alpha _{l}^{\prime }+i\alpha _{l}^{\prime \prime }\ \ 
\text{are\ \ }N+1\text{\ \ complex constants}  \label{eq111}
\end{equation}%
and $\left( \alpha ^{\prime },\alpha ^{\prime \prime }\right) $\ \ are real.%
\begin{equation}
\zeta _{l}\equiv \xi _{l}+i\eta _{l}\ \ \text{are\ \ }N+1\text{\ \
singularities}  \label{eq112}
\end{equation}%
simple poles of $\frac{\partial f}{\partial z}$ that move in time.

A detailed treatment of the conformal mapping is provided in the Appendix A. The evolution keeps
the zeros and the poles in the upper half-plane, where they are initialized. At any
moment of time the new positions of the singularities are inserted in Eq.(%
\ref{eq110}) and the conformal mapping is determined. Therefore we have the
new shape of the interface between the two physical media (expanding cloud
and environment).

\clearpage

\subsection{The small scale structure: roughening of the interface due to
random fluctuations}
Besides the deterministic evolution described by the \emph{fingering}
instability, one should consider random fluctuations that produces the
roughening of the interface. It has been found that the interface roughness
in $2D$ is algebraic on short length scale. Here we only mention from Ref. 
\cite{kineticroughening1} a part of the argument that introduces the Hilbert
transform. This is an analytical step which prepares the
discussion on the large scale features of the cloud boundary. We consider the interface between the cloud and the environment consisting,
in the horizontal plane, of a line $\Gamma $ of length $L$. The coordinate
along the interface is $s$ and along the local normal is $y$. The position
in plane of the current point on the interface is given by the distance 
\begin{equation*}
h\left( s,t\right)
\end{equation*}%
relative to a fixed reference system. The Laplacian field (the pressure) $%
\phi \left( \mathbf{x},t\right) $ is defined as%
\begin{eqnarray}
\Delta \phi &=&0\ \ \text{for}\ \ y<h\ \ \text{inside the cloud, and}
\label{eq114} \\
\phi &=&0\ \ \text{for}\ \ y>h  \notag
\end{eqnarray}
The velocity of a current point of the interface is given by the \emph{%
gradient of the scalar function} $\phi $, as%
\begin{equation}
\frac{\partial h}{\partial t}=-D\left( \frac{\partial \phi }{\partial y}-%
\frac{\partial \phi }{\partial s}\frac{\partial h}{\partial s}\right) _{x=h}
\label{eq115}
\end{equation}%
This equation describes the \emph{gradient flow}. The last term describes
the local dilation or the compression of the length of the curve $\Gamma $.
The deterministic part of the dynamics is introduced by the average motion of the
front%
\begin{equation}
h=Vt  \label{eq116}
\end{equation}%
A small perturbation of the \textquotedblleft height\textquotedblright\ $h\left( x,t\right) $ taken as%
\begin{equation}
\widetilde{h}_{q}\left( s\right) =h_{q}\left( t\right) \exp \left( iqs\right)
\label{eq117}
\end{equation}%
acting on the moving interface decays as%
\begin{equation}
\left\vert \widetilde{h}_{q}\left( s\right) \right\vert \sim \exp \left[
-\sigma \left( q\right) t\right]  \label{eq118}
\end{equation}%
so the perturbation is exponentially vanishing with the rate%
\begin{equation}
\sigma \left( q\right) =V\left\vert q\right\vert  \label{eq119}
\end{equation}
Then the relaxation of the interface \textquotedblleft height\textquotedblright\ is a simple linear decay of
the logarithm of $h_{q}$, working in the Fourier space. Since we want to
take into account the random fluctuations, it is introduced a noise source%
\begin{equation}
\frac{\partial h_{q}\left( t\right) }{\partial t}=-V\left\vert q\right\vert
h_{q}\left( t\right) +\eta _{q}\left( t\right)  \label{Langevin}
\end{equation}
This is a nonlocal equation since $\sigma \left( q\right) $ acts in Fourier
space. The noise is Gaussian%
\begin{equation}
\left\langle \eta _{q}\left( t\right) \eta _{q^{\prime }}\left( t^{\prime
}\right) \right\rangle =\frac{\Delta }{L}\delta \left( q+q^{\prime }\right)
\delta \left( t-t^{\prime }\right)  \label{eq120}
\end{equation}
We note that the dynamical equation for the noise-driven interface contains
the term $-V\left\vert q\right\vert h_{q}\left( t\right) $. It comes from taking the Fourier transform of the real-space function $h$, multiplying with the absolute value of the Fourier space variable $\left\vert q\right\vert$ and returning to real space. This is the Hilbert transformation applied on $h$ and will play an essential role in the following.

In Ref.\cite{kineticroughening1} it is shown that the width of the interface increases as the $log$ of the length $L$.%

\subsection{The large scale dynamical structuring of the interface: cusp
singularities}

We are now interested in the phase of the cloud expansion where the
convection flux coming from lower levels is progressively reduced, the
column reaching a regime of quasi-stationarity. On a large spatial scale (of
the circumference of the expanding cloud) the interface has a dynamics of
the expansion of a front as the propagation of a planar flame into a
static homogeneous medium. A model equation for the latter case has been
developed by Sivashinsky \cite{sivashinsky1}, \cite{sivashinsky2}. It treats
the advancement of a front of a flame in a chanel.

The planar flames expanding freely has an interaface that is unstable. For a
simple model in $1D$ (flames propagating in a channel of width $\widetilde{L}
$) the variable is%
\begin{equation*}
h\left( x,t\right) \equiv \text{position of the flame front above the }x%
\text{ axis}
\end{equation*}%
The equation of Sivashinsky \cite{sivashinsky1} is%
\begin{equation}
\frac{\partial h\left( x,t\right) }{\partial t}-\frac{1}{2}\left[ \frac{%
\partial h\left( x,t\right) }{\partial x}\right] ^{2}-\Lambda \left\{
h\left( x,t\right) \right\} =\nu \frac{\partial ^{2}h\left( x,t\right) }{%
\partial x^{2}}+1  \label{eq281}
\end{equation}%
in the domain%
\begin{equation*}
0<x<\widetilde{L}
\end{equation*}%
The functional $\Lambda $ is the Hilbert transform. To define its action,
first one makes the Fourier transform of the function $h\left( x,t\right) $,%
\begin{equation}
h\left( x,t\right) =\int_{-\infty }^{\infty }\exp \left( ikx\right) \widehat{%
h}\left( k,t\right) dk  \label{eq282}
\end{equation}%
then multiply $\widehat{h}\left( k,t\right) $ by the absolute value of the
Fourier variable%
\begin{equation}
\Lambda \left\{ h\left( k,t\right) \right\} =\left\vert k\right\vert 
\widehat{h}\left( k,t\right)  \label{eq283}
\end{equation}%
and returns to the real space, $\Lambda \left\{ h\left( x,t\right) \right\} $%
. We can trace the occurence of the Hilbert operator in the equation for the
front advancement from the derivation of Eq.(\ref{Langevin}).

The front is unstable and it generates singularities in finite time. The
singularities are of the \emph{cusp}\textbf{\ }type. We note that this
evolution implicitely renders the interface piecewise smoother since it
collects the smaller scale singularities into a single, \emph{giant cusp}.
This corresponds to the late phases, where the flux of cloud air is reduced
and the reserve of buoyancy is decaying.

\subsubsection{Formalism: Pole decomposition}

The origin of the \emph{cusp} profile can be understood from the evolution
of the positions of the complex singularities (poles) of the solution. The motion of the pole
singularities is controlled by the Hilbert operator. The
multiplication with the absolute value of the Fourier variable $\left\vert
k\right\vert $ is equivalent to the differential operator $i\frac{%
\partial }{\partial x}$. It makes the poles to approach the real axis, from
both sides. The accumulation of poles (\textquotedblleft
condensation\textquotedblright\ at the limit of continuous density of poles)
leads to the formation of the \emph{cusp}, a singularity of the interface
that is qualitatively similar with what is frequently seen in the evolution
of an expanding cloud. This suggests that the large scale structure at the
expansion of the circumference of a cloud as it rises can be studied using
the Sivashinsky equation, (\ref{eq281}).

Therefore the first step in an analytic development is to extend the space
coordinates to complex variables \cite{leechenHilbert}, \cite{cmplxheleshaw}, \cite{heleshawfreebound}, \cite{Joulin1},  \cite{Joulin2},  \cite{Joulin3}. The nature of processes that
take place during time evolution is: (1) attraction between the poles along
the horizontal axis ($x$ or $\theta $) leading to clusterization of poles
along vertical direction in the complex plane, and (2) dynamical evolution
of the poles towards the real axis. \ This produces singularities that look
like \emph{cusps}.

The problem is restricted to a single space variable, relative to which we
measure the spatial position $h\left( x,t\right) $ of the interface. The nature and the positions of the wrinkles
of the interface can be associated to the singularities of the interface -
function in the plane of the complexified spatial variable. A singular
profile will occur in finite time but for short time the equation is
integrable which means that it has the Painleve property and the formal
solution is meromorphic and can be written as an expansion in a set of
order-one poles.

In Ref.\cite{thualfrischhenon} it is porposed the Lagrangian version of Eq.(\ref{Langevin}) with the
white noise converted into a diffusion generated by viscosity $\nu $,
essentially constructed on the ground of Burgers equation 
\begin{equation}
\frac{\partial u}{\partial t}+u\frac{\partial u}{\partial x}=\Lambda \left[ u%
\right] +\nu \frac{\partial ^{2}u}{\partial x^{2}}  \label{TFH}
\end{equation}%
As before, $\Lambda \left[ u\right] :\widehat{u}\left( k,t\right)
\rightarrow \left\vert k\right\vert \widehat{u}\left( k,t\right) $ where the
Fourier transform of the function $u$ has been introduced $u\left(
x,t\right) =\int_{-\infty }^{+\infty }dk\ \exp \left( ikx\right) \ \widehat{u%
}\left( k,t\right) $. The solution is a meromorphic function expressed in
terms of simple poles by terms like%
\begin{equation}
p_{\alpha }=\frac{1}{x-z_{\alpha }}  \label{eq209}
\end{equation}%
The operator $\Lambda $ applied on such term results in 
\begin{eqnarray}
\Lambda p_{\alpha } &=&\Lambda \left[ \frac{1}{x-z_{\alpha }}\right]
\label{eq210} \\
&=&\mathrm{sign}\left[ \mathbf{Im}\left( z_{\alpha }\right) \right] \ i\frac{%
\partial }{\partial x}p_{\alpha }\left( x\right)  \notag
\end{eqnarray}%
The operator $\Lambda $ produces advection of the poles, along the imaginary
direction, towards the real axis (both from above and from below since the
operator multiplies with $\left\vert k\right\vert $). This enhances the
effect of the singularity, \emph{i.e.} the real function which is the
solution becomes even more perturbed there. The typical profiles are \emph{quasicusps}, which are cusps with the tip
rounded due to the presence of $\nu $. In Refs. \cite{thualfrischhenon}, \cite{frischmorf} this
is explained by showing that the singularities never reach the real axis,
but remains at a distance whose expression is given by $\nu $. This will
produce \emph{wrinkles}. The \emph{solution} of the Eq.(\ref{TFH}) can be
written, for the linear geometry as 
\begin{equation}
u\left( t,x\right) =-2\nu \sum\limits_{\alpha =1}^{2N}\frac{1}{x-z_{\alpha
}\left( t\right) }  \label{eq205}
\end{equation}%
where $z_{\alpha }\left( t\right) $ are $2N$ poles placed symmetrically, as
complex conjugated pairs ($u\left( x,t\right) $ results \emph{%
real}) relative to the real axis. The equations of motion is%
\begin{equation}
\frac{dz_{\alpha }}{dt}=-2\nu \sum\limits_{\beta \neq \alpha }\frac{1}{%
z_{\alpha }-z_{\beta }}-i\mathrm{sign}\left[ \mathbf{Im}\left( z_{\alpha
}\right) \right]  \label{eq206}
\end{equation}

The poles tend to attract themselves \emph{horizontally}, parallel with the
real axis. As noted in Ref.\cite{procaccia1} this results from the dominant
behavior of the $x$ coordinate of the poles%
\begin{equation*}
\frac{dx_{j}}{dt}\sim -\sum\limits_{k\neq j}\sin \left( x_{j}-x_{k}\right) %
\left[ ...\right]
\end{equation*}%
In the paranthesis $\left[ ...\right] $ there are terms that are invariant
to the change $x_{j}\leftrightarrow x_{k}$, for any pair $j,k=1,N$, with $%
j\neq k$. This factor is positive. When $0<x_{j}<x_{k}$ we have $dx_{j}/dt>0$
and for $0<x_{k}<x_{j}$ we have $dx_{j}/dt<0$. In both cases, $x_{j}$
evolves to be closer to $x_{k}$.

In addition there is a tendency of the poles to place themselves on a line
parallel to the imaginary axis (this is the analog of the same phenomenon
for the Burgers equation). At short range the poles tend to repel each other
vertically and the repulsion between two poles aligned on a vertical $\left(
y\right) $ line becomes infinite when they come too close. At longer
distances, the interaction become attractive.

Details on the integration of Eq.(\ref{eq206}) are in Appendix B.

\begin{figure}[h]
\includegraphics[height=10cm]{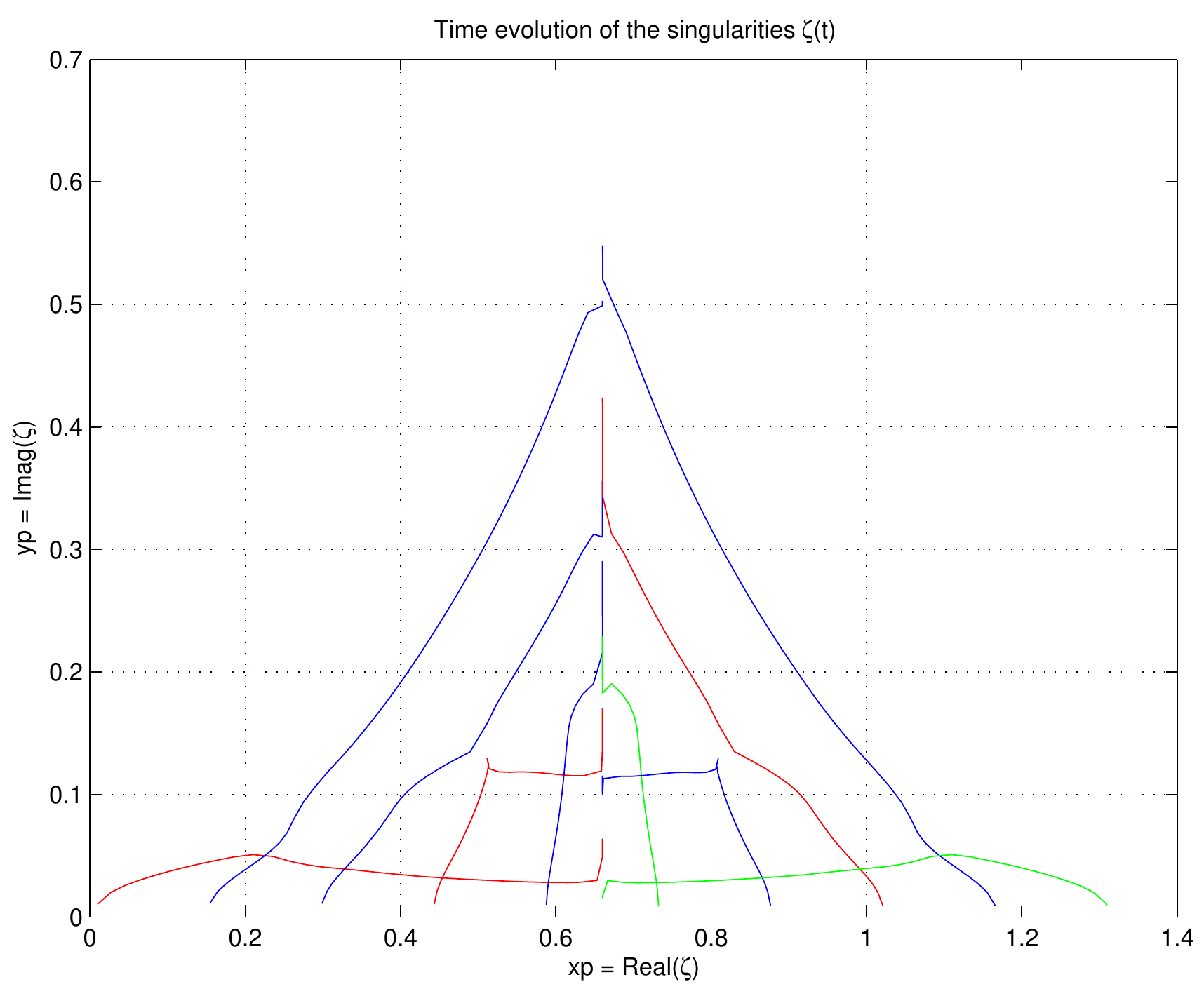}
\caption{The time evolution of the complex poles of the solution of the interface problem for the case of the \emph{cusp} singularity.}
\label{fig7}
\end{figure}
%

\begin{figure}[h]
\includegraphics[height=10cm]{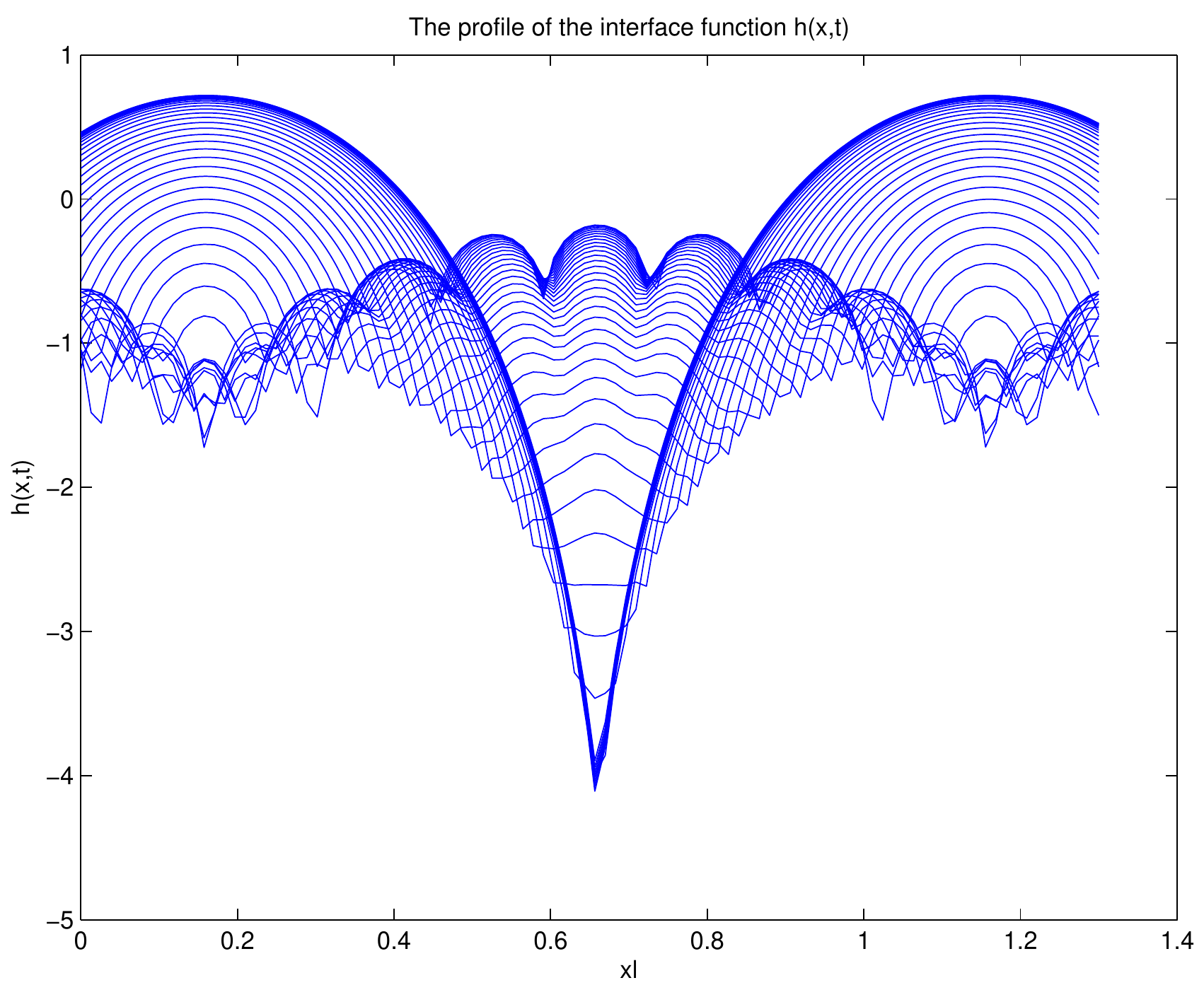}
\caption{The profile of the interface at several moments of time. The initial state has many oscillations of small amplitude and the final state presents the cusp singularity.}
\label{fig8}
\end{figure}
%

\begin{figure}[h]
\includegraphics[height=10cm]{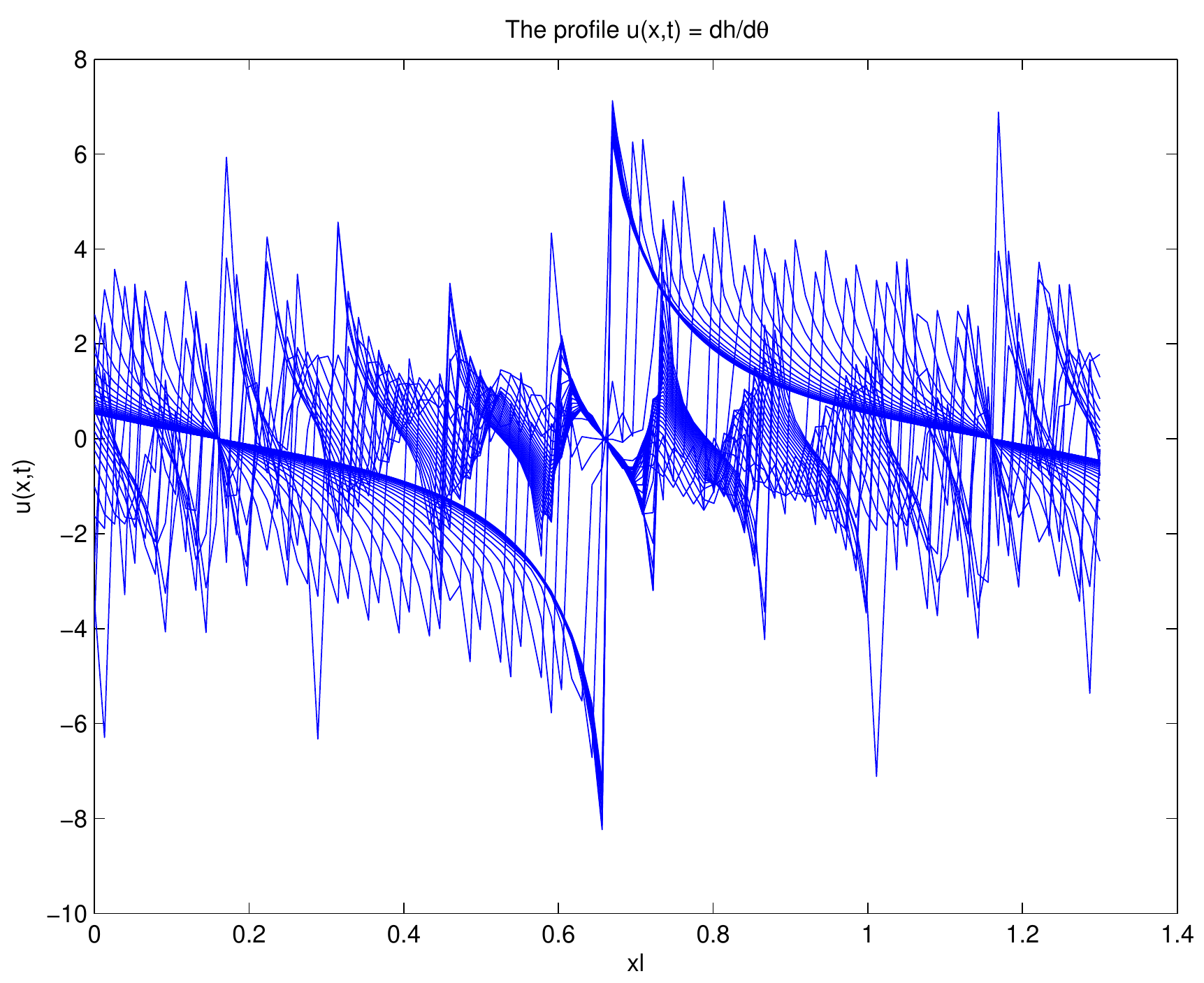}
\caption{The slope of the interface.}
\label{fig9}
\end{figure}
%

\begin{figure}[h]
\includegraphics[height=10cm]{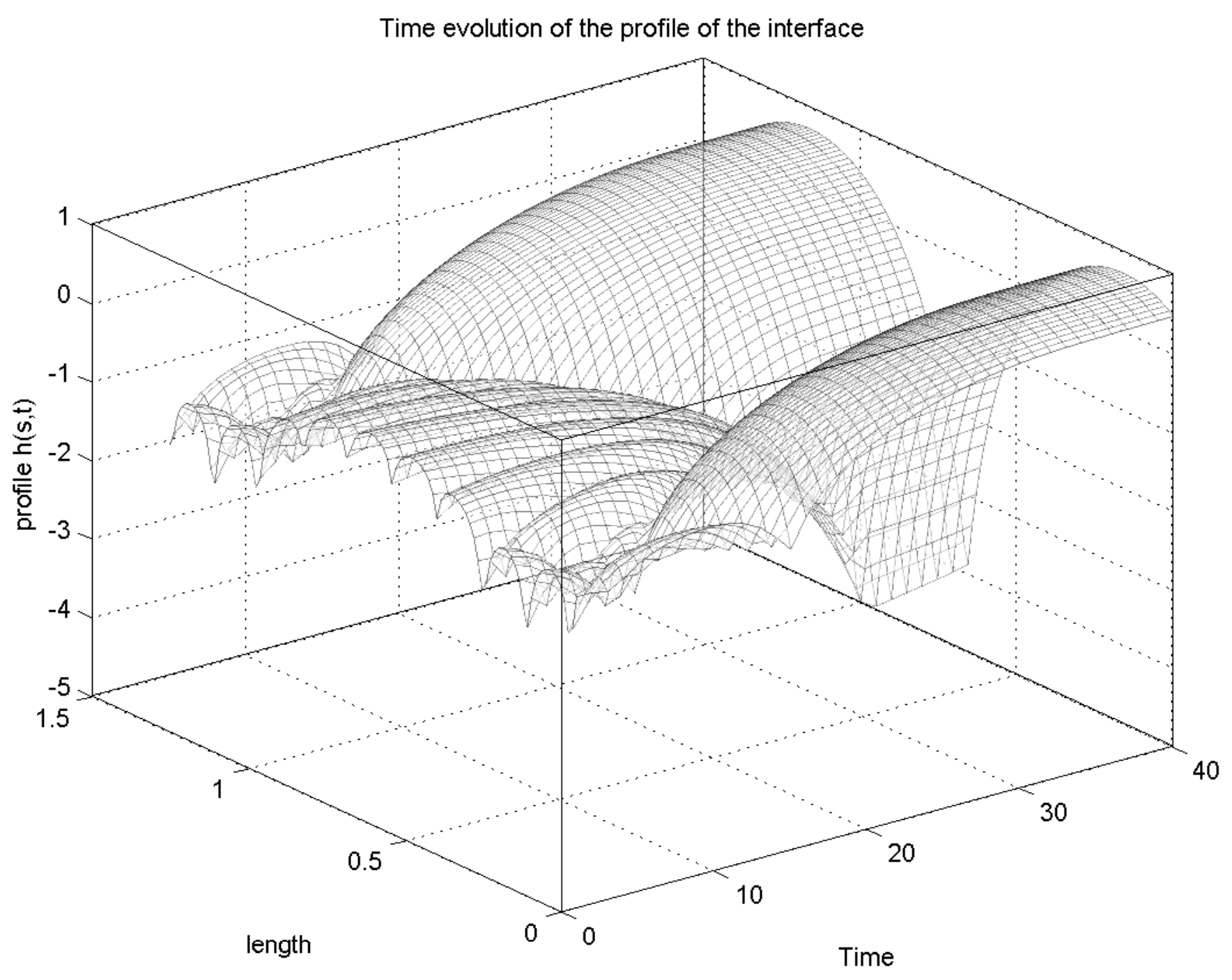}
\caption{The relief view of the evolution of the interface.}
\label{fig10}
\end{figure}

\subsubsection{Further developments}

The stability of the interface consisting of a giant cusp has been examined
in \cite{procaccia1} and \cite{procaccia2} where it is explained the role of
the noise in generation of new poles in the structure of the solution. This
may also explain the self-fractalization of the interface resulting from
self-acceleration \cite{selfacceler} of the flame front determined
experimentally \cite{gostintsev}.

\section{The breaking of the rising convection column}

\subsection{Introduction. The loss of compacity of the column}

During its rise the convective column may undergo breakup and loss of
compacity in its volume. The column will then consist of vertical streams of
cloudy air separated by volumes of environmental air, also vertical and of
irregular shape, which is either static and not yet mixed or consists of
downdrafts after evaporative cooling of parcels entrained close to the top
of the column. As recalled in the Introduction, the simplest view on the
convection is a picture of intepenetration of two gases, one being the
stream of rising air whose ascension is sustained by buoyancy and the other
being the environment. There is a wide spectrum of possible evolutions, from
the early loss of the streaming compacity to the preservation of robust
columnar features during all rise. We try to find an analytical
representation of the breaking of the rising column.

There are observations revealing that, on a horizontal plane, the
updraft inside a cloud presents substantial inhomogeneity \cite{squires}, \cite{paluch}, \cite{macpherson1977}. There are large variations
of the vertical velocity, in magnitude and sign, meaning that there are
updrafts and downdrafts and also regions of static environmental air. It
results that there is a breaking-up of the rising column. Obviously, the
surface of contact between the convective air and the environment, either
static or penetrative downdrafts, is much larger than in the case where the
column remained compact as it rises. The examination of the breaking up of
the column, with substantial increase of the \textquotedblleft
interface\textquotedblright , is a necessary step in a better representation
of the exchanges and transport processes. In particular this refers to the model of Squires \cite{squires} where it is
assumed that the environmental air is mixed with the cloud at the cloud top.
Then, due to evaporative cooling, the mixed air loses its buoyancy and
descends deep into the cloud column (kilometers) \cite{paluch}. The exchange of heat and
vapors with the cloud air continues for these \emph{internal} downdrafts and
the result is dilution of the cloud air but also increase of the buoyancy of
the air penetrating from above. The equilibrium between the penetrating
downdrafts and the cloud air surrounding it is reached at equal buoyancy.

We are interested in the process of competition between the rising air of
the convective column and the environmental air. The result is the physical
breakup of the column and the coexistence, at every horizontal level, of
regions of rising air and of environmental air. The inhomogeneity of the cloud can be pronounced: inside the clouds there are narrow updraft regions but between them there are strong negative vertical velocity flows, due to rapid downdrafts \cite{macpherson1977}. This pattern, in the horizontal plane, justifies the use of a model of interface dynamics which exhibits the {\it labyrinth} instability.

\subsection{The breaking of the convective column as a phase-competition
dynamics}

We look for a schematic analytical and/or numerical description to the real
phenomenon of breaking of the rising column into distinct vertical streams
separated by regions of environmental air. It is easier to restrict to
horizontal planes. The most elementary representation consists of the
separation of phases of a fluid in $2D$. For a binary fluid the variable is
the concentration $c$ where the two pure phases have $c=\pm 1$. The dynamics
is described by a parabolic equation where the local change of the
concentration is the Laplacian flow of the density of a functional of $c$,
which can be called \textquotedblleft chemical potential\textquotedblright .
The free energy decays to zero when there is full separation of phases, 
\emph{i.e.} the two phases occupy disjoint regions in plane. These regions
can look similar to a \emph{labyrinth} pattern \cite{goldsteinlabyrinth}. For
example, in the final stage of the slowing down of the rise of the cloud,
where substantial loss of buoyancy has resulted from mixing with
environment, the horizontal plane is mostly covered by regions of
environmental air, with few patches of cloud.

Technically, we can use the analytical approach based on one of the standard
models, in particular Cahn-Hilliard system, studied in Ref.\cite{goldsteinlabyrinth}, or cellular automata \cite{bechtoldcellular}.
However it is more useful to implement another aspect besides the
phase-separation: phase competition. In similar cases it has been adopted a representation of dynamical phase
transition which employs a system of coupled cubic map lattice \cite{kapral1}%
.

If the two phases have the same stability then the motion of the interface
is governed as in a gradient flow, by the local curvature. If the two phases
have distinct stability properties then the most stable phase will advance
irreversibly into the other and will eventually replace it. The problem
belongs to the same class as phase separation and nucleation phenomena.

The coupled lattice maps are described by the set of equations for a
discretized variable $q\left( \mathbf{x},t\right) $ which represents the
nonconserving order parameter at the points of a regular lattice%
\begin{eqnarray}
q\left( \mathbf{i},t+1\right) &=&f\left[ q\left( \mathbf{i},t\right) \right]
\label{eq301} \\
&&+\gamma \left[ \sum\limits_{\mathbf{j=}n.n}^{p}q\left( \mathbf{j}%
,t\right) -pq\left( \mathbf{i},t\right) \right]  \notag
\end{eqnarray}%
where $\mathbf{i\equiv }\left( i_{1},i_{2},...\right) $ is a set of integers
that specifies the position of a point in the lattice. In our case the
dimension of the problem is $d=2$ (plane), $\mathbf{i\equiv }\left(
i_{1},i_{2}\right) =\left( i_{x},i_{y}\right) $. The sum in the square
paranthesis extends over the \emph{nearest neighbors} ($n.n$) of the point $%
\mathbf{i}$ of the lattice. They are in number of $p$ in general and in $d=2$
they are $p=4$. The square paranthesis is actually a discretization of the
Laplace operator and this term represents the diffusion. \ The nonlinearity
of the dynamics of the order parameter $q\left( \mathbf{x},t\right) $ is
introduced by the term%
\begin{equation}
f\left[ q\left( \mathbf{i},t\right) \right] =-q^{3}+\left( 1+\varepsilon
\right) q+c  \label{eq302}
\end{equation}%
We recognize easily the meaning of choosing this nonlinearity. The
\textquotedblleft potential\textquotedblright\ $f$ has two extrema,
corresponding to non-dynamical equilibria. The order parameter can take one
or another of these two equilibria values, and they are associated to the
two states. The dynamical equation for $q\left( \mathbf{x},t\right) $ is a
discretized form of the Landau-Ginzburg equation.

\begin{figure}[h]
\includegraphics[height=10cm]{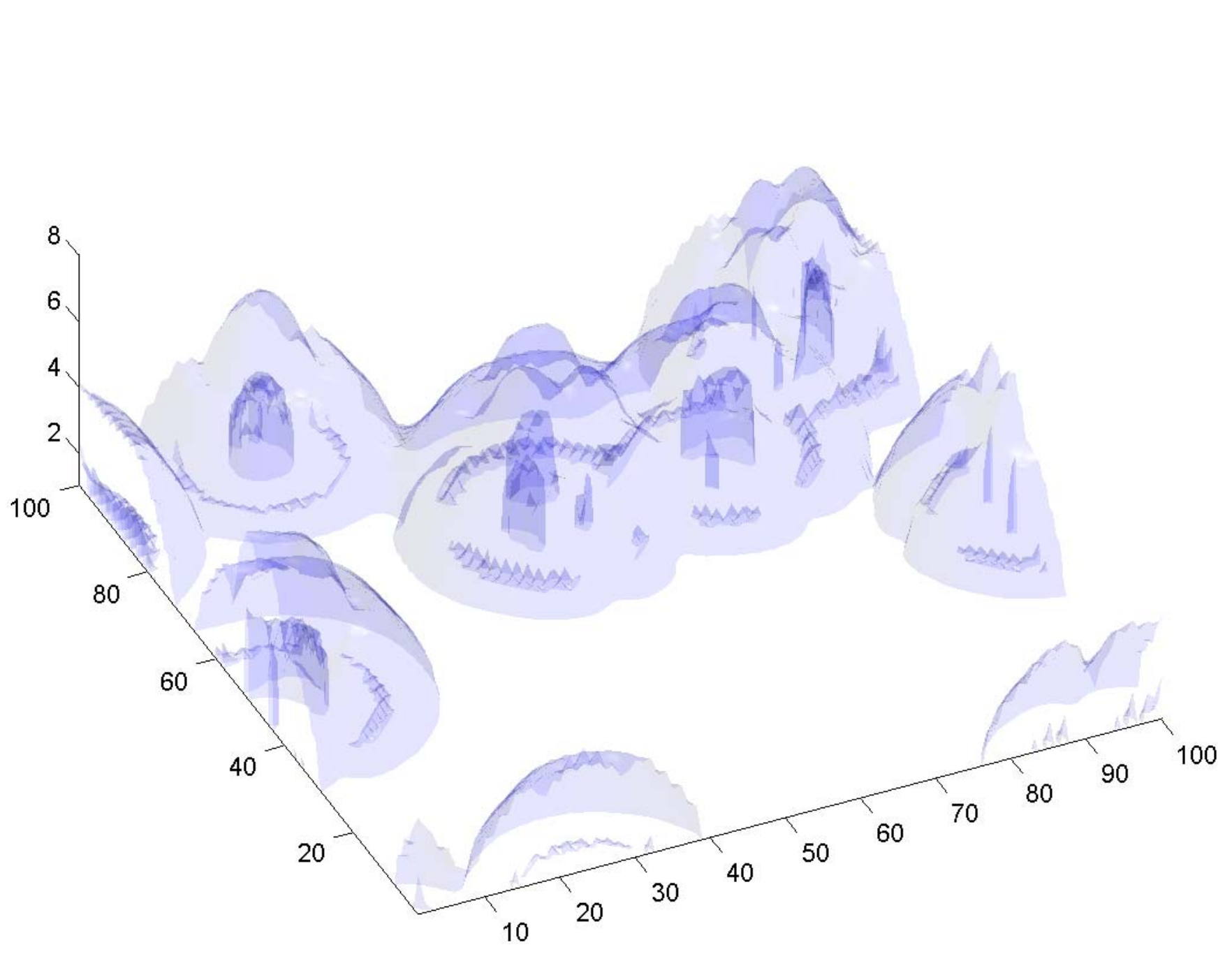}
\caption{The end stage of the competition between two phases (which we would attribute to cloud, respectively environment). The vertical axis is the height of successive planes that are labelled by the time in the iterative map Eq.(\ref{eq301}). The parameters are $a=1$, $\epsilon=0.65$, $c=0.10$ and $\gamma=0.09$.}
\label{fig11}
\end{figure}
%

\begin{figure}[h]
\includegraphics[height=10cm]{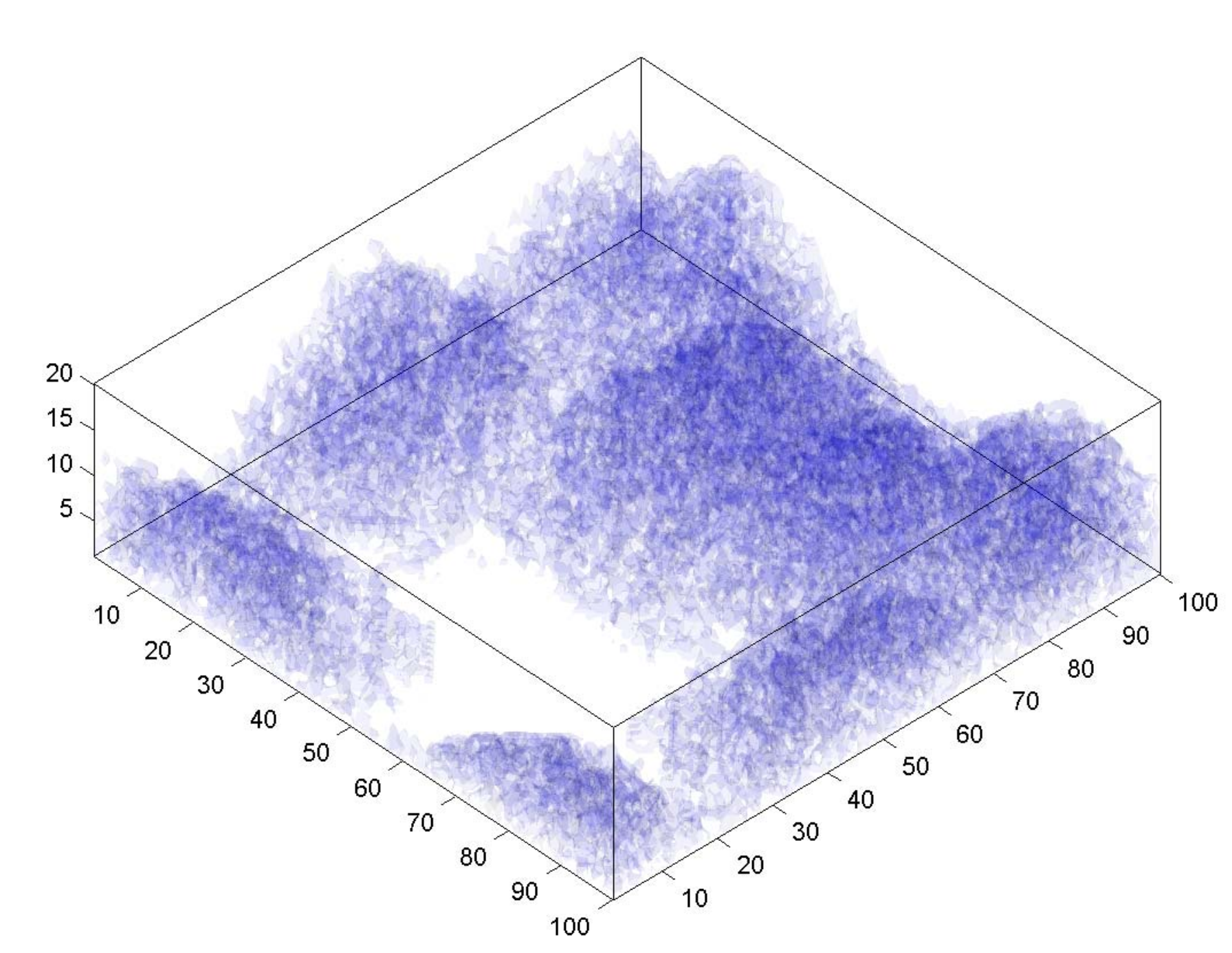}
\caption{The same as Fig.\ref{fig11}, for $a=1$, $\epsilon=0.45$, $c=0.6$ and $\gamma=0.06$.}
\label{fig12}
\end{figure}
%

\begin{figure}[h]
\includegraphics[height=10cm]{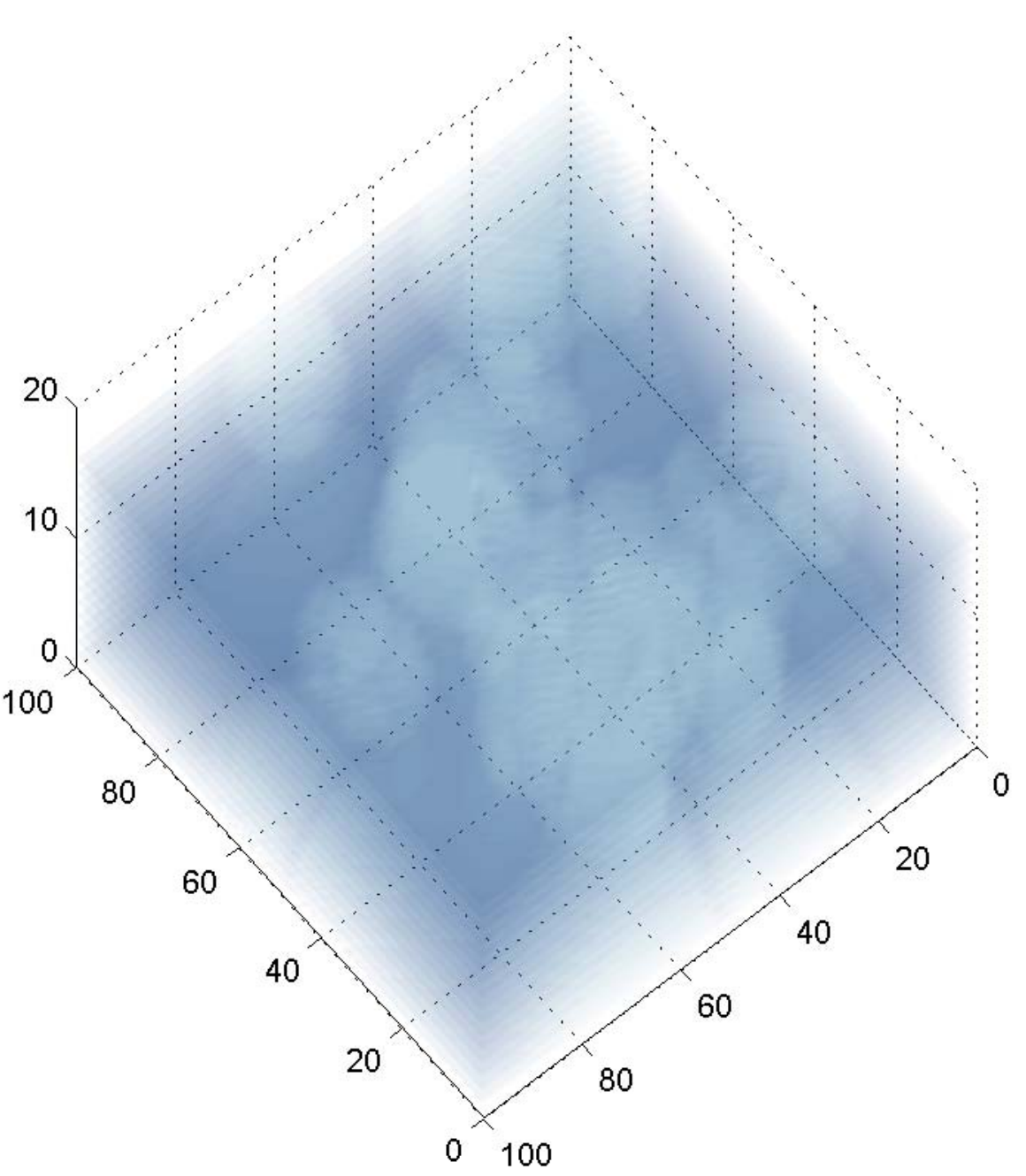}
\caption{The competition of the two phases is shown as results from iteration. The structure (barely visible) has the small scale oscillations related to the propagation mode at the level of one mesh cell length.}
\label{fig13}
\end{figure}
%

\begin{figure}[h]
\includegraphics[height=7cm]{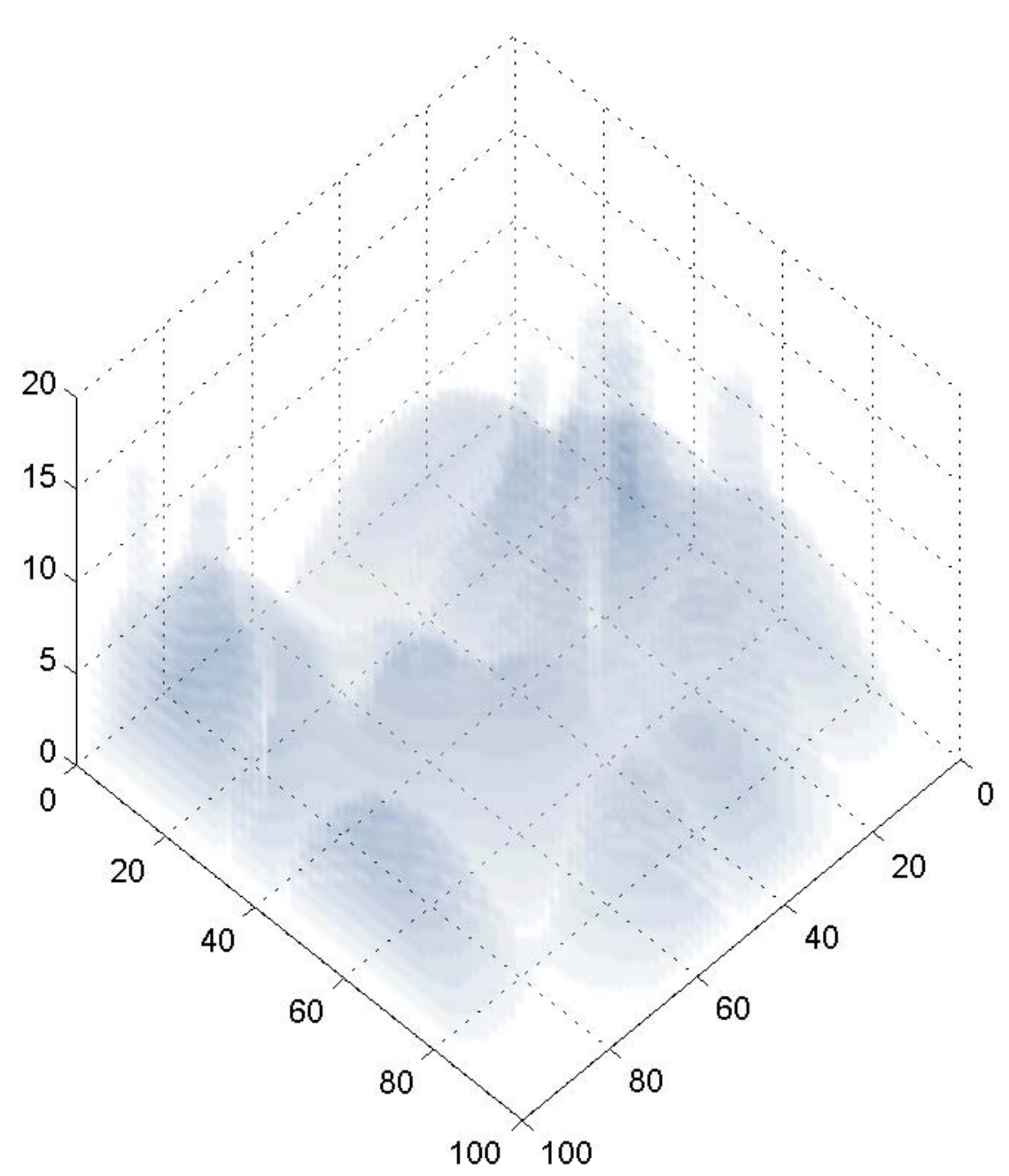}
\hfill
\includegraphics[height=6cm]{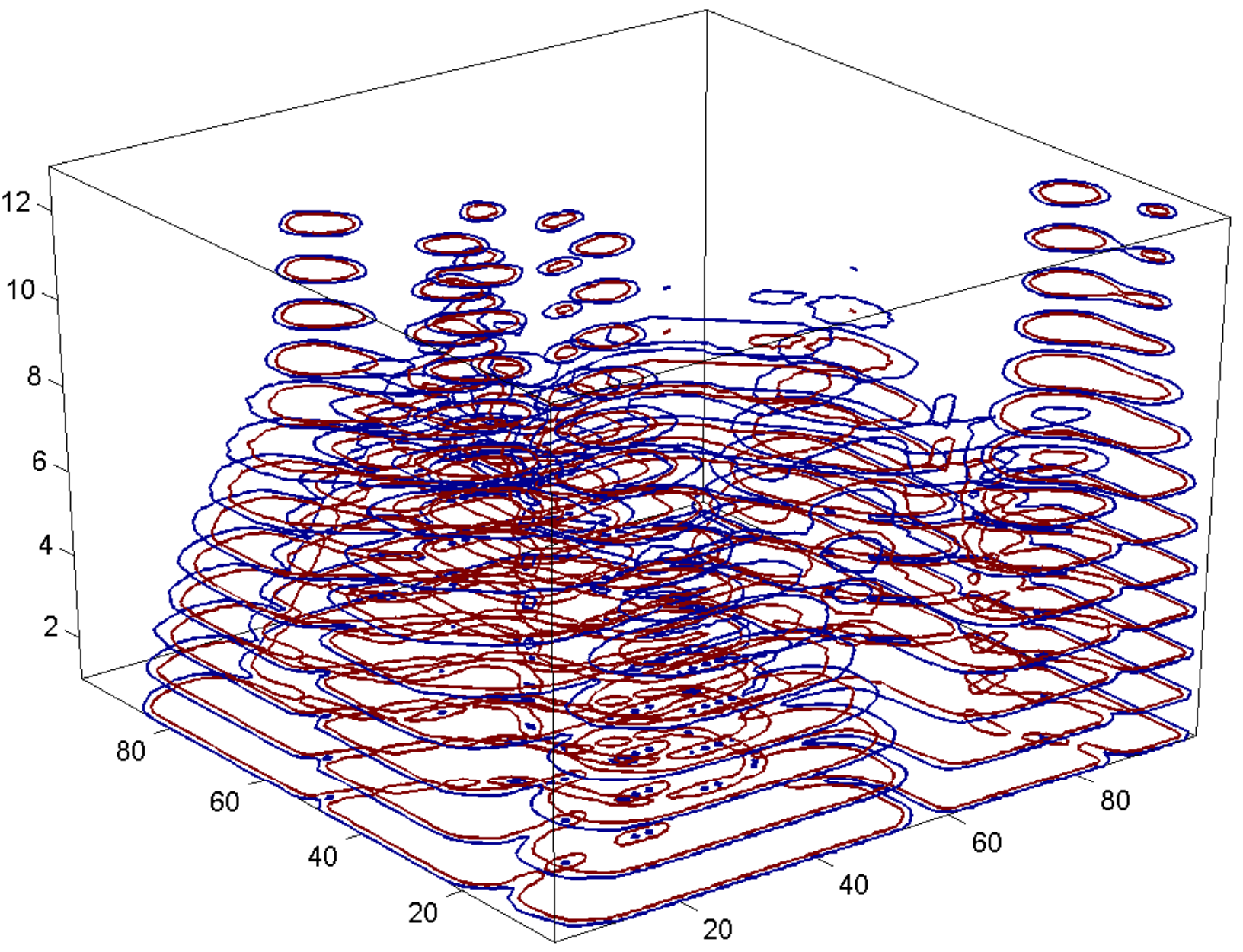}
\caption{Left: Same as Fig.\ref{fig13} after artificially removing the small scale oscillation. This makes more visible the two phases. Right: The contour slices of the left figure, that are used for calculation of the length of the circumference of horizontal patches.}
\label{fig14a15}
\end{figure}
%

\subsection{Phase separation for equally stable states}

This corresponds to 
\begin{equation}
c=0  \label{eq303}
\end{equation}%
and the initial state is 
\begin{eqnarray}
q_{0}\left( i,j\right) &=&0\ \ \left( \text{stable}\right)  \label{eq304} \\
&&+\xi \ \ \left( \text{zgomot}\right)  \notag
\end{eqnarray}

The spatial competition produces clusters of one phase inside the other. Any curved frontier between the two phases evolves with a velocity \cite{kapral1}%
\begin{equation}
v=-\gamma K  \label{eq305}
\end{equation}%
transversal on the boundary. $K$ is the curvature of the boundary. For a
disk of radius $R$ the curvature $K=R^{-1}$ and the evolution is symmetrical
with%
\begin{equation}
R^{2}\left( t\right) -R^{2}\left( 0\right) =-2\gamma t  \label{eq307}
\end{equation}

The initial situation
is a compact covering of the domain of interest with a single phase. This
initial state is perturbed by the random nucleation of the other phase on
small regions. The evolution consists of
extension of the regions associated with the Phase II, against the Phase I
and the break up of the region initially occupied by only the Phase I. The
process leads to the loss of compacity of the region of Phase I (cloud) and
to progressive reduction of the surface occupied by it in horizontal planes.

\begin{figure}[htbp]
\centering
\subfigure[]{
    \label{fig:subfig:a}
    \includegraphics[width=0.3\textwidth]{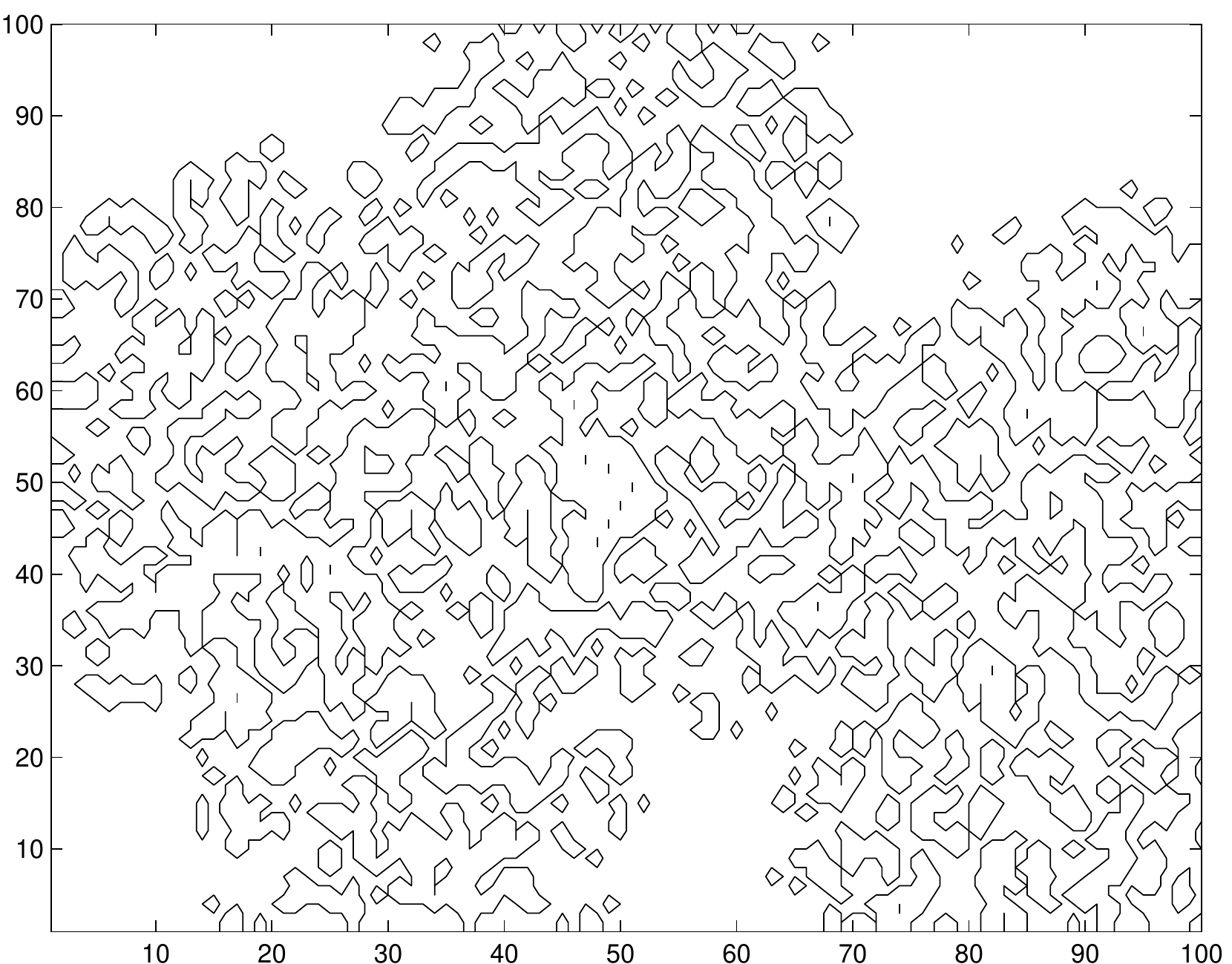}}
\hspace{0.1\textwidth} 
\subfigure[]{
    \label{fig:subfig:b}
    \includegraphics[width=0.3\textwidth]{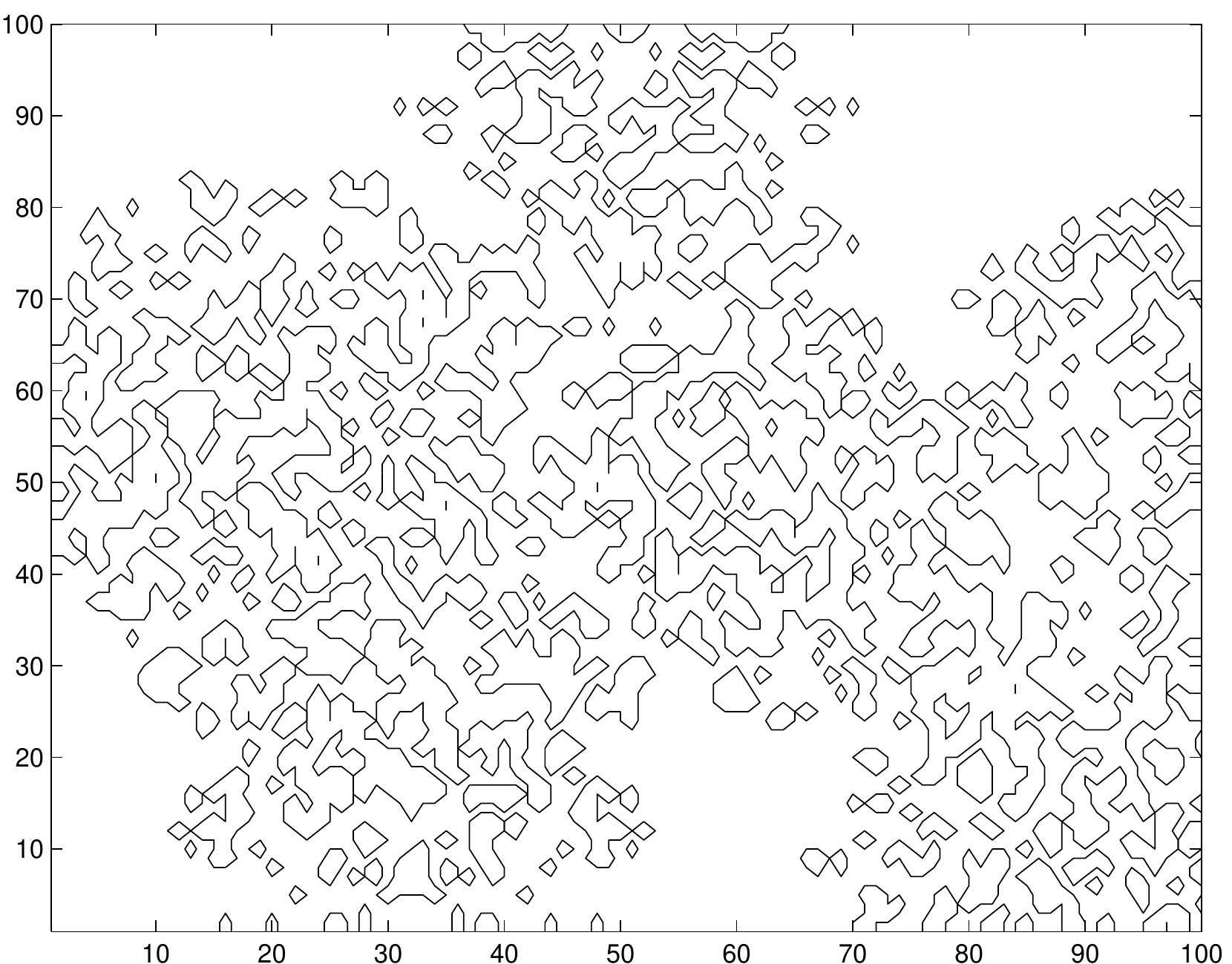}}
\subfigure[]{
    \label{fig:subfig:c}
    \includegraphics[width=0.3\textwidth]{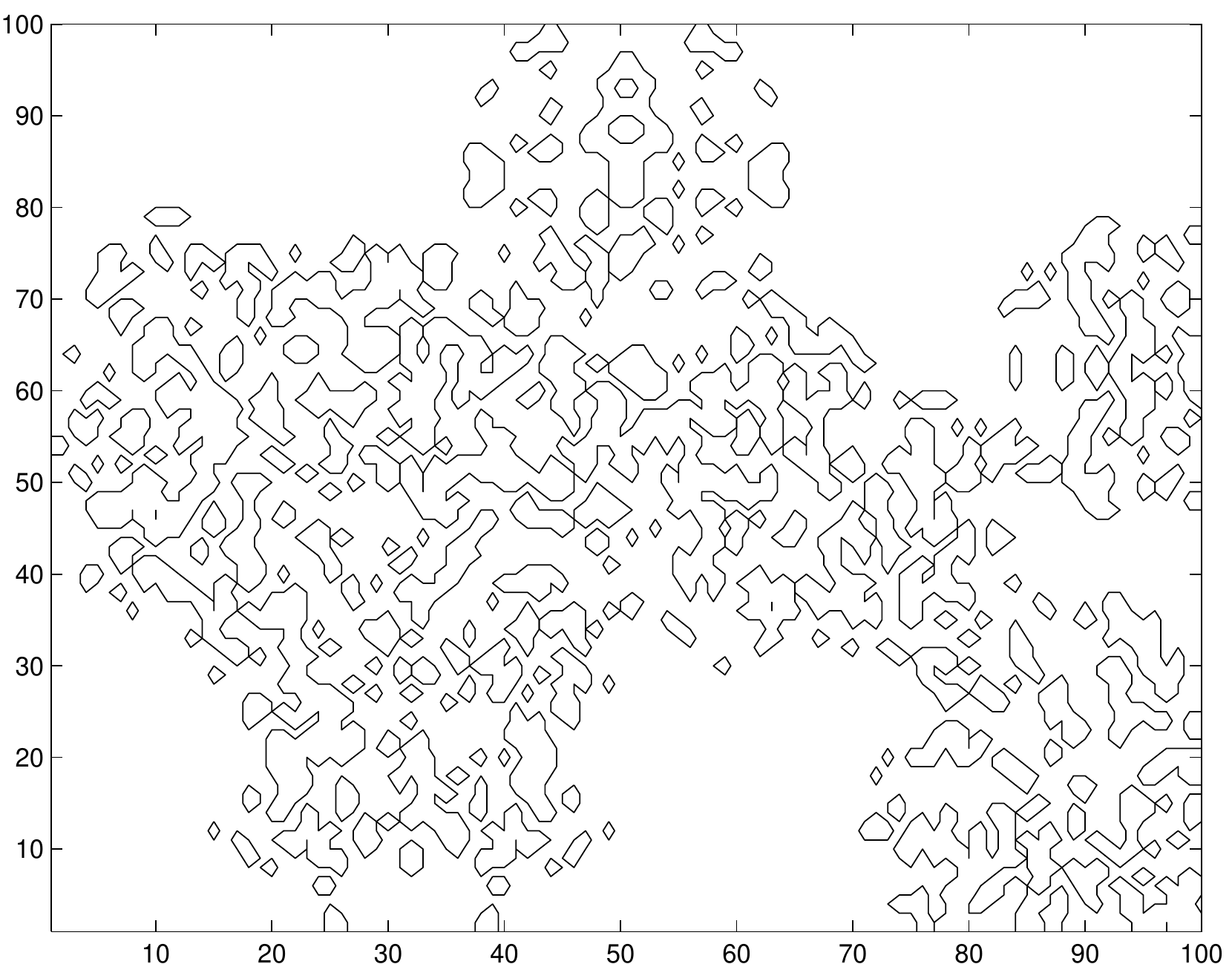}}
\hspace{0.1\textwidth} 
\subfigure[]{
    \label{fig:subfig:d}
    \includegraphics[width=0.3\textwidth]{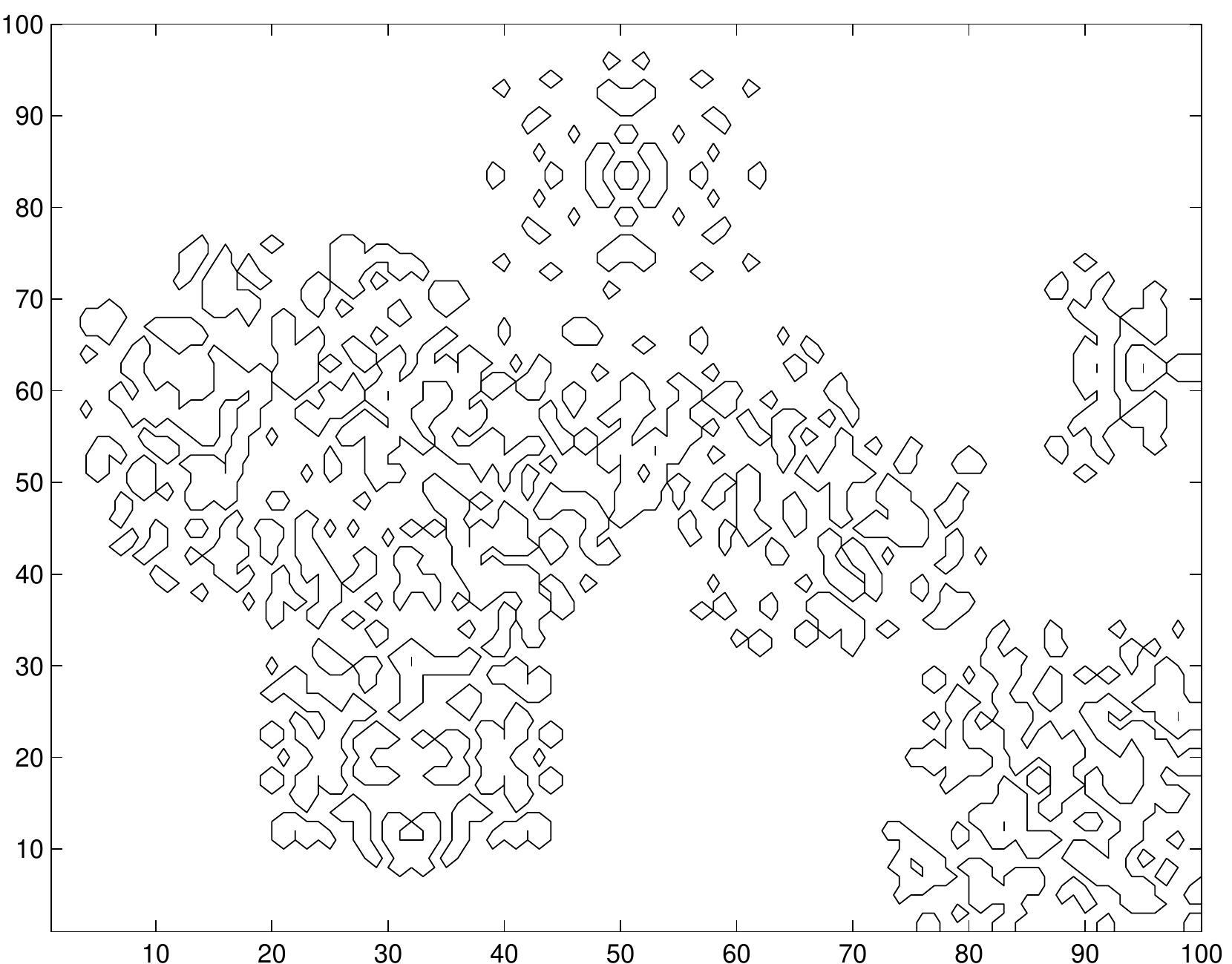}}
\subfigure[]{
    \label{fig:subfig:e}
    \includegraphics[width=0.3\textwidth]{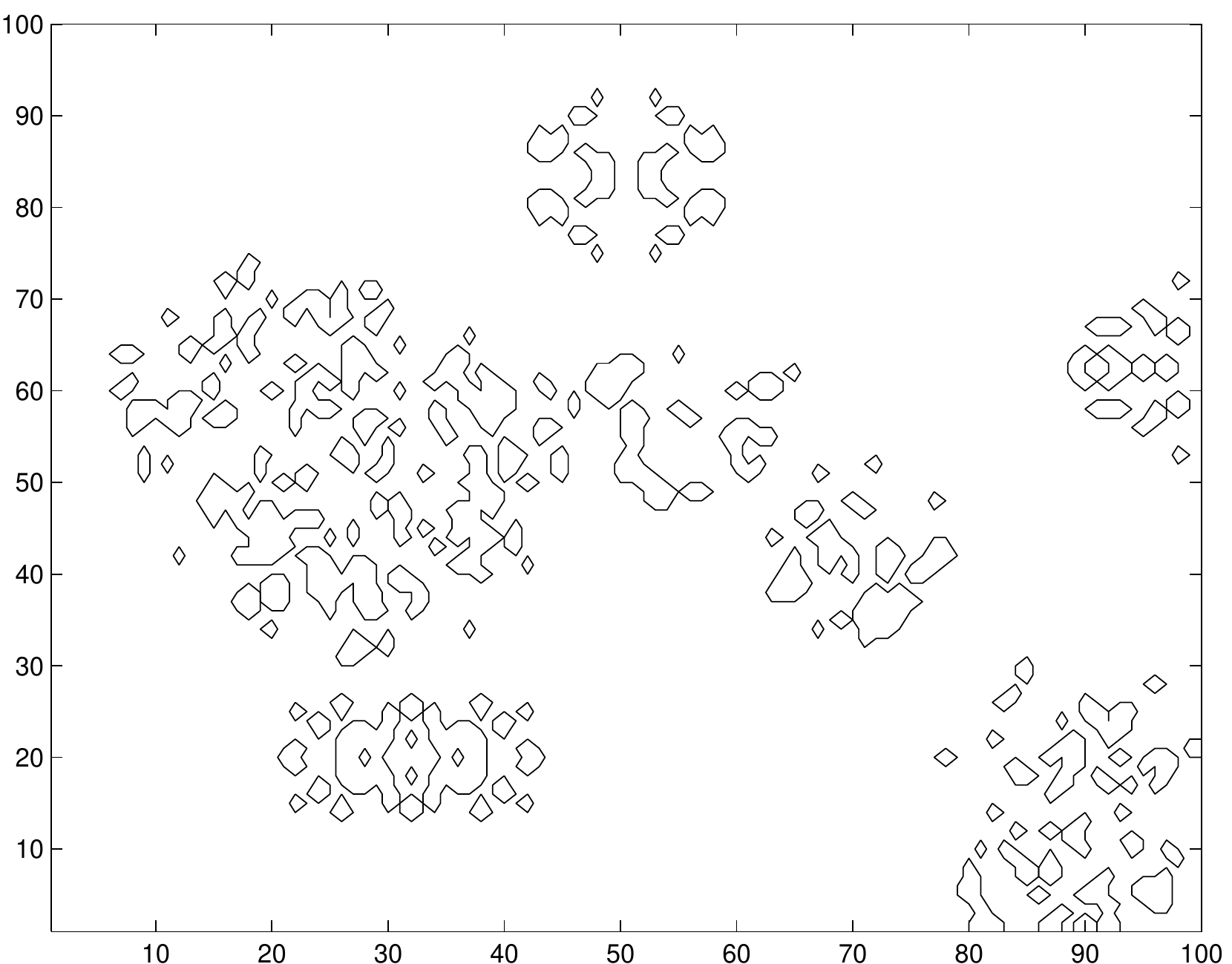}}
\hspace{0.1\textwidth} 
\subfigure[]{
    \label{fig:subfig:f}
    \includegraphics[width=0.3\textwidth]{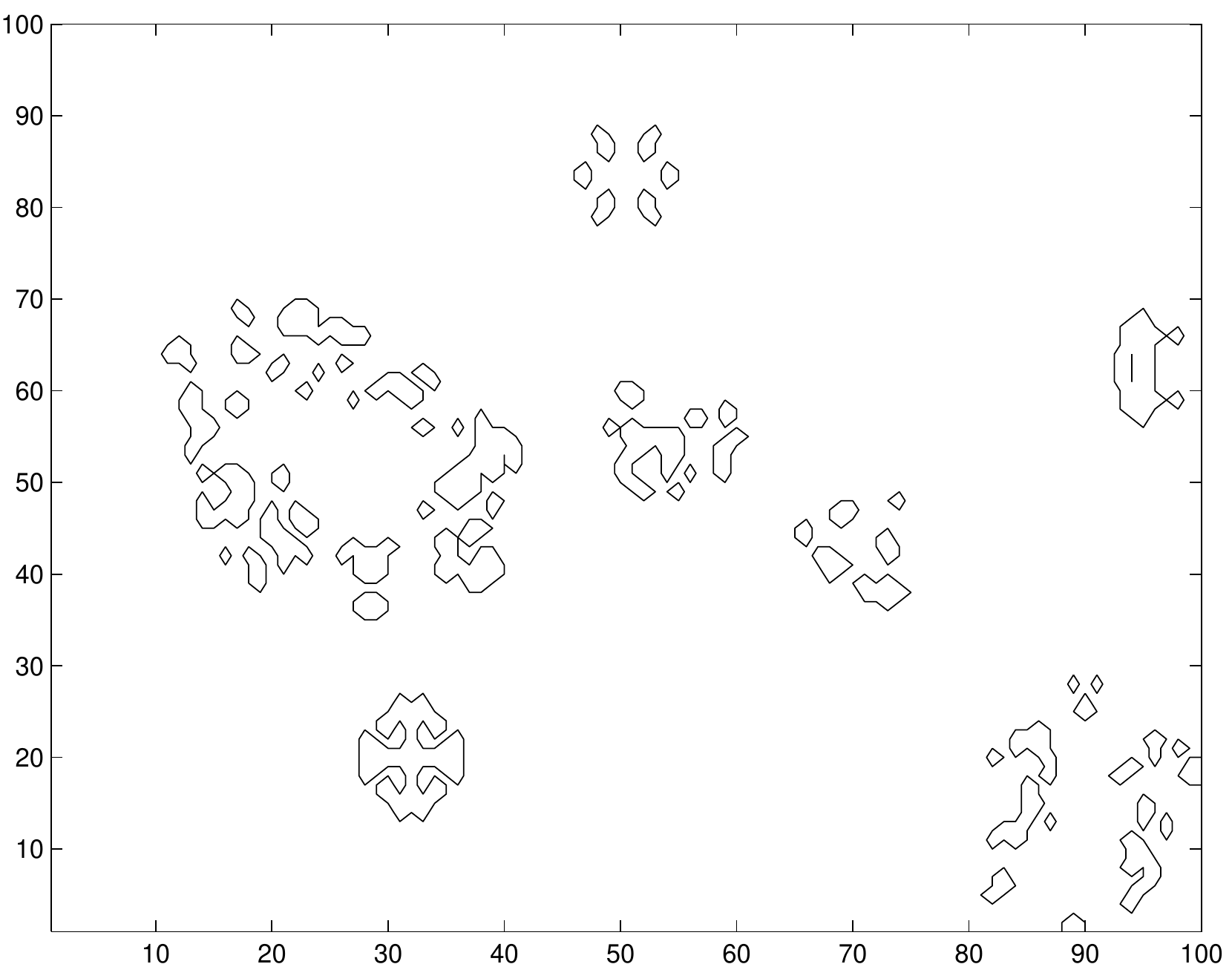}}
\subfigure[]{
    \label{fig:subfig:g}
    \includegraphics[width=0.3\textwidth]{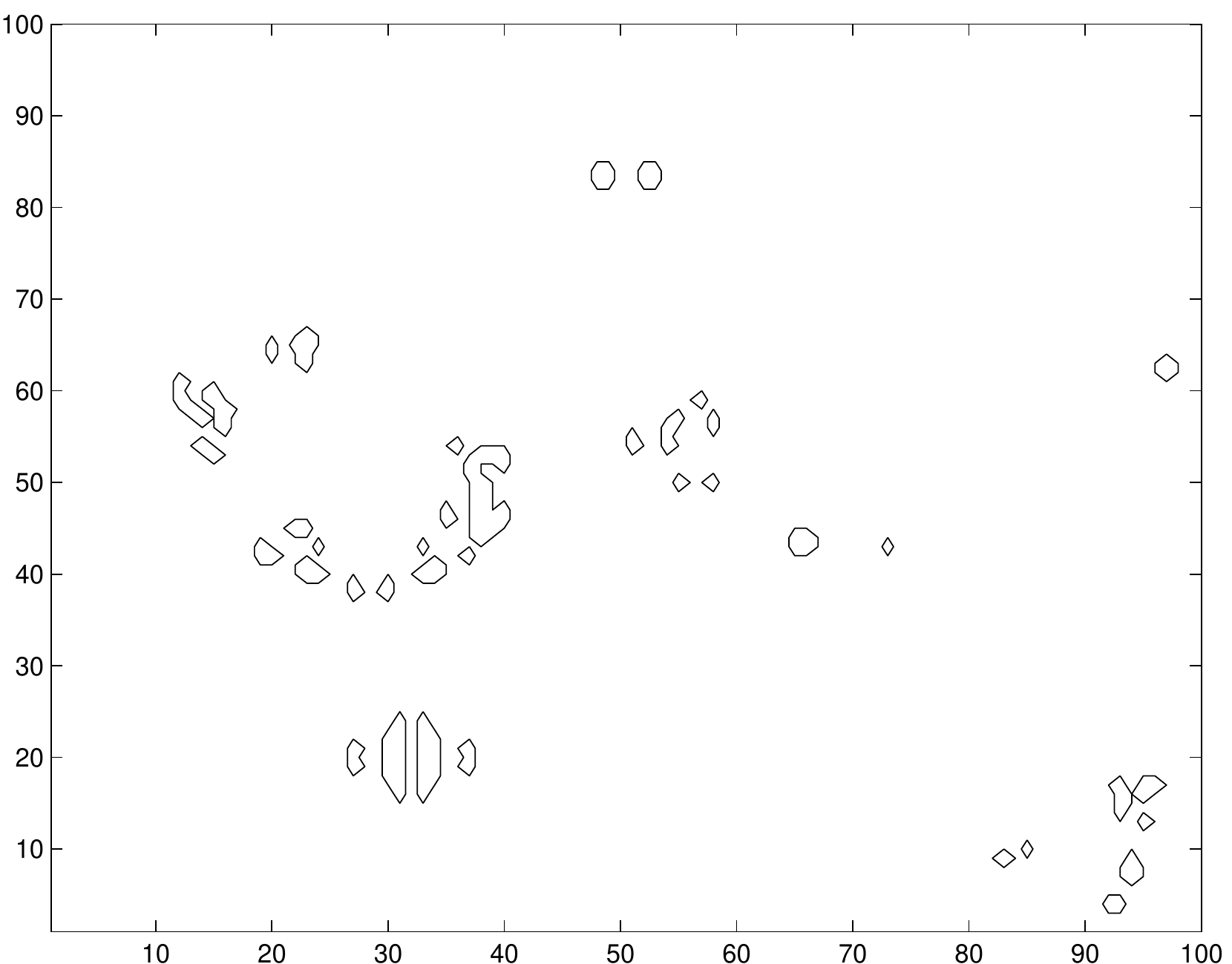}}
\caption{The contour levels corresponding to patches of compact cloudy air in several horizontal planes.}
\label{figcontours}
\end{figure}

\begin{figure}[htbp]
\centering
\subfigure[]{
    \label{fig:subfig:a}
    \includegraphics[width=0.3\textwidth]{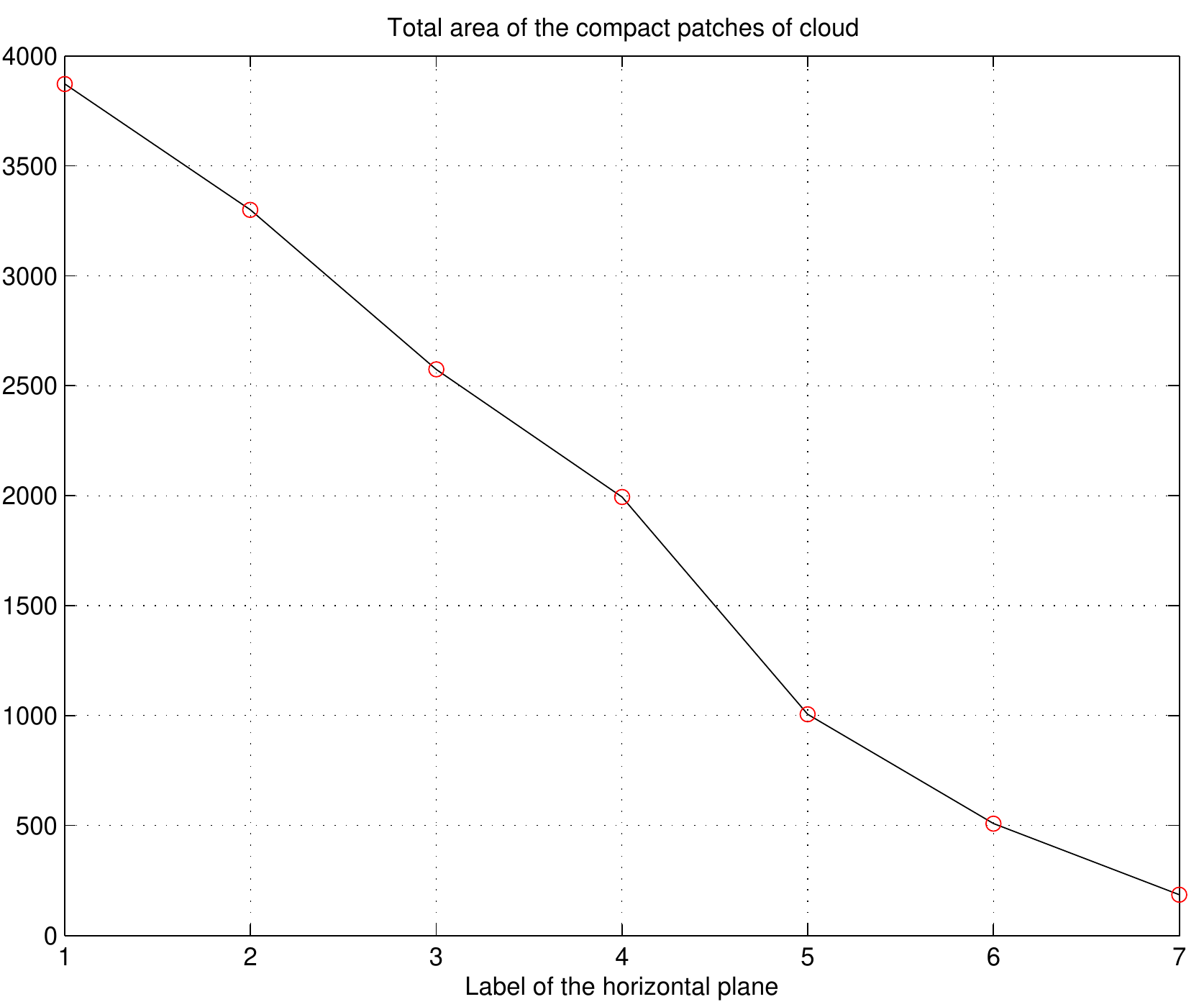}}
\hspace{0.1\textwidth} 
\subfigure[]{
    \label{fig:subfig:b}
    \includegraphics[width=0.3\textwidth]{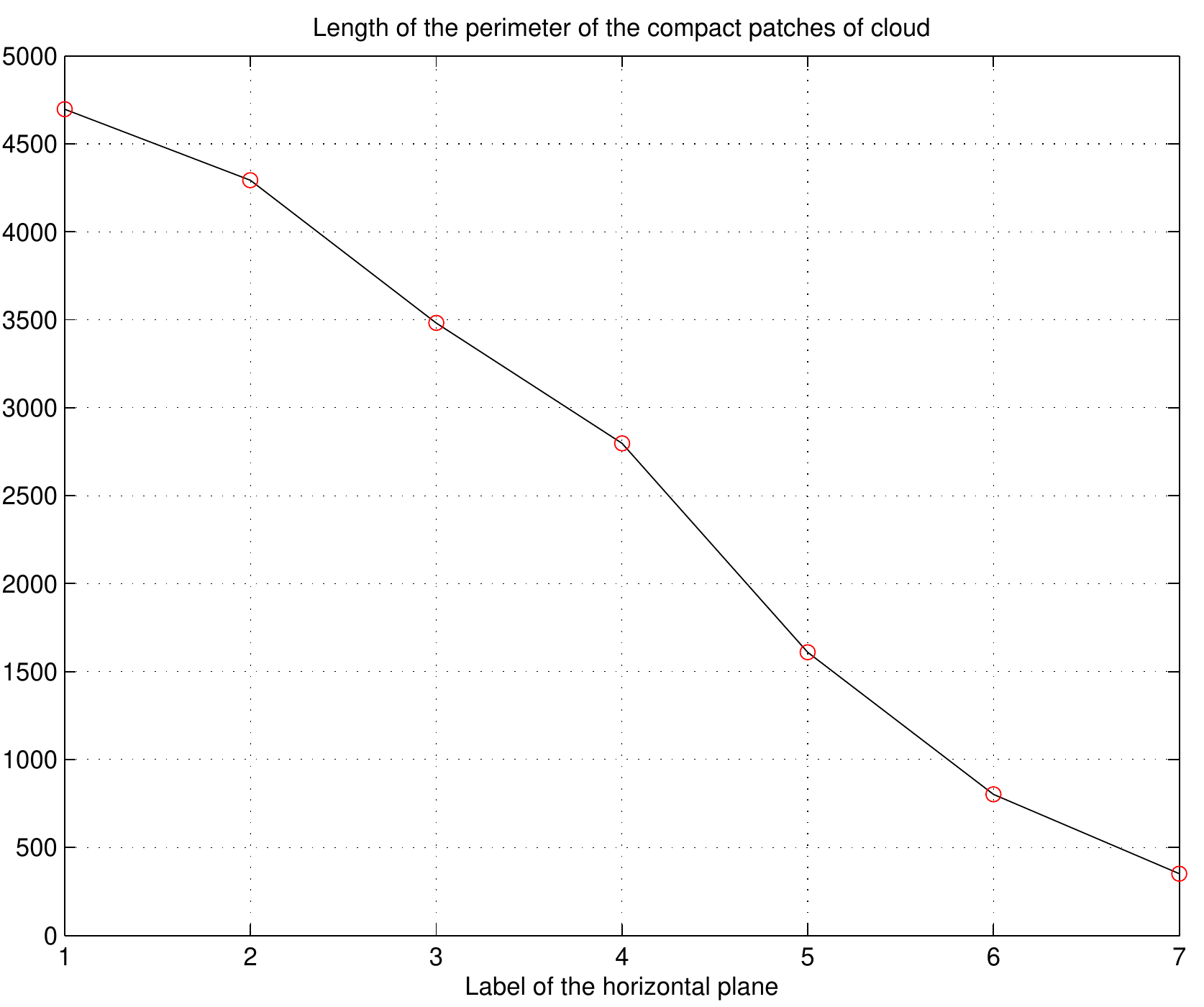}}
\caption{Area and length of the perimeter of compact patches of cloud air, with dependence on the height of the horizontal plane. The latter variable is mapped from the variable time of the iterative sequence. The parameters are: $a=1$, $ε=0.45$, $c=0.6$, $\gamma=0.06$ and time step $0.02$. The figures correspond to $200$ time steps.}
\label{figareaperim}
\end{figure}

It has been proved \cite{kapral1} that the iterative update adopted in Eq.(%
\ref{eq301}) and (\ref{eq302}) is equivalent with the curvature flow Eq.(\ref%
{eq305}). The dynamic structure function, defined as the discrete Fourier
transform of the correlation of the field $q\left( i,j,t\right) $ has at
large time a decay like $t^{1/2}$. We note however that this is a purely
geometric property, and the cloud-environment mixing is not taken into
account. We have however a lower bound to the rate of disappearence of
compact parcels of cloud in the late phases of the convection.
\FloatBarrier

\section{Conclusions}

We have examined three geometrical aspects that can be important in the quantitative studies of the exchanges of heat and vapor between the cloud and the environment.

The first model, regarding the {\it fingering} instability can be an important step in representing the fractalization of the boundary of the clouds.

The second model, intended to allow a quantitative description of the {\it cusp} singularity of the interface, can be also implemented in any study of the lateral exchanges cloud-environment. It can be developed further for the study of the process of absorbtion of parcels of environmental air inside the cloud, but this is a problem of complex function with a certain difficulty.

Finally we have considered the possibility to describe the loss of compacity of the rising column by a discrete, coupled lattice map model. At least the basic facts of decay of the convection and loss of continuity of the cloud tower, can be examined using this model.

A wide range of similar models can now be proposed and further development can be considered.

{\bf Acknowledgments}

This work is partially supported by the Contract PN 09 39 01 01.

\bigskip

\begin{appendices}

\section{Appendix. The interface developing fingers in Laplacian growth} \label{app:appendixa}
\renewcommand{\theequation}{A.\arabic{equation}} \setcounter{equation}{0}

\subsection{Interface between infinite regions}
The evolution of the front of the expanding cloud, in any horizontal plane, is represented through the conformal mapping from the lower half of the \textquotedblleft mathematical plane\textquotedblright\ on the region below $\Gamma$, the interior of the cloud. The\emph{\ Laplacian growth} is described by the \textbf{Polubarinova - Galin} (P-G) equation%
\begin{eqnarray}
\mathbf{Im}\left( \frac{\partial f\left( z,t\right) }{\partial z}\overline{%
\frac{\partial f\left( z,t\right) }{\partial t}}\right) &=&1\ \text{at}\ \
z=x-i0  \label{eqa106} \\
&&\text{(for }z\text{ just below the real }x\text{ axis)}  \notag
\end{eqnarray}%
A class of solutions is \cite{poncemineev0}, \cite{poncemineev1}, \cite{poncemineev2}%
\begin{equation}
f_{\inf }\left( z,t\right) =z-it-i\sum\limits_{l=1}^{N+1}\alpha _{l}\log 
\left[ z-\zeta _{l}\left( t\right) \right]  \label{eqa108}
\end{equation}%
where 
\begin{equation}
\alpha _{l}\equiv \alpha _{l}^{\prime }+i\alpha _{l}^{\prime \prime }\ \ 
\text{are\ \ }N+1\text{\ \ complex constants}  \label{eqa109}
\end{equation}%
and $\left( \alpha ^{\prime },\alpha ^{\prime \prime }\right) $\ \ are real.%
\begin{equation}
\zeta _{l}\equiv \xi _{l}+i\eta _{l}\ \ \text{are\ \ }N+1\text{\ \
singularities}  \label{eqa110}
\end{equation}%
simple poles of $\frac{\partial f}{\partial z}$ that move in time.

\bigskip

Let us calculate explicitly%
\begin{equation}
\frac{\partial f}{\partial z}=1-i\sum\limits_{l=1}^{N+1}\alpha _{l}\frac{1}{%
z-\zeta _{l}\left( t\right) }  \label{eqa111}
\end{equation}%
and%
\begin{equation}
\frac{\partial f}{\partial t}=-i-i\sum\limits_{l=1}^{N+1}\alpha _{l}\frac{1%
}{z-\zeta _{l}\left( t\right) }\frac{d\zeta _{l}\left( t\right) }{dt}
\label{eqa112}
\end{equation}%
\begin{equation}
\overline{\frac{\partial f}{\partial t}}=i+i\sum\limits_{l=1}^{N+1}%
\overline{\alpha _{l}}\frac{1}{\overline{z}-\overline{\zeta _{l}\left(
t\right) }}\overline{\frac{d\zeta _{l}\left( t\right) }{dt}}  \label{eqa113}
\end{equation}%
and the product is%
\begin{eqnarray}
\frac{\partial f}{\partial z}\overline{\frac{\partial f}{\partial t}}
&=&\left( 1-i\sum\limits_{l=1}^{N+1}\alpha _{l}\frac{1}{z-\zeta _{l}\left(
t\right) }\right) \left( i+i\sum\limits_{l=1}^{N+1}\overline{\alpha _{l}}%
\frac{1}{\overline{z}-\overline{\zeta _{l}\left( t\right) }}\overline{\frac{%
d\zeta _{l}\left( t\right) }{dt}}\right)  \label{eqa115} \\
&=&i  \notag \\
&&+\sum\limits_{l=1}^{N+1}\alpha _{l}\frac{1}{z-\zeta _{l}\left( t\right) }
\notag \\
&&+i\sum\limits_{l=1}^{N+1}\overline{\alpha _{l}}\frac{1}{\overline{z}-%
\overline{\zeta _{l}\left( t\right) }}\overline{\frac{d\zeta _{l}\left(
t\right) }{dt}}  \notag \\
&&+\sum\limits_{l=1}^{N+1}\sum\limits_{k=1}^{N+1}\alpha _{l}\overline{%
\alpha _{k}}\frac{1}{z-\zeta _{l}\left( t\right) }\frac{1}{\overline{z}-%
\overline{\zeta _{k}\left( t\right) }}\overline{\frac{d\zeta _{k}\left(
t\right) }{dt}}  \notag
\end{eqnarray}%
This must be taken for the line that in the mathematical (complex) plane $z$ represent
the interface%
\begin{equation}
z\rightarrow x-i0  \label{eqa116}
\end{equation}%
which means that $z$ will be replaced everywhere with real $x$. Before doing this we make a test, looking at what this expression becomes
when we replace 
\begin{equation}
z\rightarrow \overline{\zeta _{m}\left( t\right) }  \label{eqa117}
\end{equation}%
and find for the first two terms%
\begin{eqnarray}
&&+\sum\limits_{l=1}^{N+1}\alpha _{l}\frac{1}{z-\zeta _{l}\left( t\right) }%
+i\sum\limits_{l=1}^{N+1}\overline{\alpha _{l}}\frac{1}{\overline{z}-%
\overline{\zeta _{l}\left( t\right) }}\overline{\frac{d\zeta _{l}\left(
t\right) }{dt}}  \label{eqa118} \\
&\rightarrow &\sum\limits_{l=1}^{N+1}\alpha _{l}\frac{1}{\overline{\zeta
_{m}\left( t\right) }-\zeta _{l}\left( t\right) }+i\sum\limits_{l=1}^{N+1}%
\overline{\alpha _{l}}\frac{1}{\zeta _{m}\left( t\right) -\overline{\zeta
_{l}\left( t\right) }}\overline{\frac{d\zeta _{l}\left( t\right) }{dt}} 
\notag
\end{eqnarray}%
According to \cite{poncemineev1}, \cite{poncemineev2} the replacement $z\rightarrow \overline{%
\zeta _{m}}$ in the expression of the function $f$ produces \emph{constants}
due to the fact that $f$ verifies the equation \textbf{Polubarinova Galin}, as will be proved further below.
These constants are denoted $\beta _{m}$. After that we take the imaginary part%
\begin{eqnarray}
&&1  \label{eqa119} \\
&&+\mathbf{Im}\left[ \sum\limits_{l=1}^{N+1}\alpha _{l}\frac{1}{x-\zeta
_{l}\left( t\right) }\right.  \notag \\
&&+i\sum\limits_{l=1}^{N+1}\overline{\alpha _{l}}\frac{1}{x-\overline{\zeta
_{l}\left( t\right) }}\overline{\frac{d\zeta _{l}\left( t\right) }{dt}} 
\notag \\
&&\left. +\sum\limits_{l=1}^{N+1}\sum\limits_{k=1}^{N+1}\alpha _{l}%
\overline{\alpha _{k}}\frac{1}{x-\zeta _{l}\left( t\right) }\frac{1}{x-%
\overline{\zeta _{k}\left( t\right) }}\overline{\frac{d\zeta _{k}\left(
t\right) }{dt}}\right]  \notag \\
&=&1  \notag
\end{eqnarray}

\subsection{The constraints resulting from the P-G equation}
\subsubsection{The algebraic system}
The invariants of the solutions to the equations P-G are%
\begin{equation}
\beta _{k}=f_{k}\left( \overline{\zeta _{k}},t\right) =\overline{\zeta _{k}}%
-it-i\sum\limits_{l=1}^{N+1}\alpha _{l}\log \left( \overline{\zeta _{k}}%
-\zeta _{l}\right)  \label{eaq130}
\end{equation}

Here we apply the operator $d/dt$%
\begin{equation}
0=\frac{d\overline{\zeta _{k}\left( t\right) }}{dt}-i-i\sum%
\limits_{l=1}^{N+1}\alpha _{l}\frac{1}{\overline{\zeta _{k}\left( t\right) }%
-\zeta _{l}\left( t\right) }\left( \frac{d\overline{\zeta _{k}\left(
t\right) }}{dt}-\frac{d\zeta _{l}\left( t\right) }{dt}\right)  \label{eqa131}
\end{equation}

\bigskip

Or, write first the real and imaginary parts of the constants $\beta _{k}$.
We have%
\begin{eqnarray}
\log \left( \overline{\zeta _{k}}-\zeta _{l}\right) &=&\log \left[ \left(
\xi _{k}-\xi _{l}\right) +i\left( -\eta _{k}-\eta _{l}\right) \right]
\label{eqa132} \\
&=&\log \left( \left\vert \left( \xi _{k}-\xi _{l}\right) -i\left( \eta
_{k}+\eta _{l}\right) \right\vert \right)  \notag \\
&&+i\arg \left[ \left( \xi _{k}-\xi _{l}\right) -i\left( \eta _{k}+\eta
_{l}\right) \right]  \notag
\end{eqnarray}%
The modulus is%
\begin{eqnarray}
&&\log \left[ \left\vert \left( \xi _{k}-\xi _{l}\right) -i\left( \eta
_{k}+\eta _{l}\right) \right\vert \right]  \label{eqa134} \\
&=&\frac{1}{2}\log \left[ \left( \xi _{k}-\xi _{l}\right) ^{2}+\left( \eta
_{k}+\eta _{l}\right) ^{2}\right]  \notag
\end{eqnarray}%
The phase is%
\begin{equation}
\arg \left[ \left( \xi _{k}-\xi _{l}\right) -i\left( \eta _{k}+\eta
_{l}\right) \right] =\arctan \left( \frac{\eta _{k}+\eta _{l}}{\xi _{k}-\xi
_{l}}\right)  \label{eqa135}
\end{equation}%
When the abscissa $\xi _{k}-\xi _{l}$ is negative the complex argument $%
\overline{\zeta _{k}}-\zeta _{l}$ crosses the cut and we have to add $\iota
\pi $.

In addition, when two singularities are such that  $\overline{\zeta _{k}}=\zeta _{l}$ we
have%
\begin{eqnarray}
\overline{\zeta _{k}}-\zeta _{l} &=&0\rightarrow \xi _{k}-\xi _{l}=0\ \ 
\text{and}\ \ \eta _{k}+\eta _{l}\rightarrow 2\eta _{k}  \label{eqa136} \\
\text{we have\ }\arctan \left( \frac{\eta _{k}+\eta _{l}}{\xi _{k}-\xi _{l}}%
\right) &\rightarrow &\frac{\pi }{2}  \notag
\end{eqnarray}

\bigskip

Then%
\begin{eqnarray}
\beta _{k}^{\prime } &=&\xi _{k}-\sum\limits_{l\neq k}\alpha _{l}^{\prime
}\arctan \left( \frac{\eta _{k}+\eta _{l}}{\xi _{k}-\xi _{l}}\right)
\label{eqa137} \\
&&-\frac{1}{2}\pi \alpha _{k}^{\prime }-\pi \sum\limits_{l;\xi _{k}-\xi
_{l}<0}\alpha _{l}^{\prime }+\alpha _{k}^{\prime \prime }\log \left( 2\eta
_{k}\right)  \notag \\
&&+\frac{1}{2}\sum\limits_{l\neq k}\alpha _{l}^{\prime \prime }\log \left[
\left( \xi _{k}-\xi _{l}\right) ^{2}+\left( \eta _{k}+\eta _{l}\right) ^{2}%
\right]  \notag
\end{eqnarray}%
and for the imaginary part%
\begin{eqnarray}
\beta _{k}^{\prime \prime } &=&-t-\eta _{k}-\alpha _{k}^{\prime }\log \left(
2\eta _{k}\right) -\frac{1}{2}\pi \alpha _{k}^{\prime \prime }-\pi
\sum\limits_{l;\xi _{k}-\xi _{l}<0}\alpha _{l}^{\prime \prime }
\label{eqa138} \\
&&-\sum\limits_{l\neq k}\alpha _{l}^{\prime \prime }\arctan \left( \frac{%
\eta _{k}+\eta _{l}}{\xi _{k}-\xi _{l}}\right)  \notag \\
&&-\frac{1}{2}\sum\limits_{l\neq k}\alpha _{l}^{\prime }\log \left[ \left(
\xi _{k}-\xi _{l}\right) ^{2}+\left( \eta _{k}+\eta _{l}\right) ^{2}\right] 
\notag
\end{eqnarray}

Note that the in-determination induced by the function $\log $ is made
explicit by the multivalued function $\arctan $. The arguments of the $%
\arctan $ are real and the first determination $Arc\tan $ is finite for any
choice of variables.

\bigskip

The unknown functions are the derivatives with
respect to time, of the real $\left( \xi _{k}\right) $ part and imaginary $%
\left( \eta _{k}\right) $ functions of time-dependent positions of the singularities.%
\begin{equation}
\frac{d\xi _{k}}{dt}\ ,\ \frac{d\eta _{k}}{dt}  \label{eqa139}
\end{equation}

Let us introduce systematic standard notations. We note%
\begin{eqnarray}
X\left( 1\right) &\equiv &\frac{d\xi _{1}}{dt},X\left( 2\right) \equiv \frac{%
d\xi _{2}}{dt},...,X\left( N+1\right) \equiv \frac{d\xi _{N+1}}{dt},
\label{eqa140} \\
X\left( N+2\right) &\equiv &\frac{d\eta _{1}}{dt},X\left( N+3\right) \equiv 
\frac{d\eta _{2}}{dt},...,X\left( N+1+N+1\right) \equiv \frac{d\eta _{N+1}}{%
dt}  \notag
\end{eqnarray}%
The variables%
\begin{equation}
X\left( i\right) ,\ \ \text{for}\ \ i=1,2N+2  \label{eqa141}
\end{equation}%
are the unknown.

\subsubsection{The coefficients as result from the equation for the real
part ($d\protect\beta _{k}^{\prime }/dt=0$)}

We prepare the coefficients in the linear system for $\frac{d\xi _{k}}{dt}$
and $\frac{d\eta _{k}}{dt}$ by calculating the derivatives%
\begin{eqnarray}
&&\frac{d}{dt}\arctan \left( \frac{\eta _{k}+\eta _{l}}{\xi _{k}-\xi _{l}}%
\right)  \label{eqa142} \\
&=&\frac{1}{1+\left( \frac{\eta _{k}+\eta _{l}}{\xi _{k}-\xi _{l}}\right)
^{2}}\left[ \frac{1}{\xi _{k}-\xi _{l}}\left( \frac{d\eta _{k}}{dt}+\frac{%
d\eta _{l}}{dt}\right) -\frac{\eta _{k}+\eta _{l}}{\left( \xi _{k}-\xi
_{l}\right) ^{2}}\left( \frac{d\xi _{k}}{dt}-\frac{d\xi _{l}}{dt}\right) %
\right]  \notag \\
&=&\frac{1}{\left( \xi _{k}-\xi _{l}\right) ^{2}+\left( \eta _{k}+\eta
_{l}\right) ^{2}}\left[ \left( \xi _{k}-\xi _{l}\right) \left( \frac{d\eta
_{k}}{dt}+\frac{d\eta _{l}}{dt}\right) -\left( \eta _{k}+\eta _{l}\right)
\left( \frac{d\xi _{k}}{dt}-\frac{d\xi _{l}}{dt}\right) \right]  \notag
\end{eqnarray}%
Further%
\begin{eqnarray}
&&\frac{d}{dt}\log \left[ \left( \xi _{k}-\xi _{l}\right) ^{2}+\left( \eta
_{k}+\eta _{l}\right) ^{2}\right]  \label{eqa143} \\
&=&\frac{1}{\left( \xi _{k}-\xi _{l}\right) ^{2}+\left( \eta _{k}+\eta
_{l}\right) ^{2}}\left[ 2\left( \xi _{k}-\xi _{l}\right) \left( \frac{d\xi
_{k}}{dt}-\frac{d\xi _{l}}{dt}\right) +2\left( \eta _{k}+\eta _{l}\right)
\left( \frac{d\eta _{k}}{dt}+\frac{d\eta _{l}}{dt}\right) \right]  \notag
\end{eqnarray}%
and%
\begin{eqnarray}
\frac{d}{dt}\log \left( 2\eta _{k}\right) &=&\frac{d}{dt}\left[ \log \left(
\eta _{k}\right) +\log \left( 2\right) \right]  \label{eqa144} \\
&=&\frac{1}{\eta _{k}}\frac{d\eta _{k}}{dt}  \notag
\end{eqnarray}%
Then we write%
\begin{eqnarray}
&&\frac{d\beta _{k}^{\prime }}{dt}  \label{eqa145} \\
&=&\frac{d\xi _{k}}{dt}  \notag \\
&&\hspace{-1.25cm}\hspace{-1.25cm}-\sum\limits_{l\neq k}\alpha _{l}^{\prime
}\frac{1}{\left( \xi _{k}-\xi _{l}\right) ^{2}+\left( \eta _{k}+\eta
_{l}\right) ^{2}}\left[ \left( \xi _{k}-\xi _{l}\right) \left( \frac{d\eta
_{k}}{dt}+\frac{d\eta _{l}}{dt}\right) -\left( \eta _{k}+\eta _{l}\right)
\left( \frac{d\xi _{k}}{dt}-\frac{d\xi _{l}}{dt}\right) \right]  \notag \\
&&\hspace{-1.25cm}\hspace{-1.25cm}+\alpha _{k}^{\prime \prime }\frac{1}{\eta
_{k}}\frac{d\eta _{k}}{dt}  \notag \\
&&\hspace{-1.25cm}\hspace{-1.25cm}+\frac{1}{2}\sum\limits_{l\neq k}\alpha
_{l}^{\prime \prime }\frac{1}{\left( \xi _{k}-\xi _{l}\right) ^{2}+\left(
\eta _{k}+\eta _{l}\right) ^{2}}\left[ 2\left( \xi _{k}-\xi _{l}\right)
\left( \frac{d\xi _{k}}{dt}-\frac{d\xi _{l}}{dt}\right) +2\left( \eta
_{k}+\eta _{l}\right) \left( \frac{d\eta _{k}}{dt}+\frac{d\eta _{l}}{dt}%
\right) \right]  \notag
\end{eqnarray}

In Eq. $k$. For $k=1,N+1$.

Coefficient of $\frac{d\xi _{k}}{dt}$ is%
\begin{eqnarray}
&&1+  \label{eqa146} \\
&&-\sum\limits_{l\neq k}\alpha _{l}^{\prime }\frac{1}{\left( \xi _{k}-\xi
_{l}\right) ^{2}+\left( \eta _{k}+\eta _{l}\right) ^{2}}\left[ -\left( \eta
_{k}+\eta _{l}\right) \right]  \notag \\
&&+\sum\limits_{l\neq k}\alpha _{l}^{\prime \prime }\frac{1}{\left( \xi
_{k}-\xi _{l}\right) ^{2}+\left( \eta _{k}+\eta _{l}\right) ^{2}}\left[
\left( \xi _{k}-\xi _{l}\right) \right]  \notag
\end{eqnarray}

Coefficient of $\frac{d\xi _{l}}{dt}$ , for $l=1,N+1$ but $l\neq k$; it is%
\begin{eqnarray}
&&-\alpha _{l}^{\prime }\frac{1}{\left( \xi _{k}-\xi _{l}\right) ^{2}+\left(
\eta _{k}+\eta _{l}\right) ^{2}}\left[ \left( \eta _{k}+\eta _{l}\right) %
\right]  \label{eqa147} \\
&&+\alpha _{l}^{\prime \prime }\frac{1}{\left( \xi _{k}-\xi _{l}\right)
^{2}+\left( \eta _{k}+\eta _{l}\right) ^{2}}\left[ -\left( \xi _{k}-\xi
_{l}\right) \right]  \notag
\end{eqnarray}

Coefficient of $\frac{d\eta _{k}}{dt}$; this means the the subscript of $%
\eta _{l}$ is the same as the number of the line $k$. The coefficient is%
\begin{eqnarray}
&&-\sum\limits_{l\neq k}\alpha _{l}^{\prime }\frac{1}{\left( \xi _{k}-\xi
_{l}\right) ^{2}+\left( \eta _{k}+\eta _{l}\right) ^{2}}\left[ \left( \xi
_{k}-\xi _{l}\right) \right]  \label{eqa148} \\
&&+\alpha _{k}^{\prime \prime }\frac{1}{\eta _{k}}  \notag \\
&&+\sum\limits_{l\neq k}\alpha _{l}^{\prime \prime }\frac{1}{\left( \xi
_{k}-\xi _{l}\right) ^{2}+\left( \eta _{k}+\eta _{l}\right) ^{2}}\left[
\left( \eta _{k}+\eta _{l}\right) \right]  \notag
\end{eqnarray}

Coefficient of $\frac{d\eta _{l}}{dt}$; for $l=1,N+1$ with the constraint $%
l\neq k$; the subscript of $\eta _{l}$ is different from the number of the
line $k$. The coefficient is%
\begin{eqnarray}
&&-\alpha _{l}^{\prime }\frac{1}{\left( \xi _{k}-\xi _{l}\right) ^{2}+\left(
\eta _{k}+\eta _{l}\right) ^{2}}\left[ \left( \xi _{k}-\xi _{l}\right) %
\right]  \label{eqa149} \\
&&+\alpha _{l}^{\prime \prime }\frac{1}{\left( \xi _{k}-\xi _{l}\right)
^{2}+\left( \eta _{k}+\eta _{l}\right) ^{2}}\left[ \left( \eta _{k}+\eta
_{l}\right) \right]  \notag
\end{eqnarray}

Free term; it is $0$.

\bigskip

\bigskip

\paragraph{The list of types of coefficients.}

We note that there are few types.

\begin{figure}[h]
\includegraphics[height=7cm]{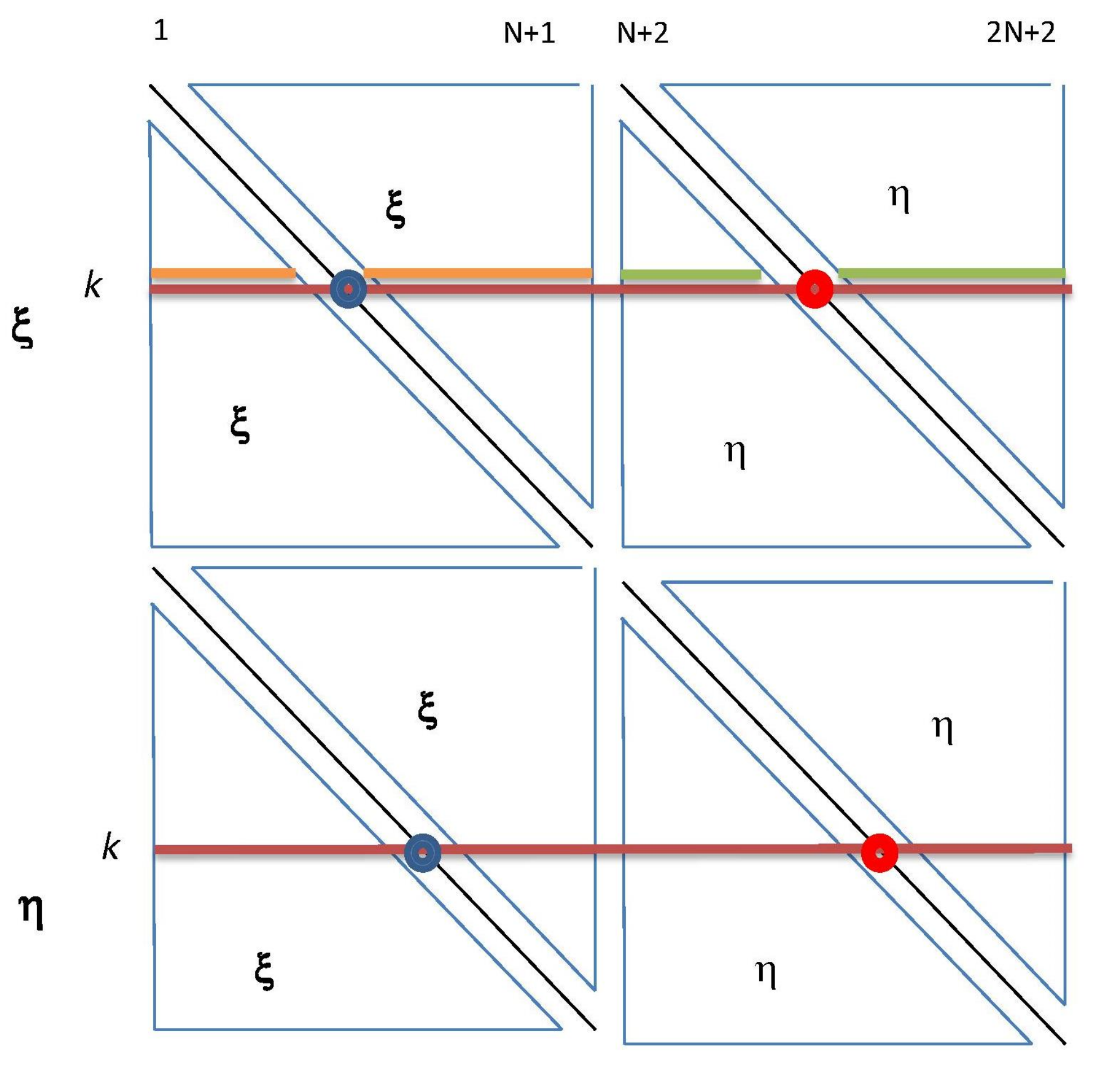}
\hfill
\includegraphics[height=6cm]{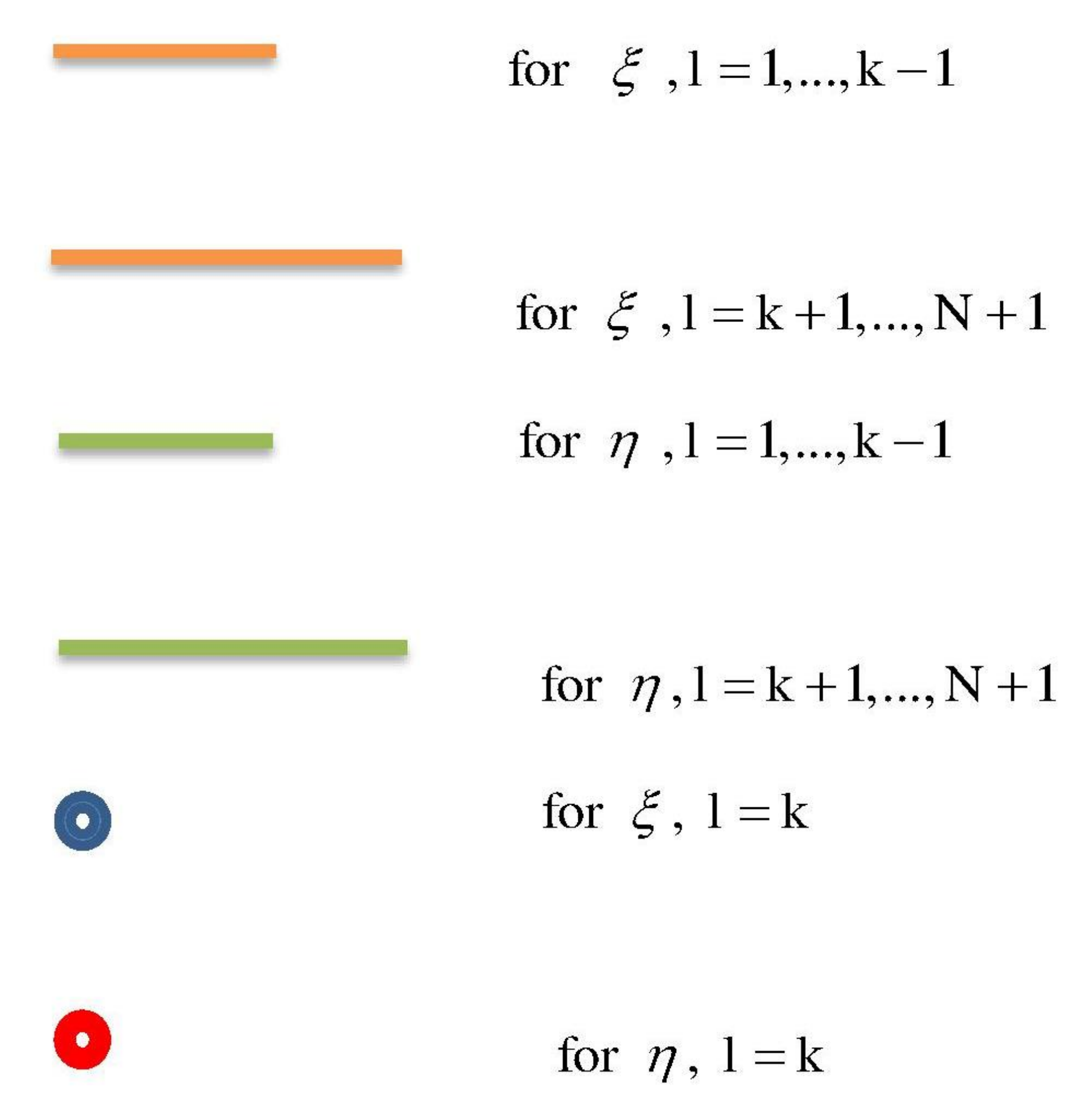}
\caption{The schematic representation of the matrix of the system.}
\label{fig2}
\end{figure}
%
%

First divide the full $\left( 2N+2\right) \times \left( 2N+2\right) $ matrix
into four squares each of $\left( N+1\right) \times \left( N+1\right) $.

\bigskip

There are $6$ regions along a line of the matrix $A(I,J)$.

the number of types is however less due to repetitions.

We note that the coefficients for line $I$ , coming from $\frac{d\xi _{J}}{dt%
}$ but for $J\neq I$ is one type. They are to be found for \textquotedblleft
columns\textquotedblright\ $J$ that are in the first Jacobi square matrix%
\begin{equation}
1\leq I\leq N+1\ ,\ 1\leq J\leq N+1  \label{eqa401}
\end{equation}

The first part is%
\begin{equation}
1\leq I\leq N+1\ ,\ 1\leq J<I  \label{eqa402}
\end{equation}%
This is the type $W1\left( k,l\right) $. The expression is%
\begin{eqnarray}
&&W1\left( k,l\right) =  \label{eqa403} \\
&&-\alpha _{l}^{\prime }\frac{1}{\left( \xi _{k}-\xi _{l}\right) ^{2}+\left(
\eta _{k}+\eta _{l}\right) ^{2}}\left[ \left( \eta _{k}+\eta _{l}\right) %
\right]  \notag \\
&&+\alpha _{l}^{\prime \prime }\frac{1}{\left( \xi _{k}-\xi _{l}\right)
^{2}+\left( \eta _{k}+\eta _{l}\right) ^{2}}\left[ -\left( \xi _{k}-\xi
_{l}\right) \right]  \notag
\end{eqnarray}%
where%
\begin{eqnarray}
k &=&I  \label{eqa404} \\
l &=&J  \notag
\end{eqnarray}

\bigskip

Then we have a different expression for the local diagonal $J=I$. 
\begin{equation}
J=I  \label{eqa405}
\end{equation}%
This type is $W2\left( k,k\right) $. The expression is%
\begin{eqnarray}
W2\left( k,k\right) &=&  \label{eqa406} \\
&&1+  \notag \\
&&-\sum\limits_{l\neq k}\alpha _{l}^{\prime }\frac{1}{\left( \xi _{k}-\xi
_{l}\right) ^{2}+\left( \eta _{k}+\eta _{l}\right) ^{2}}\left[ -\left( \eta
_{k}+\eta _{l}\right) \right]  \notag \\
&&+\sum\limits_{l\neq k}\alpha _{l}^{\prime \prime }\frac{1}{\left( \xi
_{k}-\xi _{l}\right) ^{2}+\left( \eta _{k}+\eta _{l}\right) ^{2}}\left[
\left( \xi _{k}-\xi _{l}\right) \right]  \notag
\end{eqnarray}%
where%
\begin{equation}
k=I=J  \label{eqa407}
\end{equation}

\bigskip

Along the line, for columns less than $J=N+1$ which is the end of the first
square, 
\begin{equation}
1\leq I\leq N+1\ ,\ I+1\leq J\leq N+1  \label{eqa408}
\end{equation}%
the type is again $W1\left( k,l\right) $.%
\begin{eqnarray}
&&W1\left( k,l\right) =  \label{eqa409} \\
&&-\alpha _{l}^{\prime }\frac{1}{\left( \xi _{k}-\xi _{l}\right) ^{2}+\left(
\eta _{k}+\eta _{l}\right) ^{2}}\left[ \left( \eta _{k}+\eta _{l}\right) %
\right]  \notag \\
&&+\alpha _{l}^{\prime \prime }\frac{1}{\left( \xi _{k}-\xi _{l}\right)
^{2}+\left( \eta _{k}+\eta _{l}\right) ^{2}}\left[ -\left( \xi _{k}-\xi
_{l}\right) \right]  \notag
\end{eqnarray}%
where%
\begin{eqnarray}
k &=&I  \label{eqa410} \\
l &=&J  \notag
\end{eqnarray}

\bigskip

Along the line $I$, we now go to the second square%
\begin{equation}
1\leq I\leq N+1\ ,\ \left( N+1\right) +1\leq J\leq \left( N+1\right) +N+1
\label{eqa411}
\end{equation}%
These columns come from $\frac{d\eta _{J}}{dt}$. Here again we have to make
the difference between $J\neq I$ and $J=I$.

For the first columns, before the local diagonal, 
\begin{equation}
1\leq I\leq N+1\ ,\ \left( N+1\right) +1\leq J<\left( N+1\right) +I
\label{eqa412}
\end{equation}%
the type is $W3\left( I,J\right) $. The expression is adapted to this range
of indices.%
\begin{eqnarray}
&&W3\left( k,l\right) =  \label{eqa413} \\
&&-\alpha _{l}^{\prime }\frac{1}{\left( \xi _{k}-\xi _{l}\right) ^{2}+\left(
\eta _{k}+\eta _{l}\right) ^{2}}\left[ \left( \xi _{k}-\xi _{l}\right) %
\right]  \notag \\
&&+\alpha _{l}^{\prime \prime }\frac{1}{\left( \xi _{k}-\xi _{l}\right)
^{2}+\left( \eta _{k}+\eta _{l}\right) ^{2}}\left[ \left( \eta _{k}+\eta
_{l}\right) \right]  \notag
\end{eqnarray}%
with%
\begin{eqnarray}
k &=&I  \label{eqa414} \\
l &=&J-\left( N+1\right)  \notag
\end{eqnarray}

\bigskip

Then comes the local diagonal in this square matrix%
\begin{eqnarray}
1 &\leq &I\leq N+1\ ,\text{with}~\ \ J-\left( N+1\right) =I  \label{eqa415}
\\
\ \text{for}\ \left( N+1\right) +1 &\leq &J\leq \left( N+1\right) +N+1 
\notag
\end{eqnarray}%
This is type $W4\left( k,l\right) $. The expression is%
\begin{eqnarray}
&&W4\left( k,k\right) =  \label{eqa416} \\
&&-\sum\limits_{l\neq k}\alpha _{l}^{\prime }\frac{1}{\left( \xi _{k}-\xi
_{l}\right) ^{2}+\left( \eta _{k}+\eta _{l}\right) ^{2}}\left[ \left( \xi
_{k}-\xi _{l}\right) \right]  \notag \\
&&+\alpha _{k}^{\prime \prime }\frac{1}{\eta _{k}}  \notag \\
&&+\sum\limits_{l\neq k}\alpha _{l}^{\prime \prime }\frac{1}{\left( \xi
_{k}-\xi _{l}\right) ^{2}+\left( \eta _{k}+\eta _{l}\right) ^{2}}\left[
\left( \eta _{k}+\eta _{l}\right) \right]  \notag
\end{eqnarray}%
where%
\begin{equation}
k=I=J-(N+1)  \label{eqa417}
\end{equation}

\bigskip

Next along the same line $I$, 
\begin{equation}
1\leq I\leq N+1\ ,\ \left( N+1\right) +I+1\leq J\leq \left( N+1\right) +N+1
\label{eqa418}
\end{equation}%
the type is again $W3\left( k,l\right) $. \ The expression is%
\begin{eqnarray}
&&W3\left( k,l\right) =  \label{eqa419} \\
&&-\alpha _{l}^{\prime }\frac{1}{\left( \xi _{k}-\xi _{l}\right) ^{2}+\left(
\eta _{k}+\eta _{l}\right) ^{2}}\left[ \left( \xi _{k}-\xi _{l}\right) %
\right]  \notag \\
&&+\alpha _{l}^{\prime \prime }\frac{1}{\left( \xi _{k}-\xi _{l}\right)
^{2}+\left( \eta _{k}+\eta _{l}\right) ^{2}}\left[ \left( \eta _{k}+\eta
_{l}\right) \right]  \notag
\end{eqnarray}%
where%
\begin{eqnarray}
k &=&I  \label{eqa420} \\
l &=&J-\left( N+1\right)  \notag
\end{eqnarray}%
for the range $\left( I,J\right) $ written above.

\bigskip

Let us consider application to few lines.

\paragraph{The line for the unknown $X\left( 1\right) $.}

This is $\frac{d\xi _{1}}{dt}$.

The first case consists of: the indice ($l$) of the coefficient coincides
with the number of the line (which is $k=1$).

This first case is the diagonal element. It comes from $\frac{d\xi _{1}}{dt}$%
. The model is $W2\left( k,k\right) $ for $k=I=1$. 
\begin{eqnarray}
&&A\left( 1,1\right) =  \label{eqa430} \\
&&1+  \notag \\
&&-\sum\limits_{l\neq 1}^{N+1}\alpha _{l}^{\prime }\frac{1}{\left( \xi
_{1}-\xi _{l}\right) ^{2}+\left( \eta _{1}+\eta _{l}\right) ^{2}}\left[
-\left( \eta _{1}+\eta _{l}\right) \right]  \notag \\
&&+\sum\limits_{l\neq 1}^{N+1}\alpha _{l}^{\prime \prime }\frac{1}{\left(
\xi _{1}-\xi _{l}\right) ^{2}+\left( \eta _{1}+\eta _{l}\right) ^{2}}\left[
\left( \xi _{1}-\xi _{l}\right) \right]  \notag
\end{eqnarray}

\bigskip

The other cases of entries coming from $\frac{d\xi _{l}}{dt}$ have the
indice of the matrix entry (coefficient, $l$) different of the number of the
line ($k=1$).

They are non-diagonal elements, upper to local diagonal (which in this
particular case is the main diagonal); first $N$ non-diagonal - upper -
elements. They come from $\frac{d\xi _{l}}{dt}$, for $l\neq 1$. The model is 
$W1\left( I,J\right) $, for $I=1$ and $I+1\leq J\leq N+1$. 
\begin{eqnarray}
&&A\left( 1,J\right) =  \label{eqa431} \\
&&-\alpha _{J}^{\prime }\frac{1}{\left( \xi _{1}-\xi _{J}\right) ^{2}+\left(
\eta _{1}+\eta _{J}\right) ^{2}}\left[ \left( \eta _{1}+\eta _{J}\right) %
\right]  \notag \\
&&+\alpha _{J}^{\prime \prime }\frac{1}{\left( \xi _{1}-\xi _{J}\right)
^{2}+\left( \eta _{1}+\eta _{J}\right) ^{2}}\left[ -\left( \xi _{1}-\xi
_{J}\right) \right]  \notag
\end{eqnarray}

\bigskip

Next $N+1$ entries on the first line originate from $\frac{d\eta _{l}}{dt}$.

The real position in the line of must be found by adding to its indice $l$
of $\frac{d\eta _{l}}{dt}$ the number of coefficients already considered, $%
N+1$.

They are classified according to the same criterium. The indice ($l$) is
equal or not with the number of the line ($k=1$).

The first case is $l=k=1$, the indice is equal to the number of the line. It
comes from $\frac{d\eta _{1}}{dt}$. The model is $W4\left( k,l\right) $,
adapted for the second indice to be returned to the $l=I$.%
\begin{eqnarray}
&&A\left( 1,N+1+1\right) =  \label{eqa432} \\
&&\ \ \ \ \ \ \ \ \ \ \ \ \ \ \ \ \ \alpha _{1}^{\prime \prime }\frac{1}{%
\eta _{1}}  \notag \\
&&-\sum\limits_{l\neq 1}\alpha _{l}^{\prime }\frac{1}{\left( \xi _{1}-\xi
_{l}\right) ^{2}+\left( \eta _{1}+\eta _{l}\right) ^{2}}\left[ \left( \xi
_{1}-\xi _{l}\right) \right]  \notag \\
&&+\sum\limits_{l\neq 1}\alpha _{l}^{\prime \prime }\frac{1}{\left( \xi
_{1}-\xi _{l}\right) ^{2}+\left( \eta _{1}+\eta _{l}\right) ^{2}}\left[
\left( \eta _{1}+\eta _{l}\right) \right]  \notag
\end{eqnarray}%
which means $A\left( 1,N+2\right) $.

\bigskip

The next $N$ come from $\frac{d\eta _{l}}{dt}$ . The indice $l$ in NOT equal
with the number of the line $k=1$. The type is again $W3\left( k,l\right) $
adapted for 
\begin{equation}
I=1\ \text{and}\ \ \left( N+1\right) +1\leq J\leq \left( N+1\right) +N+1
\label{eqa433}
\end{equation}%
with%
\begin{eqnarray}
&&A\left( 1,N+1+l\right) =  \label{eqa434} \\
&&-\alpha _{l}^{\prime }\frac{1}{\left( \xi _{1}-\xi _{l}\right) ^{2}+\left(
\eta _{1}+\eta _{l}\right) ^{2}}\left[ \left( \xi _{1}-\xi _{l}\right) %
\right]  \notag \\
&&+\alpha _{l}^{\prime \prime }\frac{1}{\left( \xi _{1}-\xi _{l}\right)
^{2}+\left( \eta _{1}+\eta _{l}\right) ^{2}}\left[ \left( \eta _{1}+\eta
_{l}\right) \right]  \notag
\end{eqnarray}%
for $l=2,N+1$, which means $A\left( 1,N+3\right) ,...,A\left( 1,2N+2\right) $%
.

\bigskip

\paragraph{The line for the unknown $X\left( 2\right) $.}

The first elements are non-diagonal elements; here are the lower part.

For this case the indice $l$ is NOT equal with the number of the line $k=2$.

They come from $\frac{d\xi _{l}}{dt}$ for $l=1,2-1$. This means $\frac{d\xi
_{1}}{dt}$, a unique element. The type is $W1\left( I,J\right) $ adapted for 
$I=2$ and $J=1$.%
\begin{eqnarray}
&&A\left( 2,l\right) =  \label{eqa435} \\
&&-\alpha _{l}^{\prime }\frac{1}{\left( \xi _{2}-\xi _{l}\right) ^{2}+\left(
\eta _{2}+\eta _{l}\right) ^{2}}\left[ \left( \eta _{2}+\eta _{l}\right) %
\right]  \notag \\
&&+\alpha _{l}^{\prime \prime }\frac{1}{\left( \xi _{2}-\xi _{l}\right)
^{2}+\left( \eta _{2}+\eta _{l}\right) ^{2}}\left[ -\left( \xi _{2}-\xi
_{l}\right) \right]  \notag
\end{eqnarray}%
for $l=1,...,1$, a single value.

\bigskip

Next entry consists of the case where the indice $l$ of the entry
(coefficient) coincides with the number of the line $k=2$.

The diagonal element. It comes from $\frac{d\xi _{2}}{dt}$. The type is $%
W2\left( I,J\right) $ for $J=I=2$.%
\begin{eqnarray}
&&A\left( 2,2\right) =  \label{eqa436} \\
&&1+  \notag \\
&&-\sum\limits_{l\neq 2}\alpha _{l}^{\prime }\frac{1}{\left( \xi _{2}-\xi
_{l}\right) ^{2}+\left( \eta _{2}+\eta _{l}\right) ^{2}}\left[ -\left( \eta
_{2}+\eta _{l}\right) \right]  \notag \\
&&+\sum\limits_{l\neq 2}\alpha _{l}^{\prime \prime }\frac{1}{\left( \xi
_{2}-\xi _{l}\right) ^{2}+\left( \eta _{2}+\eta _{l}\right) ^{2}}\left[
\left( \xi _{2}-\xi _{l}\right) \right]  \notag
\end{eqnarray}

\bigskip

Next cases are still coming from the set $\frac{d\xi _{l}}{dt}$ but their
indice $l$ does not coincide with the number of the line $k=2$, it is
greater than $k=2$. The type is again $W1\left( I,J\right) $ for $J>I=2$. 
\begin{eqnarray}
&&A\left( 2,l\right) =  \label{eqa437} \\
&&-\alpha _{l}^{\prime }\frac{1}{\left( \xi _{2}-\xi _{l}\right) ^{2}+\left(
\eta _{2}+\eta _{l}\right) ^{2}}\left[ \left( \eta _{2}+\eta _{l}\right) %
\right]  \notag \\
+ &&\alpha _{l}^{\prime \prime }\frac{1}{\left( \xi _{2}-\xi _{l}\right)
^{2}+\left( \eta _{2}+\eta _{l}\right) ^{2}}\left[ -\left( \xi _{2}-\xi
_{l}\right) \right]  \notag
\end{eqnarray}%
Their number is $l=2+1,...,N+1$ which is $l=3,N+1$.

\bigskip

Next entries (coefficients) are upper diagonal elements and come from $\frac{%
d\eta _{l}}{dt}$.

We must make difference between (1) $l=1,...,N+1$ but $l\neq 2$ and (2) $l=2$
, where the index $l$ only refers to $\frac{d\eta _{l}}{dt}$. The indices
for the full matrix are calculated adding $N+1$, corresponding to all the
positions occupied by $\frac{d\xi _{l}}{dt}$.

The first case consists of the entries coming from $\frac{d\eta _{l}}{dt}$
with $l$ less than the number of the line $k=2$. The type of these terms is $%
W3\left( I,J\right) $, for $J-\left( N+1\right) <I$. This group here only
contains one element, since we are at the line $2$. Then $l=1$. This
coefficient comes from $\frac{d\eta _{1}}{dt}$.%
\begin{eqnarray}
&&A\left( 2,N+1+1\right) =  \label{eqa438} \\
&&-\alpha _{1}^{\prime }\frac{1}{\left( \xi _{2}-\xi _{1}\right) ^{2}+\left(
\eta _{2}+\eta _{1}\right) ^{2}}\left[ \left( \xi _{2}-\xi _{1}\right) %
\right]  \notag \\
&&+\alpha _{1}^{\prime \prime }\frac{1}{\left( \xi _{2}-\xi _{1}\right)
^{2}+\left( \eta _{2}+\eta _{1}\right) ^{2}}\left[ \left( \eta _{2}+\eta
_{1}\right) \right]  \notag
\end{eqnarray}

\bigskip

It follows the entry coming from $\frac{d\eta _{l}}{dt}$ with the indice $l$
coinciding with the number of the line $k=2$, \emph{i.e.} $l=k=2$. It comes
from $\frac{d\eta _{2}}{dt}$. The type is $W4\left( I,I\right) $.%
\begin{eqnarray}
&&A\left( 2,N+1+l\right) =A\left( 2,N+1+2\right) =  \label{eqa439} \\
&&\ \ \ \ \ \ \ \ \ \ \alpha _{2}^{\prime \prime }\frac{1}{\eta _{2}}  \notag
\\
&&-\sum\limits_{l\neq 2}^{N+1}\alpha _{l}^{\prime }\frac{1}{\left( \xi
_{2}-\xi _{l}\right) ^{2}+\left( \eta _{2}+\eta _{l}\right) ^{2}}\left[
\left( \xi _{2}-\xi _{l}\right) \right]  \notag \\
&&+\sum\limits_{l\neq 2}^{N+1}\alpha _{l}^{\prime \prime }\frac{1}{\left(
\xi _{2}-\xi _{l}\right) ^{2}+\left( \eta _{2}+\eta _{l}\right) ^{2}}\left[
\left( \eta _{2}+\eta _{l}\right) \right]  \notag
\end{eqnarray}

\bigskip

The next group contains all other coefficients coming from $\frac{d\eta _{l}%
}{dt}$ whose indice $l$ does not coincide with the number of the line (\emph{%
i.e.} $k=2$). This means for $l=3,...,N+1$. The indice will be produced by
adding $l=3,...,N+1$ to $N+1$ (already taken by $\frac{d\xi _{l}}{dt}$). The
type is again $W3\left( I,J\right) $ for $I=2$ and $\left( N+1\right) +1<J$.%
\begin{eqnarray}
&&A\left( 2,N+1+l\right) =\ \left( \text{only for }l=3,...,N+1\right)
\label{eqa440} \\
&&-\alpha _{l}^{\prime }\frac{1}{\left( \xi _{2}-\xi _{l}\right) ^{2}+\left(
\eta _{2}+\eta _{l}\right) ^{2}}\left[ \left( \xi _{2}-\xi _{l}\right) %
\right]  \notag \\
&&+\alpha _{l}^{\prime \prime }\frac{1}{\left( \xi _{2}-\xi _{l}\right)
^{2}+\left( \eta _{2}+\eta _{l}\right) ^{2}}\left[ \left( \eta _{2}+\eta
_{l}\right) \right]  \notag
\end{eqnarray}

This fills all entries on the line $k=2$. The two exercises (for $k=1,2$) can be used as a check for the computer code.

\bigskip

\subsubsection{The coeffcients resulting from the equation for the imaginary
part ($d\protect\beta _{k}^{\prime \prime }/dt=0$)}

The time derivative of the imaginary part, $\beta _{k}^{\prime \prime }$.%
\begin{eqnarray}
&&\frac{d\beta _{k}^{\prime \prime }}{dt}  \label{eqa501} \\
&=&-1-\frac{d\eta _{k}}{dt}-\alpha _{k}^{\prime }\frac{1}{\eta _{k}}\frac{%
d\eta _{k}}{dt}  \notag \\
&&\hspace{-1.25cm}\hspace{-1.25cm}-\sum\limits_{l\neq k}\alpha _{l}^{\prime
\prime }\frac{1}{\left( \xi _{k}-\xi _{l}\right) ^{2}+\left( \eta _{k}+\eta
_{l}\right) ^{2}}\left[ \left( \xi _{k}-\xi _{l}\right) \left( \frac{d\eta
_{k}}{dt}+\frac{d\eta _{l}}{dt}\right) -\left( \eta _{k}+\eta _{l}\right)
\left( \frac{d\xi _{k}}{dt}-\frac{d\xi _{l}}{dt}\right) \right]  \notag \\
&&\hspace{-1.25cm}\hspace{-1.25cm}-\frac{1}{2}\sum\limits_{l\neq k}\alpha
_{l}^{\prime }\frac{1}{\left( \xi _{k}-\xi _{l}\right) ^{2}+\left( \eta
_{k}+\eta _{l}\right) ^{2}}\left[ 2\left( \xi _{k}-\xi _{l}\right) \left( 
\frac{d\xi _{k}}{dt}-\frac{d\xi _{l}}{dt}\right) +2\left( \eta _{k}+\eta
_{l}\right) \left( \frac{d\eta _{k}}{dt}+\frac{d\eta _{l}}{dt}\right) \right]
\notag
\end{eqnarray}

In Eq. $k$. For $k=1,N+1$.

Coefficient of $\frac{d\xi _{k}}{dt}$ is%
\begin{eqnarray}
&&-\sum\limits_{l\neq k}\alpha _{l}^{\prime \prime }\frac{1}{\left( \xi
_{k}-\xi _{l}\right) ^{2}+\left( \eta _{k}+\eta _{l}\right) ^{2}}\left[
-\left( \eta _{k}+\eta _{l}\right) \right]  \label{eqa502} \\
&&-\sum\limits_{l\neq k}\alpha _{l}^{\prime }\frac{1}{\left( \xi _{k}-\xi
_{l}\right) ^{2}+\left( \eta _{k}+\eta _{l}\right) ^{2}}\left[ \left( \xi
_{k}-\xi _{l}\right) \right]  \notag
\end{eqnarray}

Coefficient of $\frac{d\xi _{l}}{dt}$. For $l=1,N+1$, but $l\neq k$. 
\begin{eqnarray}
&&-\alpha _{l}^{\prime \prime }\frac{1}{\left( \xi _{k}-\xi _{l}\right)
^{2}+\left( \eta _{k}+\eta _{l}\right) ^{2}}\left[ \left( \eta _{k}+\eta
_{l}\right) \right]  \label{eqa503} \\
&&-\alpha _{l}^{\prime }\frac{1}{\left( \xi _{k}-\xi _{l}\right) ^{2}+\left(
\eta _{k}+\eta _{l}\right) ^{2}}\left[ -\left( \xi _{k}-\xi _{l}\right) %
\right]  \notag
\end{eqnarray}

Coefficient of $\frac{d\eta _{k}}{dt}$; it is%
\begin{eqnarray}
&&-1-\alpha _{k}^{\prime }\frac{1}{\eta _{k}}  \label{eqa504} \\
&&-\sum\limits_{l\neq k}\alpha _{l}^{\prime \prime }\frac{1}{\left( \xi
_{k}-\xi _{l}\right) ^{2}+\left( \eta _{k}+\eta _{l}\right) ^{2}}\left[
\left( \xi _{k}-\xi _{l}\right) \right]  \notag \\
&&-\sum\limits_{l\neq k}\alpha _{l}^{\prime }\frac{1}{\left( \xi _{k}-\xi
_{l}\right) ^{2}+\left( \eta _{k}+\eta _{l}\right) ^{2}}\left[ \left( \eta
_{k}+\eta _{l}\right) \right]  \notag
\end{eqnarray}

Coefficient of $\frac{d\eta _{l}}{dt}$; it is%
\begin{eqnarray}
&&-\alpha _{l}^{\prime \prime }\frac{1}{\left( \xi _{k}-\xi _{l}\right)
^{2}+\left( \eta _{k}+\eta _{l}\right) ^{2}}\left[ \left( \xi _{k}-\xi
_{l}\right) \right]  \label{eqa505} \\
&&-\alpha _{l}^{\prime }\frac{1}{\left( \xi _{k}-\xi _{l}\right) ^{2}+\left(
\eta _{k}+\eta _{l}\right) ^{2}}\left[ \left( \eta _{k}+\eta _{l}\right) %
\right]  \notag
\end{eqnarray}%
for $l=1,N+1$, but $l\neq k$.

The free term; it is, after transfering it to the right side $1$ which comes
from all equations of $\beta _{k}^{\prime \prime }$.

\paragraph{The list of types of coefficients as result from the time
derivation of the imaginary part ($d\protect\beta _{k}^{\prime \prime }/dt=0$%
)}

There are $6$ regions and the types are also only $4$.

Consider the line $I$, which must be 
\begin{equation}
\left( N+1\right) +1\leq I\leq \left( N+1\right) +N+1  \label{eqa506}
\end{equation}%
The first columns $J$ are from $\frac{d\xi _{I}}{dt}$ which are under the
local diagonal. The type is $\Omega 1\left( k,l\right) $,%
\begin{eqnarray}
\Omega 1\left( k,l\right) &=&  \label{eqa507} \\
&&-\alpha _{J}^{\prime \prime }\frac{1}{\left( \xi _{k}-\xi _{l}\right)
^{2}+\left( \eta _{k}+\eta _{l}\right) ^{2}}\left[ \left( \eta _{k}+\eta
_{l}\right) \right]  \notag \\
&&-\alpha _{J}^{\prime }\frac{1}{\left( \xi _{k}-\xi _{l}\right) ^{2}+\left(
\eta _{k}+\eta _{l}\right) ^{2}}\left[ -\left( \xi _{k}-\xi _{l}\right) %
\right]  \notag
\end{eqnarray}%
where the indice $k$ of the variables is $I$ adapted by substracting the $%
\left( N+1\right) $ lines of the first Jacobi square.%
\begin{eqnarray}
k &=&I-\left( N+1\right)  \label{eqa508} \\
l &=&J  \notag
\end{eqnarray}%
The range of the indices is%
\begin{eqnarray}
\left( N+1\right) +1 &<&I\leq \left( N+1\right) +N+1  \label{eqa509} \\
1 &\leq &J<I-\left( N+1\right)  \notag
\end{eqnarray}

\bigskip

The element that is on the local diagonal has%
\begin{equation}
I-\left( N+1\right) =J  \label{eqa510}
\end{equation}%
It corresponds to $\frac{d\xi _{J}}{dt}$ on the line $I$, with the property $%
J=I-\left( N+1\right) $. The type is $\Omega 2\left( I,J\right) $ with the
expression%
\begin{eqnarray}
&&\Omega 2\left( k,k\right) =  \label{eqa511} \\
&&-\sum\limits_{l\neq k}\alpha _{l}^{\prime \prime }\frac{1}{\left( \xi
_{k}-\xi _{l}\right) ^{2}+\left( \eta _{k}+\eta _{l}\right) ^{2}}\left[
-\left( \eta _{k}+\eta _{l}\right) \right]  \notag \\
&&-\sum\limits_{l\neq k}\alpha _{l}^{\prime }\frac{1}{\left( \xi _{k}-\xi
_{l}\right) ^{2}+\left( \eta _{k}+\eta _{l}\right) ^{2}}\left[ \left( \xi
_{k}-\xi _{l}\right) \right]  \notag
\end{eqnarray}%
with%
\begin{equation}
k=I-\left( N+1\right)  \label{eqa512}
\end{equation}%
The range of the indices is%
\begin{eqnarray}
\left( N+1\right) +1 &<&I\leq \left( N+1\right) +N+1  \label{eqa513} \\
J &=&I-\left( N+1\right)  \notag
\end{eqnarray}

\bigskip

For columns that are beyond the local diagonal but still in the third square
matrix. they come from $\frac{d\xi _{l}}{dt}$.

The type is again $\Omega 1\left( I,J\right) $ with the expression%
\begin{eqnarray}
\Omega 1\left( k,l\right) &=&  \label{eqa514} \\
&&-\alpha _{J}^{\prime \prime }\frac{1}{\left( \xi _{k}-\xi _{l}\right)
^{2}+\left( \eta _{k}+\eta _{l}\right) ^{2}}\left[ \left( \eta _{k}+\eta
_{l}\right) \right]  \notag \\
&&-\alpha _{J}^{\prime }\frac{1}{\left( \xi _{k}-\xi _{l}\right) ^{2}+\left(
\eta _{k}+\eta _{l}\right) ^{2}}\left[ -\left( \xi _{k}-\xi _{l}\right) %
\right]  \notag
\end{eqnarray}%
The range of indices is%
\begin{eqnarray}
\left( N+1\right) +1 &\leq &I\leq \left( N+1\right) +N+1  \label{eqa515} \\
I-\left( N+1\right) &<&J\leq N+1  \notag
\end{eqnarray}%
which are translated into%
\begin{eqnarray}
k &=&I-\left( N+1\right)  \label{eqa516} \\
l &=&J  \notag
\end{eqnarray}

\bigskip

Now we continue along the line no. $I$ into the fourth square.

For an arbitrary line $I$, the first columns in the fourth square come from $%
\frac{d\eta _{l}}{dt}$. They are of the type $\Omega 3\left( I,J\right) $
with expression%
\begin{eqnarray}
&&\Omega 3\left( k,l\right) =  \label{eqa517} \\
&&-\alpha _{l}^{\prime \prime }\frac{1}{\left( \xi _{k}-\xi _{l}\right)
^{2}+\left( \eta _{k}+\eta _{l}\right) ^{2}}\left[ \left( \xi _{k}-\xi
_{l}\right) \right]  \notag \\
&&-\alpha _{l}^{\prime }\frac{1}{\left( \xi _{k}-\xi _{l}\right) ^{2}+\left(
\eta _{k}+\eta _{l}\right) ^{2}}\left[ \left( \eta _{k}+\eta _{l}\right) %
\right]  \notag
\end{eqnarray}%
where%
\begin{eqnarray}
k &=&I-\left( N+1\right)  \label{eqa518} \\
l &=&J-\left( N+1\right)  \notag
\end{eqnarray}%
with the range of the parameters%
\begin{eqnarray}
\left( N+1\right) +1 &\leq &I\leq \left( N+1\right) +N+1  \label{eqa519} \\
\left( N+1\right) +1 &\leq &J<I  \notag
\end{eqnarray}

\bigskip

The element on the local diagonal comes from $\frac{d\eta _{I}}{dt}$ and the
type is $\Omega 4\left( I,J\right) $. the expression is%
\begin{eqnarray}
&&\Omega 4\left( k,k\right) =  \label{eqa520} \\
&&\ \ \ \ \ \ \ \ \ \ \ \ \ \ \ \ -1-\alpha _{k}^{\prime }\frac{1}{\eta _{k}}
\notag \\
&&-\sum\limits_{l\neq k}\alpha _{l}^{\prime \prime }\frac{1}{\left( \xi
_{k}-\xi _{l}\right) ^{2}+\left( \eta _{k}+\eta _{l}\right) ^{2}}\left[
\left( \xi _{k}-\xi _{l}\right) \right]  \notag \\
&&-\sum\limits_{l\neq k}\alpha _{l}^{\prime }\frac{1}{\left( \xi _{k}-\xi
_{l}\right) ^{2}+\left( \eta _{k}+\eta _{l}\right) ^{2}}\left[ \left( \eta
_{k}+\eta _{l}\right) \right]  \notag
\end{eqnarray}%
with%
\begin{equation}
k=I-\left( N+1\right) =J-\left( N+1\right)  \label{eqa521}
\end{equation}%
and the range 
\begin{eqnarray}
\left( N+1\right) +1 &\leq &I\leq \left( N+1\right) +N+1  \label{eqa522} \\
J &=&I  \notag
\end{eqnarray}

\bigskip

Finally we have along the same line $\left( I\right) $ the group of columns
that are upper the local diagonal in the fourth square.

They come from $\frac{d\eta _{I}}{dt}$. They are of the $\Omega 3\left(
I,J\right) $ type again. The expression is%
\begin{eqnarray}
&&\Omega 3\left( k,l\right) =  \label{eqa523} \\
&&-\alpha _{l}^{\prime \prime }\frac{1}{\left( \xi _{k}-\xi _{l}\right)
^{2}+\left( \eta _{k}+\eta _{l}\right) ^{2}}\left[ \left( \xi _{k}-\xi
_{l}\right) \right]  \notag \\
&&-\alpha _{l}^{\prime }\frac{1}{\left( \xi _{k}-\xi _{l}\right) ^{2}+\left(
\eta _{k}+\eta _{l}\right) ^{2}}\left[ \left( \eta _{k}+\eta _{l}\right) %
\right]  \notag
\end{eqnarray}%
The indices are%
\begin{eqnarray}
k &=&I-\left( N+1\right)  \label{eqa524} \\
l &=&J-\left( N+1\right)  \notag
\end{eqnarray}%
and the range of indices is%
\begin{eqnarray}
\left( N+1\right) +1 &\leq &I\leq \left( N+1\right) +N+1  \label{eqa525} \\
I &<&J\leq \left( N+1\right) +N+1  \notag
\end{eqnarray}

\bigskip

\paragraph{The free term in the linear system of equations}

The free term is%
\begin{equation}
F\left( I\right) =\left\{ 
\begin{array}{ccc}
0 & \text{for} & 1\leq I\leq N+1 \\ 
+1 & \text{for} & \left( N+1\right) +1\leq I\leq \left( N+1\right) +N+1%
\end{array}%
\right.  \label{eqa526}
\end{equation}

\subsubsection{The numerical implementation of the time evolution of the
positions of the singularities}

Assume that we work with $N+1$ singularities (in the circular case, there are $N$ singularities, in the infinite case there are $N+1$ singularities).%
\begin{equation}
\xi _{k}+i\eta _{k}\ \ \text{for}\ \ k=1,...,N+1  \label{eqa527}
\end{equation}

The equations that we intend to solve are of general form%
\begin{eqnarray}
\frac{d\xi _{k}}{dt} &=&\Xi \left( \xi _{l},\eta _{l}\right)   \label{eqa528}
\\
\frac{d\eta _{k}}{dt} &=&\Psi \left( \xi _{l},\eta _{l}\right)   \notag
\end{eqnarray}%
for $k=1,...,N+1$. Putting together the real and imaginary parts $\left( \xi _{k},\eta
_{k}\right) $ as independent unknown variables, the system in the form%
\begin{eqnarray}
\sum\limits_{j=1,2N+2}A\left( I,J\right) X\left( J\right)  &=&F\left(
I\right)   \label{eqa529} \\
\text{for}\ I &=&1,2N+2  \notag
\end{eqnarray}

The expressions of $A\left( I,J\right) $ depend of $\left( \xi _{k},\eta
_{k}\right) $. Then we can see the sequence

\begin{enumerate}
\item we start with a set of variables%
\begin{equation}
\left( \xi _{k},\eta _{k}\right) \ \ ,\ \ k=1,2N+2  \label{eqa530}
\end{equation}
\item calculate the matrix%
\begin{equation}
A\left( I,J\right)  \label{eqa532}
\end{equation}%
and the free term%
\begin{equation}
F\left( I\right)  \label{eqa533}
\end{equation}
\item solve the linear system%
\begin{equation}
X\left( I\right) =A^{-1}\left( I,J\right) F\left( J\right)  \label{eqa534}
\end{equation}
\item integrate in time using Runge-Kutta method the system of equations%
\begin{equation}
\frac{d\zeta _{k}}{dt}=X\left( I\right)  \label{eqa535}
\end{equation}%
and find the new set $\left( \xi _{k},\eta _{k}\right) $ at the time $%
t+\delta t$.
\item find the current interface $\Gamma(t)$
\item iterate to point 2.
\end{enumerate}

\bigskip

Numerical studies consists of the inversion of the matrix equation%
\begin{equation}
AX=F  \label{eqa537}
\end{equation}%
followed by time advancement using the Runge Kutta method (the \textbf{d02pdf%
} NAG routine).

\bigskip

\subsection{The time evolution of the mapping function $f\left( z,t\right) $}
Integrating Eqs.(\ref{eqa535}) we determine the positions of the singularities $\zeta _{k}\left(
t\right) $, $k=1,...,N+1$, as function of time. Now we calculate%
\begin{eqnarray}
f\left( z,t\right)  &=&z-it  \label{eqa593} \\
&&-i\sum\limits_{l=1}^{N+1}\left( \alpha _{l}^{\prime }+i\alpha
_{l}^{\prime \prime }\right)   \notag \\
&&\times \left\{ \log \left( \left\vert x-\xi _{l}+iy-i\eta _{l}\right\vert
\right) +i\arg \left( x-\xi _{l}+iy-i\eta _{l}\right) \right\}   \notag
\end{eqnarray}%
It has been shown \cite{poncemineev0}, \cite{poncemineev1}, \cite{poncemineev2} that: if at $t=0$\ all imaginary parts of the singularities of
the $\log $ function are positive%
\begin{equation}
\eta _{l}>0\ \ \text{for}\ \ \forall l  \label{eqa598}
\end{equation}%
and all constants $\alpha _{l}$ are real and positive%
\begin{eqnarray}
\alpha _{l} &\in &\mathbf{R}  \label{eqa599} \\
\alpha _{l} &>&0\ \ \text{for}\ \ \forall l  \notag
\end{eqnarray}%
then the imaginary parts of the zeros $y_{l}$ and of the singularities $\eta
_{l}$ remain positive for all time $t>0$. This preserves the holomorphicity of $f$.

\bigskip

The \emph{interface line} in the mathematical plane $\left( z\right) $ is
the straight line which separates the two half-planes%
$-\infty  <x<\infty$, $y=0$
and it is mapped through the complex function $f\left( z,t\right)$
into a \emph{curved line} $\Gamma \left( t\right) $ in the real plane of
physical variables $X \equiv \mathbf{Re}\left[ f\left( z,t\right) \right] $, %
$Y \equiv \mathbf{Im}\left[ f\left( z,t\right) \right]$. We can identify the curve that corresponds to the complex line $z=x+i0 $ %
by taking $x \in \mathbf{R}$ and $y=0$ in the expressions of $X,Y$.

The general formula for the Real part is%
\begin{eqnarray}
&&\mathbf{Re}\left[ f\left( z,t\right) \right]  \label{eqa610} \\
&=&x  \notag \\
&&+\sum\limits_{l=1}^{N+1}\alpha _{l}^{\prime \prime }\frac{1}{2}\log \left[
\left( x-\xi _{l}\right) ^{2}+\left( y-\eta _{l}\right) ^{2}\right]  \notag
\\
&&+\sum\limits_{l=1}^{N+1}\alpha _{l}^{\prime }\arg \left( x+iy-\xi
_{l}+i\eta _{l}\right)  \notag
\end{eqnarray}%
where we choose $\alpha _{l}^{\prime \prime }\equiv 0$ for all $l$, then $y=0$.
\begin{eqnarray}
&&X_{0}\left( x,t\right)  \label{eqa611} \\
&=&x-\sum\limits_{l=1}^{N+1}\alpha _{l}^{\prime }\arg \left( x-\xi
_{l}+i\eta _{l}\right)  \notag
\end{eqnarray}

Similarly, we have the general form%
\begin{eqnarray}
&&\mathbf{Im}\left[ f\left( z,t\right) \right]  \label{eqa612} \\
&=&y-t  \notag \\
&&-\sum\limits_{l=1}^{N+1}\alpha _{l}^{\prime }\frac{1}{2}\log \left[
\left( x-\xi _{l}\right) ^{2}+\left( y-\eta _{l}\right) ^{2}\right]  \notag
\\
&&+\sum\limits_{l=1}^{N+1}\alpha _{l}^{\prime \prime }\arg \left( x+iy-\xi
_{l}+i\eta _{l}\right)  \notag
\end{eqnarray}%
and take $\alpha _{l}^{\prime \prime }\equiv 0$ and $y=0$%
\begin{eqnarray}
Y_{0}\left( x,t\right) &=&-t  \label{eqa614} \\
&&-\sum\limits_{l=1}^{N+1}\alpha _{l}^{\prime }\frac{1}{2}\log \left[
\left( x-\xi _{l}\right) ^{2}+\left( y-\eta _{l}\right) ^{2}\right]  \notag
\end{eqnarray}

\bigskip

The interface is formally defined as%
\begin{equation}
\Gamma \left( t\right) =Y_{0}\left( X_{0},t\right)   \label{eqa615}
\end{equation}%
which means that  the variable $x$ is eliminated between $X_{0}\left( x,t\right) $ and $Y_{0}\left( x,t\right) $.

The phase of the logarithm must be contained in%
\begin{equation}
0\leq \arg \left( x-\xi _{l}+i\eta _{l}\right) \leq \pi   \label{co1}
\end{equation}%
because $\eta _{l}>0\ \ $for$\ \ \forall l$. There is nothing special with $x
$ traversing the point $\xi _{l}$ if we think in terms of the vector in
complex plane based in the origin $\left( 0,0\right)$ and pointing to $\left(
x-\xi _{l},\eta _{l}\right) $. When $x-\xi _{l}\equiv -\left\vert
\varepsilon \right\vert <0$ the vector is almost aligned with the imaginary
axis and the angle it makes with the abscissa (the phase of the argument of
the logarithm) is slightly greater than $\pi /2$. When $x$ moves to become
greater than $\xi _{l}$ , $x-\xi _{l}\equiv \left\vert \varepsilon
\right\vert >0$, the vector is still almost vertical but the angle with the
abscissa is slightly less than $\pi /2$. Therefore when $x$ traverses the
position $\xi _{l}$ the phase of the logarithm smoothly changes around $\pi
/2$. When the argument of the complex number $\left( x-\xi _{l}+i\eta
_{l}\right) $ is calculated numerically, we must take into account that the
function $\arctan \equiv \left( \tan \right) ^{-1}$ is determined between $%
-\pi /2$ and $\pi /2$. Consider%
\begin{eqnarray}
\eta _{l} &>&0\ \ \left( \text{which is allways true}\right)   \label{co2} \\
x-\xi _{l} &=&+\left\vert \varepsilon \right\vert \ \ \left( \text{very small%
}\right)   \notag
\end{eqnarray}%
which means that $x$ approaches $\xi _{l}$ from the right, such that $x-\xi
_{l}>0$. Then $\arg \left( x-\xi _{l}+i\eta _{l}\right) \lesssim \frac{\pi }{%
2}$ which is obtained with the $\arctan $ function, since%
\begin{equation}
\arctan \left( \frac{\eta _{l}}{x-\xi _{l}}\right) =\arctan \left( \frac{%
\eta _{l}}{\left\vert \varepsilon \right\vert }\right) \approx \arctan
\left( +\infty \right) =\frac{\pi }{2}  \label{co4}
\end{equation}%
and we can write for this case%
\begin{equation}
\arg \left( x-\xi _{l}+i\eta _{l}\right) =\arctan \left( \frac{\eta _{l}}{%
x-\xi _{l}}\right)   \label{co5}
\end{equation}

For the opposite case, where $x$ approaches $\xi _{l}$ from the left,%
\begin{eqnarray}
\eta _{l} &>&0  \label{co7} \\
x-\xi _{l} &=&-\left\vert \varepsilon \right\vert   \notag
\end{eqnarray}%
the value of $\arg \left( x-\xi _{l}+i\eta _{l}\right) $ is again close to $%
\pi /2$, $\arg \left( x-\xi _{l}+i\eta _{l}\right) \gtrsim \frac{\pi }{2}$,
but in this case the function $\arctan $ gives%
\begin{equation}
\arctan \left( \frac{\eta _{l}}{-\left\vert \varepsilon \right\vert }\right)
\approx \arctan \left( -\infty \right) =-\frac{\pi }{2}  \label{co8}
\end{equation}%
and if we want to use $\arctan $ to obtain the $\arg $ then we must correct
the result, by adding $\pi $ 
\begin{equation}
\arg \left( x-\xi _{l}+i\eta _{l}\right) =\arctan \left( \frac{\eta _{l}}{%
x-\xi _{l}}\right) +\pi   \label{co9}
\end{equation}

The two functions become%
\begin{eqnarray*}
&&X_{0}\left( x,t\right) \\
&=&x \\
&&-\sum\limits_{l=1}^{N+1}\alpha _{l}^{\prime }\arctan \left( \frac{\eta
_{l}}{x-\xi _{l}}\right) -\pi \sum\limits_{l;x-\xi _{l}<0}\alpha
_{l}^{\prime }
\end{eqnarray*}%
\begin{eqnarray*}
Y_{0}\left( x,t\right) &=&-t \\
&&-\sum\limits_{l=1}^{N+1}\alpha _{l}^{\prime }\frac{1}{2}\log \left[
\left( x-\xi _{l}\right) ^{2}+\left( y-\eta _{l}\right) ^{2}\right]
\end{eqnarray*}

\subsection{Note on the stagnation points}

We find that the calculation of the imaginary part 
\begin{equation*}
Y_{0}=-t-\sum\limits_{l=1}^{N+1}\alpha _{l}^{\prime }\frac{1}{2}\log \left[
\left( x-\xi _{l}\right) ^{2}+\left( y-\eta _{l}\right) ^{2}\right] 
\end{equation*}%
makes $Y_{0}\left( x,t\right) $ strongly (linarly) dependent of time $t$.
Then apparently the \emph{lines} representing the interface are always substantially
shifted one relative to the other and the \emph{stagnation points} \cite{poncemineev1} do not apper. However it can be proved that the linear term $-t$ is
cancelled by an opposite term arising from the sum in the second term.

The only possibility is that%
\begin{eqnarray*}
x-\xi _{l} &\rightarrow &0 \\
y &=&0\ \ \text{this is the interface} \\
\eta _{l} &\rightarrow &\exp \left( -\frac{t}{\alpha _{l}^{\prime }}\right)
\end{eqnarray*}

This would be realized if the set of equations of time evolution of the $%
\left( \xi _{l},\eta _{l}\right) $ would give - in the second half, for $%
\eta _{l}$, the general form%
\begin{equation*}
A\left( I,J\right) \left( 
\begin{array}{c}
\frac{d\xi _{l}}{dt} \\ 
\frac{d\eta _{l}}{dt}%
\end{array}%
\right) =\left( 
\begin{array}{c}
0 \\ 
1%
\end{array}%
\right)
\end{equation*}%
or%
\begin{equation*}
\frac{d\eta _{l}}{dt}=-\frac{1}{\alpha _{l}^{\prime }}\eta _{l}+...
\end{equation*}

The diagonal of the matrix is%
\begin{eqnarray*}
&&\Omega 4\left( k,k\right) = \\
&&\ \ \ \ \ \ \ \ \ \ \ \ \ \ \ \ -1-\alpha _{k}^{\prime }\frac{1}{\eta _{k}}
\\
&&-\sum\limits_{l\neq k}\alpha _{l}^{\prime \prime }\frac{1}{\left( \xi
_{k}-\xi _{l}\right) ^{2}+\left( \eta _{k}+\eta _{l}\right) ^{2}}\left[
\left( \xi _{k}-\xi _{l}\right) \right] \\
&&-\sum\limits_{l\neq k}\alpha _{l}^{\prime }\frac{1}{\left( \xi _{k}-\xi
_{l}\right) ^{2}+\left( \eta _{k}+\eta _{l}\right) ^{2}}\left[ \left( \eta
_{k}+\eta _{l}\right) \right]
\end{eqnarray*}%
and we can approximate%
\begin{eqnarray*}
\frac{d\eta _{l}}{dt} &\sim &\left[ \Omega 4\left( l,l\right) \right] ^{-1}
\\
&=&\frac{1}{-1-\alpha _{l}^{\prime }\frac{1}{\eta _{l}}-...}
\end{eqnarray*}%
We note that%
\begin{eqnarray*}
\eta _{l} &\sim &\text{very small} \\
&\rightarrow &\left\vert \alpha _{l}^{\prime }\frac{1}{\eta _{l}}\right\vert
\gg 1
\end{eqnarray*}%
and the first term $-1$ can be neglected. Then%
\begin{equation*}
\frac{d\eta _{l}}{dt}\sim -\frac{1}{\alpha _{l}^{\prime }}\eta _{l}+...
\end{equation*}%
and the solution is indeed%
\begin{equation*}
\eta _{l}\left( t\right) \sim \eta _{l}^{\left( 0\right) }\exp \left( -\frac{%
t}{\alpha _{l}^{\prime }}\right)
\end{equation*}%
(Here we see why $\alpha _{l}^{\prime }$ must be $>0$).

We know that the singularities have evolved, qualitatively, in this way

\begin{itemize}
\item the real parts are approaching one the other but this is a slow
process. Approximately the initial positions $\xi _{l}$ are almost unchanged.

\item the imaginary part decreases rapidly, remains positive but
approaches zero exponentially fast.
\end{itemize}

Then we find that when we approach with \thinspace $x$ the position $\xi _{l}
$ of one of the singularities 
\begin{equation*}
x-\xi _{l}\rightarrow 0
\end{equation*}%
the term in the sum%
\begin{equation*}
Y_{0}=-t-\sum\limits_{l=1}^{N+1}\alpha _{l}^{\prime }\frac{1}{2}\log \left[
\left( x-\xi _{l}\right) ^{2}+\left( y-\eta _{l}\right) ^{2}\right] 
\end{equation*}%
becomes approximately%
\begin{eqnarray*}
Y_{0} &\approx &-t-\alpha _{l}^{\prime }\log \left[ \sqrt{\left( y-\eta
_{l}\right) ^{2}}\right] \ \ \text{where}\ \ y=0\ \text{on the interface} \\
&\approx &-t-\alpha _{l}^{\prime }\left( -\frac{t}{\alpha _{l}^{\prime }}%
\right) -\alpha _{l}^{\prime }\log \left( \eta _{l}^{\left( 0\right)
}\right) +... \\
&\approx &\text{const}>0
\end{eqnarray*}%
and this means that the tip $\left[ X_{0}\left( t\right) ,Y_{0}\left(
t\right) \right] $ for $t\rightarrow \infty $, of the curve 
\begin{equation*}
\left[ \xi _{l}\left( t\right) ,\eta _{l}\left( t\right) \right] 
\end{equation*}%
is almost fixed. These are the \emph{stagnation points} \cite{poncemineev1}, 
\cite{poncemineev2}.

\bigskip 

We note that the line $\mathbf{Re}\left[ f\left( z,t\right) \right] $ shows
small but abrupt changes. These are not singularities however, but they are
the result of the finite precision of the representation of the two lines
that are mapped by the conformal transformation. Essentially it is the
degree of the detail when one approaches a singularity of the complex $\log $
function which introduces these jumps.

\begin{figure}[h]
\includegraphics[height=10cm]{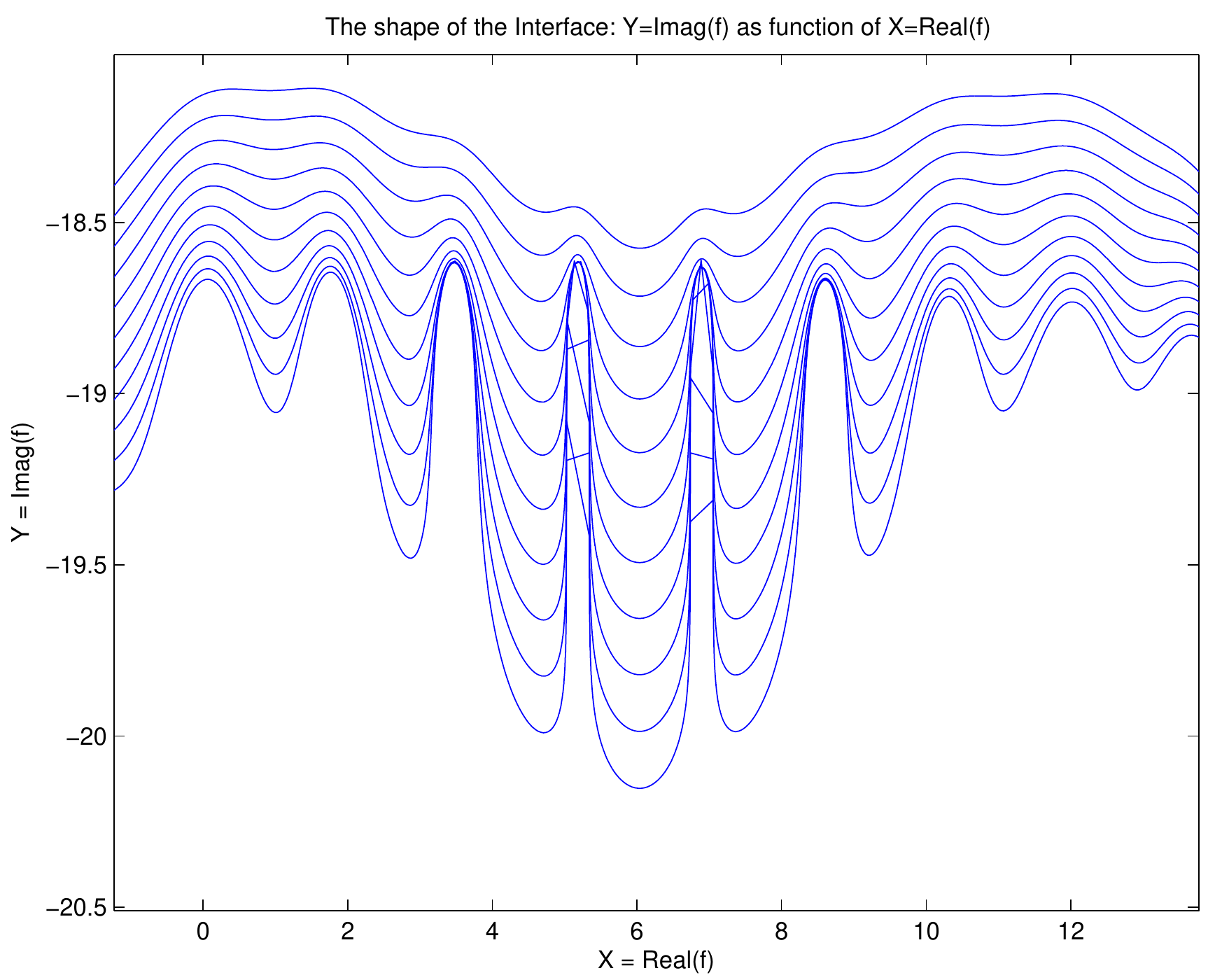}
\caption{The time variation of the shape of the interface $\left[ X\left( t\right)
,Y\left( t\right) \right] $ for the latest $11$ time moments, $t=90,...,100$%
. The effect of imprecisions generated by $\left( x-\xi _{l},\eta
_{l}\right) \rightarrow \left( 0,0\right) $ consist of spurious lines that
connect the fingers. They are made visible in Fig\ref{fig103}.}
\label{fig100}
\end{figure}

We explore the regions around the quasi-singularities with {\it adapted mesh refinement}. The spatial interval is %
$x\in \left[ x_{\min },x_{\max }\right]$ %
and we choose a number of \emph{average} mesh intervals, $M$. %
The \emph{average} mesh interval is %
$\overline{\delta x}=(x_{\max }-x_{\min })/M$. Now we choose a function $f\left( x;\xi _{l}\right)$
to modulate $\overline{\delta x}$ around a
point $x$. We take%
\begin{equation*}
f\left( x;\xi _{l}\right) =c+\sum\limits_{l=1}^{NN}a_{l}\exp \left[ -\frac{%
\left( x-\xi _{l}\right) ^{2}}{2b_{l}^{2}}\right] 
\end{equation*}%
where $c\equiv \text{constant}$, $a_{l}=\text{amplitude of }f$ for  $l=1,NN=N+1$, the number of singularities $\zeta_{l}$.
$b_{l} = $ half-width at inflection point. 
For example, for $M=1000$, $c = 1$, $a_{l} = 40$, $b_{l} = (\xi _{l+1}-\xi _{l-1})/d_{l}$%
and $d_{l} = \text{constant factor}=10$, the refined mesh has 9700 intervals. The strongly non-uniform mesh allows a good precision in the regions where $x-\xi-{l} \sim 0$ but are not able to remove the imprecisions when also $\eta_{l} \sim 0$. We conclude that special precautions must be taken for the late phase of the time evolution since in this regime $\eta_{l}$ becomes very small. In Figs.\ref{fig104}, \ref{fig105} and \ref{fig106} we have avoided this region.

\begin{figure}[h]
\includegraphics[height=10cm]{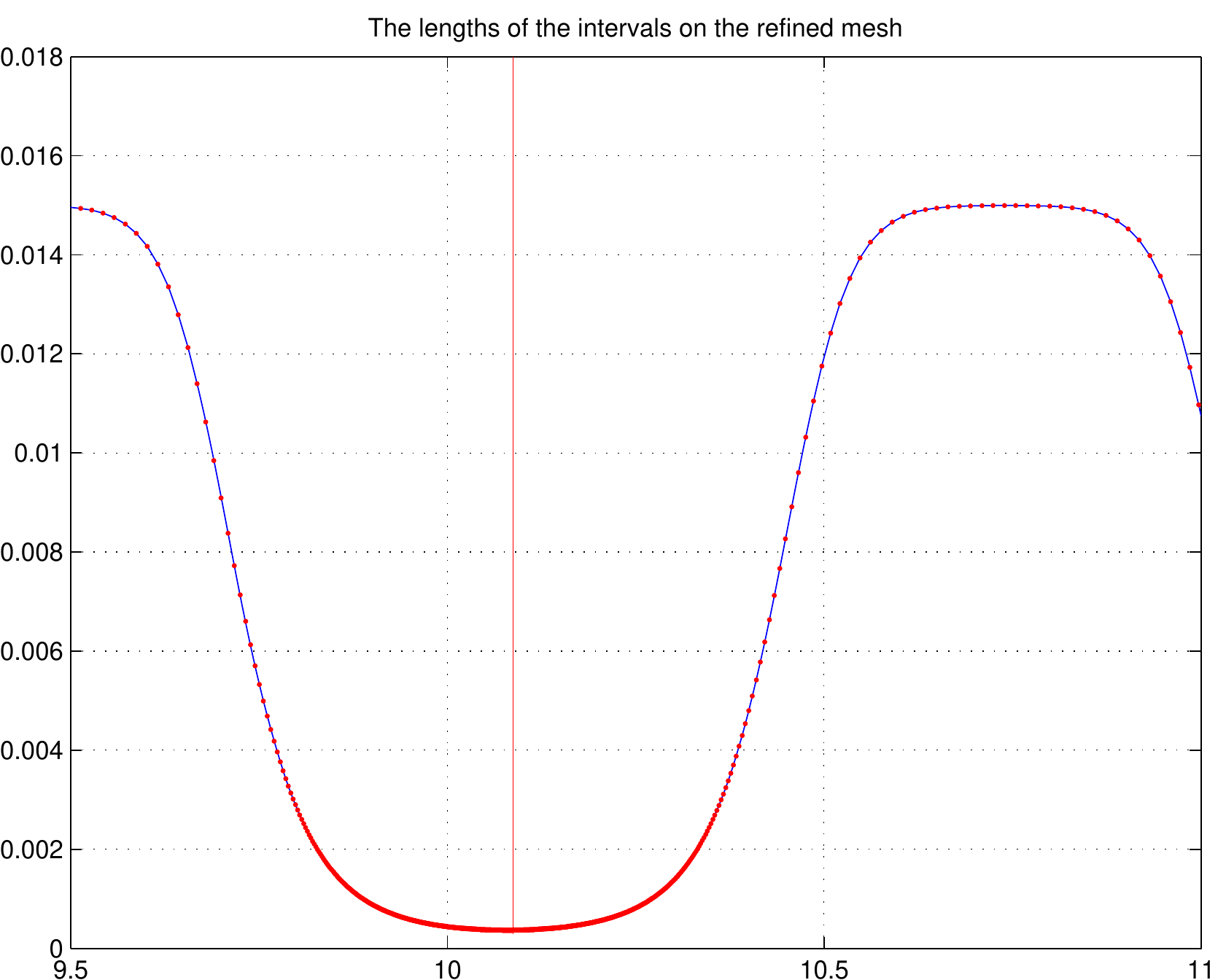}
\caption{The lengths of the intervals of the refined mesh on the line $x=\mathbf{Re}%
\left( z\right) $ in the \textquotedblleft mathematical plane\textquotedblright\ . The non-uniform distribution
is imposed by the need to explore carefully the regions around the
quasi-singularities: $\left( x-\xi _{l}\right) \sim 0$, $\eta _{l}\sim 0$.}
\label{fig110}
\end{figure}

\begin{figure}[h]
\includegraphics[height=10cm]{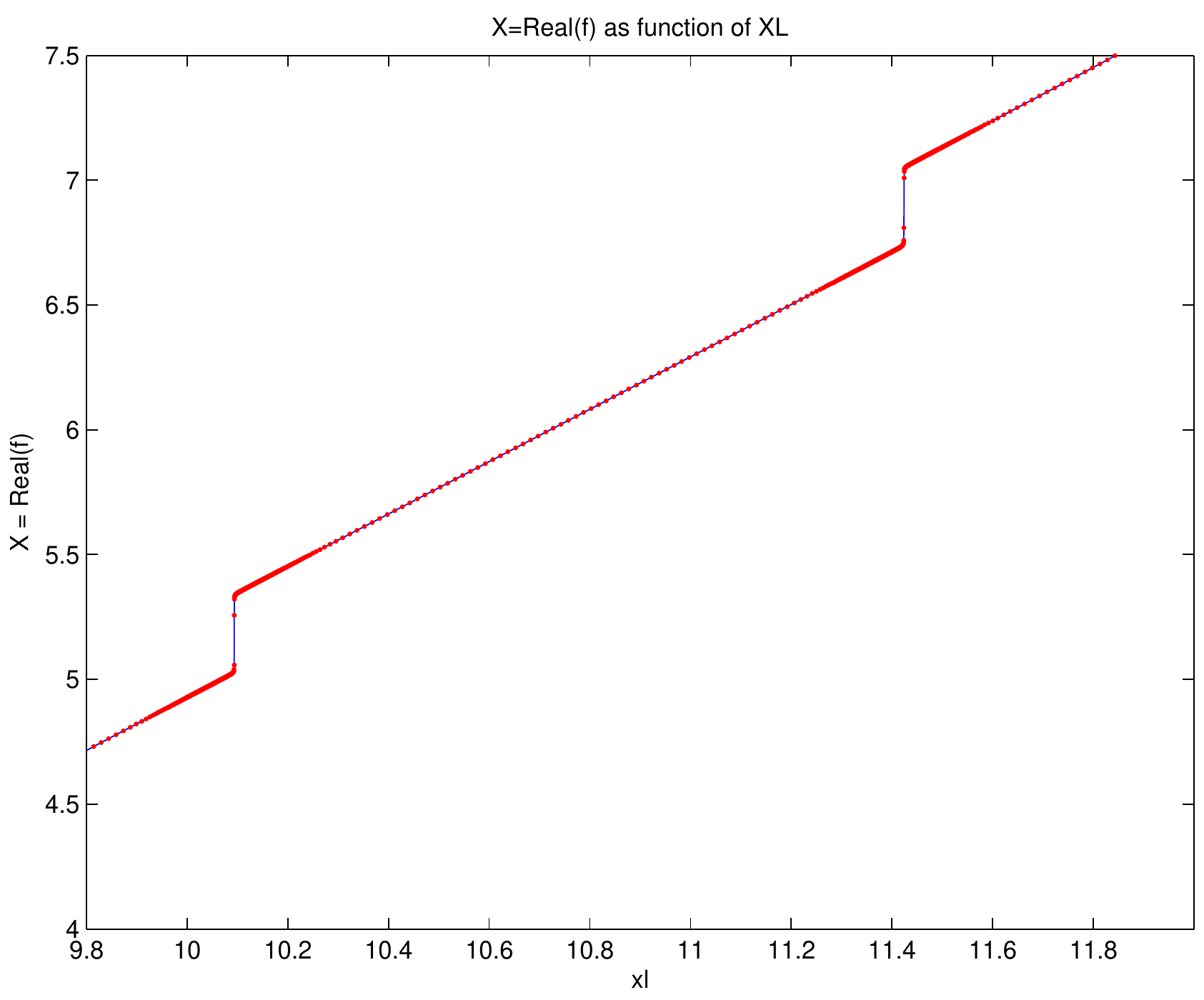}
\caption{The $X$ coordinate of the points on the interface (\emph{i.e.} $X=\mathbf{Re}%
\left[ f\left( z\right) \right] $) as function of $x$, the coordinate of the
abscissa on the mathematical plane $z=x+iy$. A small $x-$interval is
plotted, to show the quasi-singular variation of $\mathbf{Re}\left[ f\left(
z\right) \right] $ which is due to two situations where we have both $x-\xi
_{l}\rightarrow 0$ and $\eta _{l}\approx 0$. These cases produce the errors
of the interface profile. The result here is for $t=94$.}
\label{fig101}
\end{figure}

\begin{figure}[h]
\includegraphics[height=10cm]{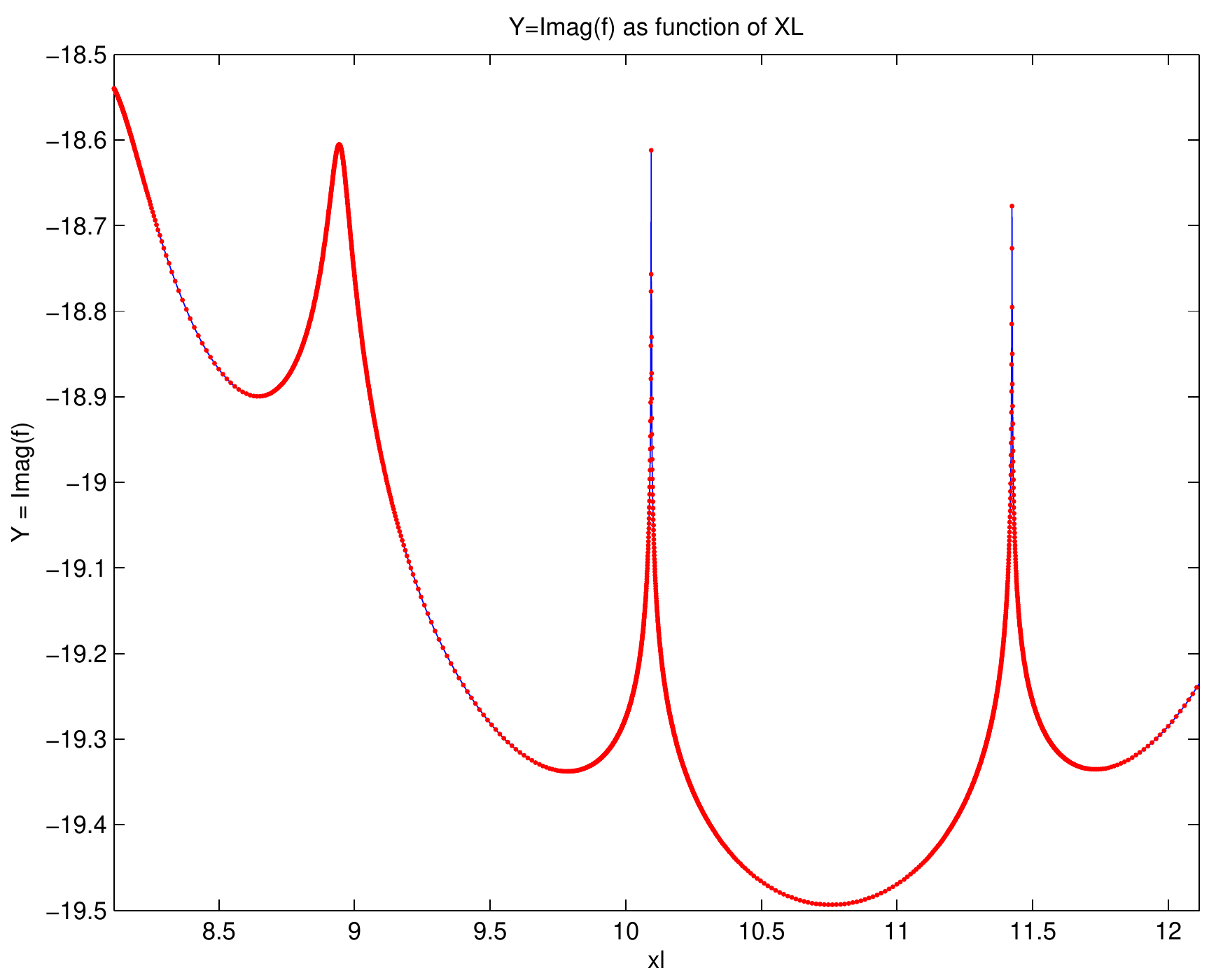}
\caption{Same as Fig\ref{fig100} but here $Y=\mathbf{Im}\left[ f\left( z\right) \right] 
$ is plotted. The quasi-singular behavior has the same origin.}
\label{fig102}
\end{figure}

\begin{figure}[h]
\includegraphics[height=10cm]{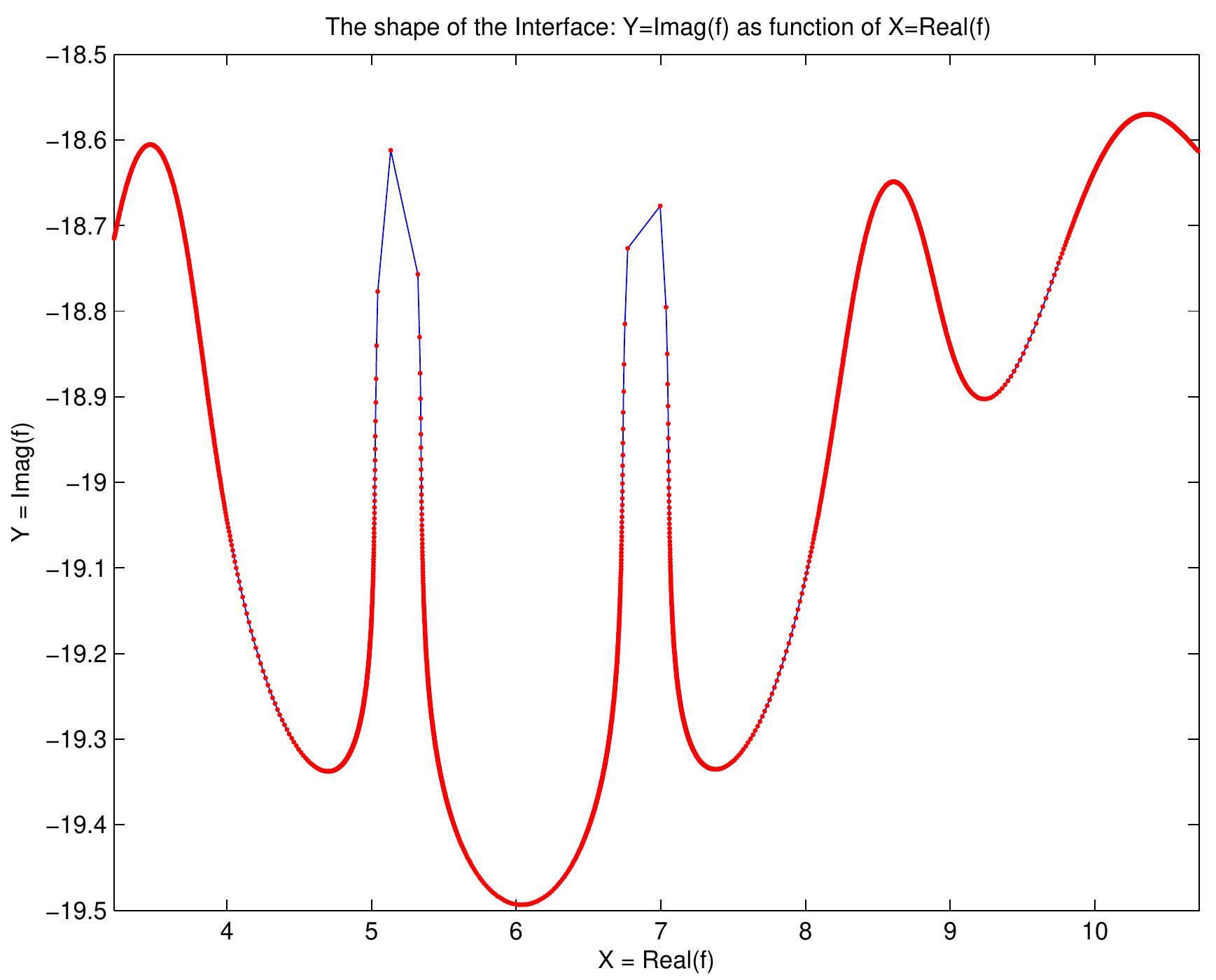}
\caption{A small region of the interface, from the same results as Figs.\ref{fig101}
and \ref{fig102}. Although a high degree of mesh refinement is used, the space variation of the interface is not correctly resolved around the two points shown in Fig.\ref{fig101} and spurious lines are introduced, the same that are seen in Fig.\ref{fig100}.}
\label{fig103}
\end{figure}

\clearpage

\bigskip

\section{Appendix. Wrinkled fronts and cusp singularities} \label{app:appendixb}
\renewcommand{\theequation}{B.\arabic{equation}} \setcounter{equation}{0}

\bigskip

For the examination of the cusp profiles, the equation of Sivashinsky type is solved in terms of a set of singularities (poles). The time dependence of the solution is encoded in the dynamics of the poles. The equations verified by the poles are \cite{procaccia1}, \cite{procaccia2}%
\begin{eqnarray}
-L^{2}\frac{dz_{j}}{dt} &=&\nu \sum\limits_{k=1,k\neq j}^{2N}\cot \left( 
\frac{z_{j}-z_{k}}{2}\right)   \label{b100} \\
&&+i\frac{L}{2}\mathrm{sign}\left[ \mathbf{Im}\left( z_{j}\right) \right]  
\notag
\end{eqnarray}%
Here the poles are counted all, with the first $N$ indices $j=1,...,N$ for
poles and the last $N$ indices $j=N+1,..,2N$ for the conjugated poles.%
\begin{equation}
z_{j+N}=\overline{z_{j}}  \label{b101}
\end{equation}

\bigskip

The equations are written for the real and imaginary parts of the poles%
\begin{equation}
z_{j}\left( t\right) =x_{j}\left( t\right) +iy_{j}\left( t\right)
\label{b102}
\end{equation}%
\begin{eqnarray}
-L^{2}\frac{dx_{j}}{dt} &=&\nu \sum\limits_{k=1,k\neq l}^{N}\sin \left(
x_{j}-x_{k}\right)  \label{b103} \\
&&\times \left[ \frac{1}{\cosh \left( y_{j}-y_{k}\right) -\cos \left(
x_{j}-x_{k}\right) }\right.  \notag \\
&&\left. +\frac{1}{\cosh \left( y_{j}+y_{k}\right) -\cos \left(
x_{j}-x_{k}\right) }\right]  \notag
\end{eqnarray}%
and%
\begin{eqnarray}
L^{2}\frac{dy_{j}}{dt} &=&\nu \sum\limits_{k=1,k\neq j}^{N}\left[ \frac{%
\sinh \left( y_{j}-y_{k}\right) }{\cosh \left( y_{j}-y_{k}\right) -\cos
\left( x_{j}-x_{k}\right) }\right.  \label{b104} \\
&&\left. +\frac{\sinh \left( y_{j}+y_{k}\right) }{\cosh \left(
y_{j}+y_{k}\right) -\cos \left( x_{j}-x_{k}\right) }\right]  \notag \\
&&+\nu \coth \left( y_{j}\right)  \notag \\
&&-L  \notag
\end{eqnarray}

These are the equations that are solved numerically.

\bigskip

\section{Appendix. A note on the discrete model for the breaking of a rising convective column} \label{app:appendixc}
\renewcommand{\theequation}{C.\arabic{equation}} \setcounter{equation}{0}

The coupled lattice map model that is used to represent the physical process of {\it phase competition} is implemented numerically. Essentially the discrete nature of the problem (in particular the representation of the Laplacian operator) induces a certain stability of the phases, examined by Oppo and Kapral \cite{kapral1}. 
On a two-dimensional square lattice we initialize the field in one of the phases and add a small amplitude noise with smooth profile. They are perturbations of Gaussian $2D$ shape, with positions, widths and amplitudes generated randomly. We then start the iteration and note the progress of the phase II into the region occupied initially almost completely by the phase I. 

One easily see formation of a spatial oscillation pattern whose characteristic extension is close to the unit cell of the lattice. This is connected with the choice of the diffusion coefficient and of the time advancement and are natural element of the model. We must however remove it if we want to use the evolving pattern of the phases to measure either the length of the interface or the area of the phases. This is a simple numerical operation but introduces a certain imprecision, which should not affect the application of this iterative discrete model to the problem of loss of compacity of rising convective columns.

One can measure various quantities of interest, like the connectivity: how many compact patches of the initial convective column still exist in a horizontal plane; or, the area occupied by one of the phases and respectively, the length of the perimeter of the patches of one phase. For this we use the {\it contour} function (either Matlab or Fortran) and extract the set of closed curves that are defined by the same level. The dependence with the parameters of the iterative map can be studied, as shown in Fig. \ref{figareaperim}.

\end{appendices}

\bigskip
 \newcommand{\noop}[1]{}

\end{document}